\documentclass[paper]{JHEP3}
\pdfoutput=1
\usepackage{amssymb}
\usepackage{amsmath}
\usepackage{cite}
\usepackage{epsfig}
\def\nc{\newcommand}
\nc{\half}{\frac{1}{2}}
\nc{\shalf}{\ensuremath{\textstyle \frac{1}{2}}}

\nc{\deldag}{\mathbin{\partial\mkern-10.5mu\big/}}
\nc{\deldagss}{\mathbin{\partial\mkern-10.5mu/}}
\nc{\kdag}{\mathbin{k\mkern-10mu\big/}}   
\nc{\skdag}{\mathbin{k\mkern-10mu/}}
\nc{\udag}{\mathbin{u\mkern-10mu\big/}}
\nc{\kdagss}{\mathbin{k\mkern-10mu/}}
\nc{\Pdag}{\mathbin{P\mkern-10mu\big/}}
\nc{\pp}{{\scriptscriptstyle ||}}
\nc{\stwo}{{\scriptscriptstyle 2}}
\nc{\pham}{{\phantom{-}}}

\def\lsim{\mathrel{\raise.3ex\hbox{$<$\kern-.75em\lower1ex\hbox{$\sim$}}}}
\def\gsim{\mathrel{\raise.3ex\hbox{$>$\kern-.75em\lower1ex\hbox{$\sim$}}}}

\def\Slashnew#1{#1\kern-0.55em\raise.05ex\hbox{/}}
\def\slashnew#1{#1\kern-0.5em\raise.05ex\hbox{{$\scriptstyle /$}}}
\def\sfrac#1#2{{\textstyle\frac#1#2}}

\def\ie{{\em i.e.~}}

\def\eg{{\em e.g.~}}

\nc{\beq} {\begin{equation}}
\nc{\eeq} {\end{equation}}
\nc{\beqa}{\begin{eqnarray}}
\nc{\eeqa}{\end{eqnarray}}

\def\Slashnew#1{#1\kern-0.55em\raise.05ex\hbox{/}}
\def\slashnew#1{#1\kern-0.5em\raise.05ex\hbox{{$\scriptstyle /$}}}
\def\emph#1{{\em #1}}
\def\hepph#1{hep-ph/#1}

\def\sfrac#1#2{\ensuremath{\textstyle \frac{#1}{#2}}}

\title{Flavoured quantum Boltzmann equations from cQPA}

\author{Christian Fidler$^a$,
        Matti Herranen$^{a,}$\footnote{Alexander-von-Humboldt Fellow}\,\,,
        Kimmo Kainulainen$^{b,c}$ and
        Pyry Matti Rahkila$^{b,c}$ 
        \\
		\\ $^a$Institut f\"ur Theoretische Teilchenphysik und Kosmologie,
		\\ RWTH Aachen University, D--52056 Aachen, Germany
        \\ $^{b}$ Department of Physics, P.O.~Box 35 (YFL), 
        \\ FIN-40014 University of Jyv\"askyl\"a, Finland, 
        \\ $^{c}$ Helsinki Institute of Physics, P.O.~Box 64, 
  	    \\ FIN-00014 University of Helsinki, Finland.\\
        \\e-mail: \email{fidler@physik.rwth-aachen.de, 
                         herranen@physik.rwth-aachen.de,   
                         kimmo.kainulainen@jyu.fi, 
                         pyry.rahkila@jyu.fi}}

\abstract{
We develop a Boltzmann-type quantum transport theory for interacting fermion and scalar fields 
including both flavour and particle-antiparticle mixing. Our formalism is based on the {\em coherent quasiparticle approximation} 
(cQPA) for the 2-point correlation functions, whose extended phase-space structure contains new spectral shells for flavour- and particle-antiparticle coherence. We derive explicit cQPA propagators and Feynman rules for the transport theory. 
In particular the nontrivial Wightman functions can be written as composite operators $\sim {\cal A} F {\cal A}$, which generalize the usual Kadanoff-Baym ansatz. Our numerical results show that particle-antiparticle coherence can strongly influence CP-violating flavour mixing even for relatively slowly-varying backgrounds. Thus, unlike recently suggested, these correlations cannot be neglected when studying 
asymmetry generation due to time-varying mass transition, for example in electroweak-type baryogenesis 
models. Finally, we show that the cQPA coherence solutions are directly related to squeezed states in the more familiar operator formalism.}

\preprint{TTK-11-35}

\begin{document}

%
\section{Introduction}
%

A new quantum transport formalism for coherent, interacting quantum fields based on a distributive expansion of the dynamical 2-point correlation functions was recently introduced in refs.~\cite{HKR1,HKR2,HKR3,Thesis_Matti,Glasgow,HKR4}. Indeed, it was observed in~\cite{HKR1} that in systems with certain space-time symmetries the information of nonlocal coherence is encoded in new spectral shell solutions for the dynamical 2-point functions at particular off-shell momenta. In the leading-order expansion in gradients this singular structure is called the coherent quasiparticle approximation, or cQPA. It allows writing down a self-consistent network of Boltzmann-like transport equations for moments of the correlation functions, or equivalently, for the phase-space distribution functions.
 
So far a complete formulation of the theory has been given only for non-mixing scalar and fermionic fields~\cite{HKR4}. However, in nature most of the interesting applications involve mixing fields, \ie~fields whose mass-eigenstates do not coincide with interaction eigenstates. This certainly is the case for neutrino oscillations in the early universe~\cite{neutrinos}, electroweak baryogenesis (EWBG)~\cite{EWBG,ClassForce,ClassForceomat,SemiClassSK,PSW,EWBG_mixing,CLR-MT10,CLT11}, models for spontaneous (or coherent) baryogenesis \cite{coherent_bg,HKR4}, and for variants of leptogenesis~\cite{leptogenesis,resonant_lepto,flavour-CTP_lepto,ABDM10}. In this paper we extend our formalism to the case of flavour mixing. The result will be a quantum transport theory including both coherent flavour-mixing and coherent particle-antiparticle-mixing effects in temporally varying backgrounds. In the companion paper \cite{HKR5}, we present a flavour-covariant formulation of the formalism, including also nonzero dispersive self-energy corrections giving rise to true quasiparticle dispersion relations. Other recent works in a similar context include \eg \cite{CLR-MT10,CLT11,Koksma,Serreau}.

In refs.~\cite{HKR1,HKR2} the fermionic transport theory was formulated using a less familiar spin-projected 2-component notation. Scalar fields were considered in~\cite{HKR3} and the need for resummation of the collision integrals involving the coherence functions was pointed out in refs.~\cite{Thesis_Matti,Glasgow}. A complete formulation of the theory in terms of the familiar 4-component Dirac notation and with a compact set of Feynman rules for non-mixing fields was presented in ref.~\cite{HKR4}. While many aspects of the current derivation are similar to the non-mixing case, the added complication of the resulting equations is nontrivial. Also new physical phenomena emerge, such as new mixing sources in the quantum Boltzmann equations (qBE) coming from the gradients of the diagonalizing matrices. As in ref.~\cite{HKR4} we will only consider the spatially homogeneous and isotropic problems in this paper.

We will derive a multi-field extension of the momentum space Feynman rules presented in~\cite{HKR4}. Moreover, we express these rules in a new form which is superior to the old one in clarity and ease of use, also in the case non-mixing fields.  In particular the vertex rules of~\cite{HKR4} involving complicated energy signatures are now absent. The only non-standard Feynman rule concerns the Wightman functions which acquire a composite form: $G^{<,>}_{ab} \sim {\cal A}_aF^{<,>}_{ab}{\cal A}_b$. Here ${\cal A}$ is the spectral function and the effective 2-point interaction vertex $F^{<,>}_{ab}$ encodes all coherence information between the mass-shell states $a$ and $b$ of possibly different flavours and/or particle-antiparticle nature. These rules enable a systematic diagrammatic evaluation of the self-energies and eventually the collision terms of the Boltzmann-like transport equations for flavour-coherent fermionic and scalar fields, valid for arbitrary types of interactions. Whenever a specific model is needed, we will use the Yukawa theory as an example. In the present work, the homogeneous background is always modeled by a time-dependent mass matrix. See ref.~\cite{HKR5} for dispersive self-energy corrections and true time-dependent quasiparticle dispersion relations.

As a numerical application we solve our cQPA induced qBE's for two mixing fermionic fields with a complex, time-varying mass matrix in the presence of decohering interactions.  We study in particular the generation of the particle-antiparticle asymmetry through the CP-violating terms in the mass matrix. 
A similar mixing-scenario for scalar fields was recently considered in ref.~\cite{CLR-MT10}, where the particle-antiparticle mixing solutions were neglected in the qBE's and the resummation procedure of the collision terms was not carried out. Surprisingly, and in contrast with the claim in  ref.~\cite{CLR-MT10}, we find that neglecting the particle-antiparticle correlations is not warranted despite the rapid oscillation frequency $\sim 2\omega$ in this sector, even in the regime of relatively slowly varying mass-profile. We show explicitly that particle-antiparticle mixing leads to a significant contribution to the large momentum modes of the flavour off-diagonal correlator. These contributions persists to late times even though the particle-antiparticle-mixing correlations are present only for a short time during the transition period (when the mass matrix is varying). The integrated contribution from these modes is typically dominant over the asymmetry created without the particle-antiparticle mixing, and therefore these effects should be very important for quantitative calculations of EWBG-type scenarios.

Finally, we show that the singular cQPA coherence solutions are related to squeezed states \cite{squeezing} in the operator formalism language. Indeed, by using a complex scalar field as an example, we first show that flavoured squeezing corresponds to flavoured Bogolyubov transformations (providing a generalization of the standard non-flavoured case) and then that there is a one-to-one correspondence between the on-shell functions $f^m_{ij\pm}$, $f^c_{ij\pm}$ and the independent parameters of the flavoured Bogolyubov transformations. A similar connection between squeezing and coherence was recently found in ref.~\cite{Koksma}, where entropy production in out-of-equilibrium quantum field theory was analyzed. 

This paper is organized as follows. In section~\ref{sec:fermions}, we review the derivation of the fermionic cQPA shell structure and the equations of motion which follow from using the cQPA structure as an ansatz in the full dynamical Kadanoff-Baym equations. We also introduce a convenient parametrization of the flavour-coherent propagators using the familiar 4-component Dirac notation with projection operators. In section~\ref{sec:fermicollision}, we present the resummation method giving rise to a consistent expansion of the fermionic collision integrals, and we give the generic forms of these integrals for spatially homogeneous and isotropic problems with arbitrary types of interactions. In section \ref{sec:scalar} we repeat the analysis for scalar fields, and in section~\ref{sect:effective-feynman} we present the generalized Feynman rules for the computation of the resummed self-energies for the Yukawa-theory. In section \ref{sec:applications}, we present the numerical example of two-fermion mixing, and in section \ref{sec:operator_form} we show the connection between cQPA flavour mixing and squeezing in the operator formalism. Finally, section~\ref{sec:discussion} contains our conclusions and outlook.

%
\section{cQPA for fermions}
\label{sec:fermions}
%

We shall first briefly review the basic theoretical setup here. We will use the Schwinger-Keldysh (or Closed Time Path (CTP)) approach~\cite{SK-formalism} to nonequilibrium quantum field theory (see \eg \cite{CalHu08}). The main quantities of interest for us are the Wightman functions $iS^<(u,v) = \langle \bar\psi(v)\psi(u) \rangle$ and $iS^>(u,v) = \langle \psi(u) \bar\psi(v)\rangle$\footnote{Note that our definition of $S^<$ differs by sign from the more standard convention}, which in the Wigner representation become:
\begin{equation}
S(k,x) \equiv \int {\rm d}^{4} r \, e^{ik\cdot r} S(x + \sfrac{r}{2},x-\sfrac{r}{2}) \,,
\label{wigner1}
\end{equation}
where $x\equiv (u+v)/2$ is the average coordinate, and $k$ is the internal momentum conjugate to the relative coordinate $r\equiv u-v$.  These functions obey the Kadanoff-Baym equations \cite{KadBay62} (see also \cite{HKR2,PSW}):
\begin{equation}
(\kdag + \frac{i}{2} \deldag_x - \hat m_0
       - i\hat m_5 \gamma^5) S^{<,>}
  -  e^{-i\Diamond}\{ \Sigma^h \}\{ S^{<,>} \}
  -  e^{-i\Diamond}\{ \Sigma^{<,>} \}\{ S^h \}
  = \pm {\cal C}_{\rm coll} \,,
\label{DynEqMix}
\end{equation}
where $S^h = S^t - (S^> - S^<)/2$ and $\Sigma^h = \Sigma^t - (\Sigma^> - \Sigma^<)/2$, where $S^t$ and $\Sigma^t$ denote the time ordered Green's function and the corresponding self-energy, and the collision term is given by
\begin{equation}
{\cal C}_{\rm coll} = \frac{1}{2}e^{-i\Diamond}
                             \left( \{\Sigma^>\}\{S^<\} -
                                    \{\Sigma^<\}\{S^>\}\right)\,.
\label{collintegral}
\end{equation}
The $\Diamond$-operator is the following generalization of the
Poisson brackets:
\begin{equation}
\Diamond\{f\}\{g\} = \frac{1}{2}\left[
                   \partial_x f \cdot \partial_k g
                 - \partial_k f \cdot \partial_x g \right]\,,
\label{diamond}
\end{equation}
and the mass operators $\hat m_0$ and $\hat m_5$ are defined as:
\begin{equation}
\hat m_{\rm 0,5} S(k,x) \equiv m_{h,a}(x) e^{-\frac{i}{2}
       \stackrel{\leftarrow}{\partial_x} \cdot \partial_k} S(k,x)\,.
\label{massoperators}
\end{equation}
Here $m_h=(m + m^\dagger)/2$ and $m_a = (m - m^\dagger)/(2i)$ are hermitian and antihermitian parts of a complex and possibly spacetime dependent $N \times N$ mass matrix $m$ in flavour indices. The components of self-energy $\Sigma$ are in general complicated functionals of the correlators $S^{<,>}$, which need to be computed using some truncation scheme, \eg 2PI effective action \cite{2PI} in loop expansion. The spectral function ${\cal A^{\psi}}=\frac{i}{2}(S^>+S^<)$ obeys an almost identical equation to Eq.~(\ref{DynEqMix}), but without the collision term on the RHS, and $\Sigma^{<,>}$ replaced by  $\Gamma = \frac{i}{2}(\Sigma^> + \Sigma^<)$ in the last term on the LHS of Eq.~(\ref{DynEqMix}). Thus, in the non-interacting limit the equations of motion for the spectral function and for the dynamical Wightman functions are the same. However, the spectral function is further subjected to the spectral sum-rule:
\begin{equation}
  \int \frac{{\rm d}k_0}{\pi}
                  {\cal A^{\psi}}(k,x) \gamma^0 = 1\,,
\label{sumrule}
\end{equation}
which follows for example from the equal-time anticommutation relations of the fermionic fields. We will see below that this condition is strong enough to completely determine the form of the free-theory spectral function.

%
\subsection{Approximations}
\label{sec:approx}
%

Compared to standard kinetic approach \cite{kinetic,CalHu08,PSW} the key idea of coherent quasiparticle approximation (cQPA) \cite{HKR1,HKR2,HKR3,Thesis_Matti,Glasgow,HKR4} is to give up the assumption of nearly translation invariant 2-point correlators. This allows new singular shell solutions, which are oscillatory in spacetime at quantum scales $\sim k$, and describe nonlocal quantum coherence between the usual quasiparticle excitations. In order to find cQPA we make the following approximations for the KB-equations (\ref{DynEqMix}):
\begin{itemize}
\item[1) ]
We neglect the terms $\propto S^h$.
\item[2) ]
We neglect the terms $\propto \Sigma^h$. 
\item[3) ] We consider only the spatially homogeneous and isotropic case.
\end{itemize}
The first approximation is made in the standard kinetic theory as well, and it should apply in the limit of weak interactions when the finite (scattering) width effects can be neglected. The second approximation is made here only for simplicity. These terms would give rise to modified dispersion relations for the quasiparticles, but not change the generic structure of the theory. We consider the case with a nonzero dispersive self-energy $\Sigma^h$ in the companion paper \cite{HKR5}. The third approximation with $\vec\partial_x S^{<,>}(k,x) = \vec\partial_x m(x) = 0$ is essential, since the simple spectral structures for the coherence solutions arise only in systems with particular space-time symmetries~\cite{HKR1}. With the assumptions above, the KB-equations (\ref{DynEqMix}) can be reduced and decomposed into the following hermitian (H) and antihermitian (AH) parts:
\begin{eqnarray}
{\rm (H)}:\quad 2k_0 \bar S^< &=& \hat H \bar S^< + \bar S^< \hat H^\dagger + i\gamma^0 ({\cal C}_{\rm coll} - {\cal C}_{\rm coll}^\dagger)\gamma^0\,,
\label{Hermitian22}
\\ 
{\rm (AH)}:\quad i \partial_t \bar S^< &=& \hat H \bar S^< - \bar S^< \hat H^\dagger + i\gamma^0 ({\cal C}_{\rm coll} + {\cal C}_{\rm coll}^\dagger)\gamma^0\,,
\label{AntiHermitian22}
\end{eqnarray}
where we have defined a hermitian Wightman function $\bar S^< \equiv iS^<\gamma^0$ and the operator 
\beq
\hat H \equiv \mathbf{k} \cdot \vec\alpha + \gamma^0 \hat m_0  + i \gamma^0\gamma^5 \hat m_5 \,, 
\eeq
which can be viewed as a local free-field Hamiltonian operator in the presence of time-varying mass. Note that only the AH-equation contains an explicit time derivative acting on $S^<$. Thus Eq.~(\ref{AntiHermitian22}) is mainly controlling the time evolution of the Wightman function $S^<$, and it will be referred to as the {\em ``kinetic equation}''. The H-equation (\ref{Hermitian22}) on the other hand, becomes algebraic in the collisionless limit and in the lowest order of time-gradients. The role of this {\em ``constraint equation}'' will be to restrict the phase space structure of $S^<$.

There are two more approximations to be made before we arrive to the cQPA spectrum and equations of motion:
\begin{itemize}
\item[4) ] We expand the constraint equation (\ref{Hermitian22}) to the  
    zeroth order in the scattering width $\Gamma$ and in the mass gradients $\partial_t m$. This gives rise to the singular cQPA phase-space structure.

\item[5) ] We insert the singular cQPA structure as an ansatz to the dynamical    		 
    equation (\ref{AntiHermitian22}), and expand this equation to the lowest nontrivial order in $\Gamma$ and the mass gradients.\footnote{The actual dimensionless expansion parameters are roughly 
$\partial_t m/\omega^2 $ and $\Gamma / \omega$. See ref.~\cite{HKR5} for further discussion.}
\end{itemize}
The set of approximations 1-5 comprises just the standard approximations in the kinetic theory.  However, in cQPA we do not assume (approximate) translation invariance of the 2-point functions $S(k,t)$. We shall see in the next section how these assumptions give rise to the singular cQPA phase-space structure with the on-shell distribution functions $f^{m <}_{ij h\pm}$ and $f^{c <}_{ij h\pm}$ containing $8 N^2$ independent real components for $N \times N$ mixing in flavour space. When this structure is fed into the dynamical equation~(\ref{AntiHermitian22}) and the equation is integrated over $k_0$, we find a closed set of extended quantum Boltzmann equations for the on-shell functions $f_\alpha \equiv f^{m,c <}_{ij h\pm}$.

%
\subsection{Phase-space shell structure}
\label{sec:shell}
%
%
Let us now analyze the constraint equation (\ref{Hermitian22}) in the approximation defined above, \ie to the lowest order in $\partial_t m$ and neglecting the collision terms, which gives:
\begin{equation}
2k_0 \bar S^< = \{ H , \bar S^<\}\,,
\label{constraint}
\end{equation}
where $\{a,b\} \equiv ab+ba $ is the usual anticommutator and 
\begin{equation}
H = \mathbf{k} \cdot \vec\alpha + \gamma^0 m_h  + i \gamma^0\gamma^5 m_a\,.
\label{Ham_flavour}
\end{equation}
We choose to work in the mass eigenbasis, and so we first diagonalize $m$ by a biunitary transformation:
\begin{equation}
m \to m_{\rm d} = U m V^\dagger\,,
\end{equation}
where $U$ and $V$ are the unitary matrices that diagonalize the product matrices $mm^\dagger$ and $m^\dagger m$, respectively, chosen such that $m_{\rm d}^\dagger = m_{\rm d}$ have real mass eigenvalues $m_i$.  With these definitions the 
direct product matrix (here $\otimes$ separates the Dirac and flavour indices) 
\begin{equation}
Y = P_L \otimes U + P_R \otimes V , 
\label{mixing_matrix}
\end{equation}
diagonalizes the Hamiltonian:
\begin{equation}
H_{\rm d} = Y H Y^\dagger = \mathbf{k} \cdot \vec\alpha + \gamma^0 m_{\rm d}\,.
\label{eq:diaghamil}
\end{equation}
Multiplying Eq.~(\ref{constraint}) from the left by $Y$ and from the right by $Y^\dagger$ we then get an identical equation in the mass eigenbasis:
\begin{equation}
2k_0 \bar S^<_{\rm d} = \{H_{\rm d} , \bar S^<_{\rm d} \} \,,
\quad {\rm where} \quad \bar S^<_{\rm d} = Y \bar S^< Y^\dagger \,.  
\label{constraint2}
\end{equation}
Next, we notice that the most general spatially homogeneous and isotropic 2-point function $\bar S^<_{\rm d}(k,t)$ can be parametrized in terms of helicity projectors: $P_h = \frac12(1 + h\,\mathbf{\hat k} \cdot \gamma^0 \vec \gamma \gamma^5)$ with $\mathbf{\hat k} \equiv \mathbf{k}/|\mathbf{k}|$ and $h = \pm 1$, as
\begin{equation}
\bar S^<_{\rm d}(k,t)  = \frac12 \sum_h P_h \big[g_{h0} - h \mathbf{\hat k} \cdot \vec\alpha\,g_{h3} + \gamma^0 g_{h1} -i\gamma^0\gamma^5 g_{h2}\big]\,,
\label{bloch_prop}
\end{equation}
where $g_{h\alpha}(k,t)$ are hermitian $N \times N$ matrices in flavour indices. This is a convenient parametrization for the problem, because the helicity operator commutes with the Hamiltonian $H$ (and with the transformation matrix $Y$), implying that helicity is conserved in a collisionless theory. Using the chiral representation of the Dirac algebra: $\gamma^0 = \sigma^1 \otimes 1$, $\vec \gamma = i\sigma^2 \otimes \vec \sigma$ and $\gamma^5 = -\sigma^3 \otimes 1$, the helicity block-diagonal Wightman function (\ref{bloch_prop}) becomes
\begin{equation}
  \bar S^<_{\rm d}  = \sum_h g_{h}^< \otimes 
  \frac12(1 + h \mathbf{\hat k}\cdot \vec \sigma)\,,\qquad {\rm with} \qquad 
g^<_h \equiv \frac12 \left(g_{h0} +  \vec g_{h} \cdot \vec\sigma \right)\,,
  \label{chiral_basis}
\end{equation}
corresponding to the standard Bloch representation for the chiral part $g^<_{h}(k,t)$.
Inserting this decomposition into the constraint equation (\ref{constraint2}) gives two (for $h=\pm 1$) homogeneous matrix equations for $g_{h\alpha}$ :
\begin{equation}
\sum_\beta (B_{h,\alpha\beta})_{ij}\,(g_{h\beta})_{ij} = 0\,,
    \qquad\quad 
    (B_h)_{ij} = \left( \begin{array}{cccc}
    k_0  &  h|\mathbf{k}|  &  -\bar m_{ij}  &  0 \\
    h|\mathbf{k}| &  k_0   &  0  & -i\Delta m_{ij} \\
    -\bar m_{ij} &  0  &  k_0   &  0   \\
    0  & i\Delta m_{ij} &  0     &  k_0 
    \end{array} \right)\,,
\label{bloch_constraint}
\end{equation}
where the dependence on the Bloch indices (index ordering is here defined as $\alpha, \beta = 0,3,1,2$) is written as an explicit $4 \times 4$ matrix, and we denote $\bar m_{ij} \equiv (m_i + m_j)/2$ and $\Delta m_{ij} \equiv (m_i - m_j)/2$. These equations may have nonzero solutions only if the determinant of $(B_h)_{ij}$ vanishes. This $4\times4$-determinant is easily evaluated for each flavour element $ij$, giving rise to $N^2$ independent constraints:
\begin{equation}
\det\big[(B_h)_{ij}\big] = (k_0^2 - {\mathbf{k}}^2 - M_{ij}^2)k_0^2 + (\Delta m_{ij} \bar m_{ij})^2 = 0\,,
\label{constraint_det}
\end{equation}
where we further denote $M_{ij}^2 \equiv (m_i^2 + m_j^2)/2$. The roots of Eq.~(\ref{constraint_det}) define the nontrivial dispersion relations:
\begin{equation}
  k_0 = \pm \frac12(\omega_i + \omega_j) \equiv \pm\bar\omega_{ij} \,,
  \qquad {\rm or} \qquad 
  k_0 = \pm \frac12(\omega_i - \omega_j) \equiv \pm\Delta\omega_{ij}\,,
\label{dispersion}
\end{equation}
where $\omega_i \equiv ({\mathbf{k}}^2 + m_i^2)^{1/2}$. These conditions give rise to a {\em singular shell structure}: $g_{h\alpha} \sim \delta(k_0\mp\bar\omega_{ij})$ or $g_{h\alpha} \sim \delta(k_0\mp\Delta\omega_{ij})$. The exact matrix form of the singular solutions can then be worked out by using the matrix equation (\ref{bloch_constraint}). We will consider flavour diagonals and off-diagonals separately. 

%
\subsubsection{Flavour diagonals}  
%

The analysis for flavour diagonals is equivalent to the single-flavour case, considered in ref.~\cite{HKR4}. Here we present it for completeness, and we also choose a different normalization for the mass-shell distribution functions for later convenience.  
For $i=j$ the dispersion relation (\ref{dispersion}) reduces to $k_0=\pm \omega_i$ and a double root at $k_0=0$. For $k_0=\pm \omega_i$ Eq.~(\ref{bloch_constraint}) gives relations (we suppress the flavour indices for clarity in the intermediate steps):
\begin{equation}
g_{h3}  = -h\frac{|\mathbf{k}|}{k_0} g_{h0}\,, \qquad 
g_{h1} =   \frac{m_i}{k_0} g_{h0} \,, \qquad 
g_{h2} = 0\,.
\label{ms_relations}
\end{equation}
The remaining equation for $g_{h0}$ then becomes just $(k_0^2 - {\mathbf{k}}^2 - m_i^2)g_{h0} = 0$, which has the spectral solution:
\begin{equation}
(g_{h0})_{ii} = 2\pi \sum_\pm \pm \frac{m_i}{\omega_i} f^{m <}_{iih\pm} \delta(k_0 \mp \omega_i)\,.
\end{equation}
For $k_0 = 0$ we get the conditions
\begin{equation}
g_{h3} = h\frac{m_i}{|\mathbf{k}|} g_{h1}\,, \qquad 
g_{h0} = 0\,,
\end{equation}
while $g_{h1}$ and $g_{h2}$ are free functions at $k_0 =0$. This solution can then be parametrized as
\begin{equation}
(g_{h1})_{ii} = 2\pi \frac{{\mathbf{k}}^2}{\omega_i^2} 
                \sum_\pm f^{c <}_{iih\pm} \delta(k_0)\,,        \qquad 
(g_{h2})_{ii} = 2\pi \frac{ih|\mathbf{k}|}{\omega_i}
                \sum_\pm \pm f^{c <}_{iih\pm} \delta(k_0)\,,
\label{coherence_onshell}
\end{equation}
where the new subscript $c$ refers to coherence. The reason for this particular parametrization is that in this way the on-shell functions $f^{c<}_{iih\pm}(\mathbf{k},t)$ will satisfy simple oscillatory zeroth-order equations of motion. Using Eq.~(\ref{bloch_prop}) and Eqs.~(\ref{ms_relations}-\ref{coherence_onshell}) the combined $4\times4$-spinor matrix solution is finally found to be (an identical expression holds for $S^>_{ii}$)
\begin{equation}
iS^<_{ii}(k,t) = 2\pi\sum_{h\pm} \pm \frac{1}{2\omega_i}P_h (\kdag_{i\pm} + m_i)\Big[ \pm \frac{m_i}{\omega_i}
 f^{m <}_{iih\pm}\delta(k_0 \mp \omega_i) + \Big(\gamma^0 \mp \frac{m_i}{\omega_i}\Big)f^{c <}_{iih\pm}\delta(k_0)\Big]\,,
\label{full_wightman_deg}
\end{equation}
where we defined $k^{\mu}_{i\pm} \equiv (\pm\omega_i,\mathbf{k})$. We have dropped the index $d$, with the understanding that we are always working in the mass eigenbasis when the flavour-indices $i,j$ are explicitly written down. The first term in the brackets is the standard mass-shell solution with dispersion relation $k_0 = \pm \omega_i$, while the second term represents the new $k_0=0$-shell coherence solutions. This solution corresponds to the propagator for the single flavour case derived in ref.~\cite{HKR4}, except here we have used a non-standard normalization for the mass-shell distribution functions $f^{m <}_{iih\pm}(\mathbf{k},t)$ for later convenience. These distribution functions correspond to the phase-space densities for particles and antiparticles with mass $m_i$ and helicity $h$, via Feynman-Stuckelberg interpretation:
\begin{equation}
n_{i\mathbf{k} h}(t) = \frac{m_i}{\omega_i} f^{m<}_{iih+}(\mathbf{k},t)\,,\qquad 
\bar n_{i\mathbf{k} h}(t) = 1+\frac{m_i}{\omega_i}f^{m<}_{iih-}(\mathbf{k},t)\,,
\label{part-number}
\end{equation}
while the new $k_0=0$-shell functions $f_{iih \pm}^{c <}(\mathbf{k},t)$ measure nonlocal quantum coherence between the (flavour-diagonal) positive an negative energy modes. Finally, we observe that in the limit of thermal equilibrium: $\pm \tfrac{m_i}{\omega_i} f_{iih\pm}^{m <} \to f_{\rm eq}(\pm \omega_i)= (e^{\pm \beta \omega_i} + 1)^{-1}$ and $f_{iih\pm}^{c <} \to 0$, Eq.~(\ref{full_wightman_deg}) reduces to the standard thermal correlator:
\begin{equation}
iS^<_{ii,{\rm eq}}(k) = 
       2\pi\,{\rm sgn}(k_0) (\kdag +  m_i) \,
        f_{\rm eq}(k_0) \,\delta(k^2 - m_i^2)\,.
\label{thermal_full}
\end{equation}
%

%
\subsubsection{Flavour off-diagonals}
%

By assuming $m_i \neq m_j$\footnote{We will find below that the limit of degenerated masses: $m_i \to m_j$, is well defined and smooth.} the dispersion relations (\ref{dispersion}) give unique solutions for equation (\ref{bloch_constraint}), because now one can always divide by $k_0 \neq 0$. Equation (\ref{ms_relations}) then generalizes to:
\begin{equation}
g_{h3} = -h\frac{k_0^2-{\bar m_{ij}}^2}{|\mathbf{k}|k_0} g_{h0}\,, \quad 
g_{h1} = \frac{\bar m_{ij}}{k_0} g_{h0}\,, \quad 
g_{h2} = ih\frac{\Delta m_{ij}(k_0^2-{\bar m_{ij}}^2)}{|\mathbf{k}|k_0^2}
           g_{h0}\,.
\label{relations_nondeg}
\end{equation}
The function $g_{h0}$ now gets support on all branches of the dispersion relations (\ref{dispersion}). It can be parametrized as
\begin{equation}
(g_{h0})_{ij} = \pi \sum_{\pm} \pm \Big[\Big( \frac{m_i}{\omega_i} + \frac{m_j}{\omega_j}\Big) 
f^{m <}_{ijh\pm} \delta(k_0 \mp \bar\omega_{ij}) + \Big( \frac{m_i}{\omega_i} - \frac{m_j}{\omega_j}\Big)f^{c <}_{ijh\pm}
\delta(k_0\mp \Delta\omega_{ij})\Big]\,,
\label{eq:gh0mix}
\end{equation}
where the particular normalizations were chosen to facilitate the comparison with the diagonal limit. Inserting the results (\ref{relations_nondeg}-\ref{eq:gh0mix}) back into Eq.~(\ref{bloch_prop}) we find the full $4\times 4$-spinor solution to be (again an identical expression can be written for $S^>_{ij}$):
\begin{eqnarray}
iS^<_{ij}(k,t) &=& 
2\pi \sum_{h\pm} \frac{1}{4\omega_i \omega_j} P_h (\kdag_{i\pm} + m_i) \Big[(\kdag_{j\pm} + m_j) f^{m<}_{ijh\pm} \delta(k_0 \mp \bar\omega_{ij})
\nonumber\\
&&\qquad\qquad\qquad\qquad\qquad\;\;\; -\,(\kdag_{j\mp} + m_j) f_{ijh\pm}^{c <} \delta(k_0 \mp \Delta\omega_{ij})\Big]\,.
\label{full_wightman_nondeg}
\end{eqnarray}
\begin{figure}[t]
\centering
\includegraphics[width=0.8\textwidth]{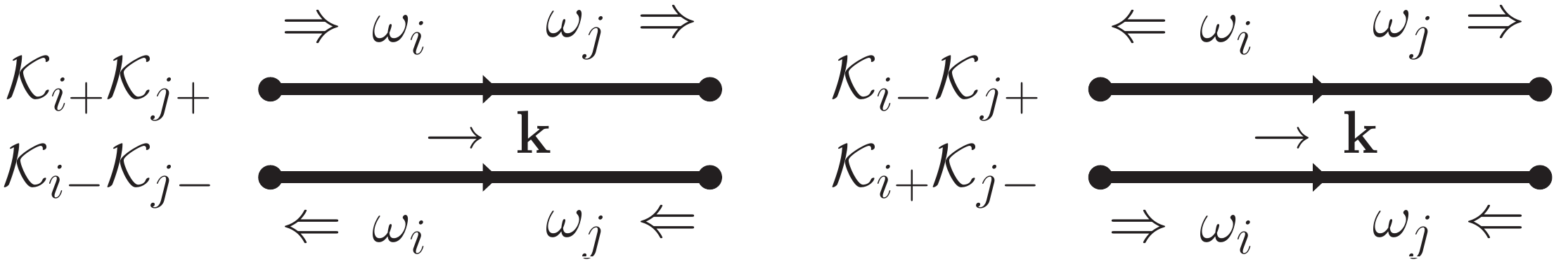}
    \caption{Graphical representation of the Dirac structures of different energy-components of the Wightman functions $iS^{<,>}_{ij}(k,t)$. Three-momentum and fermion number flow from left to right.
    Arrows below and above the fermion lines represent on-shell energies. Here ${\cal K}_{i\pm} \equiv \kdag_{i\pm} + m_i$ and all normalizations have been omitted.}
    \label{fig:Kij:s}
\end{figure}
The spinor structure of the propagator (\ref{full_wightman_nondeg}), illustrated in Fig.~\ref{fig:Kij:s}, is very suggestive: in addition to explicit helicity projectors $P_h$ it consists of projection operators $\kdag_\pm + m$ on states with positive and negative energies for both indices $i$ and $j$. It is clear that $S^<_{ij}$ mixes different energy and flavour eigenstates. The on-shell distributions $f^{m<}_{ijh\pm}$ with $i\neq j$ and $f^{c<}_{ijh\pm}$ are then naturally interpreted as describing the amount of flavour and particle-antiparticle coherence between the mass eigenstates with energies $\pm \omega_i$ and $\pm \omega_j$, respectively. 
In what follows, we shall use the following naming convention for the different contributions to the cQPA Wightman function $S^<$:
\begin{table}[h]
\begin{tabular}{l l}
 \quad\qquad ``mass-shell"   &                                     
      $f^{m<}_{iih\pm}$, \;\; $k_0 = \pm \omega_{i}$ \\
 \quad\qquad ``particle or antiparticle flavour-coherence"  &
      $f^{m<}_{ijh\pm}$, \;\; $k_0 = \pm \bar \omega_{ij}$,\; $i\neq j$ \\ 
 \quad\qquad ``particle-antiparticle coherence"                 &
      $f^{c<}_{iih\pm}$, \;\; $k_0 = 0$ \\
 \quad\qquad ``particle-antiparticle flavour-coherence"         &
      $f^{c<}_{ijh\pm}$, \;\; $k_0 = \pm\Delta \omega_{ij}$,\; $i\neq j$ 
\end{tabular}
\end{table}

\begin{figure}[t]
\centering
\includegraphics[width=0.9\textwidth]{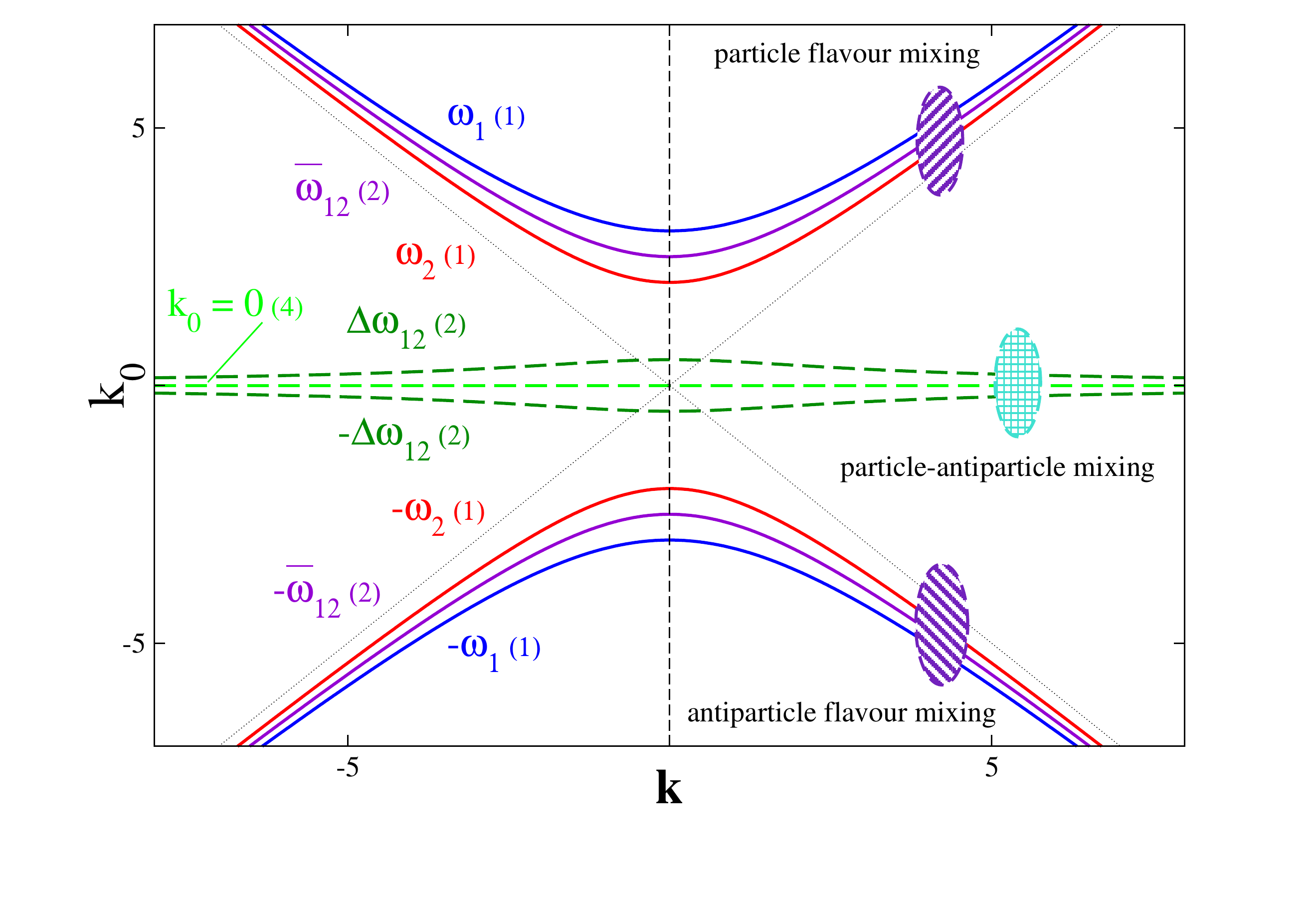}
\vskip-0.7truecm
    \caption{Shown is the cQPA shell structure for the case of two-flavour mixing. Heavy state 1 (blue) has mass $m_1= 3$ (in arbitrary units) and light 2 (red) $m_2=2$. Each curve is labeled by its energy eigenvalue and all particle-antiparticle coherence solutions are shown with dashed (green) lines. The number in parenthesis beside each eigenvalue gives the (real) degeneracy of the corresponding solution.}
    \label{fig:shells}
\end{figure}

\noindent The complete dispersion relations corresponding to this shell structure are illustrated in Fig.~\ref{fig:shells} for the case of two-flavour mixing. Upper and lower blobs on the right side indicate usual flavour-mixing scenario for particles and antiparticles, including the mass-shell states and the particle (or antiparticle) flavour-coherence states. Middle blob corresponds to particle-antiparticle flavour-coherence states.
In section \ref{sec:operator_form} we show how the different coherence solutions emerge from the operator formalism.

Finally, we note that in the limit of mass degeneracy: $m_i \to m_j$, the solution (\ref{full_wightman_nondeg}) actually reduces to the flavour-diagonal propagator (\ref{full_wightman_deg}). The limit of the propagator is thus correct despite the fact that the first step in the derivation is not a legitimate operation for $k_0 = 0$. Hence, Eq.~(\ref{full_wightman_nondeg}) is the complete solution for the Wightman function $S^<_{ij}(k,t)$ in the cQPA for all energies and all flavour indices. 
Now also the reason for choosing the particular normalizations is evident. Furthermore, the hermiticity of $\bar{S}^{<}_{ij}(k,t)$ implies that the complete flavour matrices obey the hermiticity conditions
\begin{equation}
f^{m<\dagger}_{h\pm} = f^{m<}_{h\pm} \quad {\rm and} \quad f_{h\pm}^{c < \dagger} = f_{h\mp}^{c <} \,.
\label{eq:hermiticitycond}
\end{equation} 
%

%
\subsection{Spectral function and the pole propagators}
%

The spectral function ${\cal A}^{\psi}=\frac{i}{2}(S^>+S^<)$ satisfies an identical constraint equation to Eq.~(\ref{constraint}) and thus the solution for a mass basis $ij$-element is identical to Eq.~(\ref{full_wightman_nondeg}). However, the spectral function must in addition obey the sum rule (\ref{sumrule}), which in the mass basis reads:
\begin{equation}
  \int \frac{{\rm d}k_0}{\pi}
                  {\cal A}^{\psi}_{ij}(k,t) \gamma^0 = \delta_{ij}\,.
\label{sumrule_mass}
\end{equation}
This additional constraint fixes the spectral on-shell functions to $f^{m \cal A}_{ijh\pm} =  \pm \frac{\omega_i} {2 m_i} \delta_{ij}$ and $f^{c \cal A}_{ijh\pm} = 0$, such that the spectral function reduces to the usual vacuum expression:
\begin{equation}
 {\cal A}^{\psi}_{ij}(k,t) \equiv \pi\,{\rm sgn}(k_0) (\kdag + m_i)
            \delta(k^2 - m_i^2)\delta_{ij} \,.
\label{full_spectral}
\end{equation}
This result emphasizes the fact that quantum coherence is a {\em dynamical} phenomenon; it is entirely contained within dynamical correlation functions, while the static phase space structure of mass eigenstates defined by the spectral function, remains unchanged. This division is crucial also for understanding why the (flavoured) {\em Kadanoff-Baym ansatz}:
\begin{equation}
iS^<_{ij}(k,t) \propto 2 f^{m<}_{ik}(k,t) {\cal A}^{\psi}_{kj}(k,t)\,,   
\label{kadanoffbaym}
\end{equation}
which is typically assumed in kinetic theory, completely fails in describing nonlocal quantum coherence. Indeed, the KB-ansatz explicitly forces $iS^<_{ij}$ to have the same phase-space structure as the spectral function, setting all coherence functions $f^{c <}_{ijh}$ and $f^{m <}_{i\neq jh}$ to zero. This assumption is simply not warranted in general, because $S^<_{ij}$ is in no way subjected to the spectral sum rule.  However, the relation $-2i{\cal A}^{\psi}=S^>+S^<$ together with expressions (\ref{full_wightman_nondeg}) and (\ref{full_spectral}) does give additional constraints between the distribution functions:
\begin{equation}
f_{ijh\pm}^{m>} = 
       \pm \frac{\omega_i} {m_i}\delta_{ij} - f^{m<}_{ijh\pm}\,,\qquad\quad 
f^{c>}_{ijh\pm} = -f^{c<}_{ijh\pm}\,.
\end{equation}
These conditions, unlike the Kadanoff-Baym ansatz, do not exclude the coherence solutions, but merely express $S^>_{ij}$ in terms of $S^<_{ij}$.

Let us conclude this section by noting that the retarded and advanced propagators are directly related to the spectral function: $S^{r,a}(u,v) = \mp 2i\,\theta\big[\pm(u_0-v_0)\big]{\cal A}(u,v)$, and so, in this approximation scheme, they are given by the standard vacuum expressions ($\epsilon >0$):
\begin{equation}
S^{r,a}_{ij}(k,t) = \frac{\kdag +  m_i}{k^2 - m_i^2 \pm i{\rm sgn}(k_0) \epsilon}\delta_{ij}\,.
\label{pole-props}
\end{equation}
However, the Feynman and anti-Feynman propagators have nonvanishing coherence parts, which are explicitly given by the relations
\begin{eqnarray}
S^t_{ij}(k,t) &=& S^r_{ij}(k,t) - S^<_{ij}(k,t)\,,
\nonumber\\
S^{\bar t}_{ij}(k,t) &=& -S^a_{ij}(k,t) - S^<_{ij}(k,t)\,.
\label{full_feynman}
\end{eqnarray}
Based on Eqs.~(\ref{full_spectral}-\ref{full_feynman}) we conclude that the on-shell functions $f^{m,c <}_{ijh\pm}$ form a complete set of dynamical variables for the cQPA, and we will only need the kinetic equation for $S^<$ to derive a closed set of equations of motion for them. From now on we will set $f^{m,c <}_{ijh\pm}\to f^{m,c}_{ijh\pm}$ whenever there is no risk of confusion.

%
\section{Equations of motion for fermions}
%

To derive the equations of motion for the on-shell functions $f^{m,c}_{ijh\pm}$, we need to transform the kinetic equation (\ref{AntiHermitian22}) to the mass-diagonal basis. Because the mixing matrices $U$ and $V$ are in general time dependent, this procedure introduces derivative terms:
\begin{equation}
Y (\partial_t \bar S^<) Y^\dagger  
  = \partial_t \bar S^<_{\rm d} - i[\Xi^\prime\,,\bar S^<_{\rm d}] 
  \equiv D_t \bar S^<_{\rm d}\,,
\label{eq:DDerivative}
\end{equation}
where $\Xi^\prime(t) \equiv iY(t)\partial_tY^\dagger(t)$; we use the explicit superscript $^\prime$ to indicate that $\Xi^\prime$ is proportional to the time derivatives of mass matrix $m$.  Using this notation we find the transformed kinetic equation for $ij$-element of the mass basis (we again drop the explicit $d$-indices): 
\begin{equation}   
D_t \bar S^<_{ij} = -i\big(\hat H_i \bar S^<_{ij} - \bar S^<_{ij} \hat H_j^\dagger\big) + \gamma^0 ({\cal C}_{ij} + {\cal C}_{ij}^\dagger)\gamma^0\,,
\label{eq:ekamultikinetic}
\end{equation}
where ${\cal C}_{ij} \equiv (X{\cal C}_{\rm coll}X^\dagger)_{ij}$ with $X \equiv \gamma^0 Y \gamma^0$, and the mass eigenbasis Hamiltonian 
\begin{equation}
\hat H_i = \mathbf{k} \cdot \vec\alpha + \gamma^0 m_i e^{-\frac{i}{2} \stackrel{\leftarrow}{D_t}\,\partial_{k_0}} 
\end{equation}
now involves the covariant derivative $D_t$ in the exponent instead of the standard $\partial_t$. We now insert the spectral solution (\ref{full_wightman_nondeg}) into Eq.~(\ref{eq:ekamultikinetic}) and integrate over $k_0$, which gives
\begin{equation}   
\partial_t \bar{{\cal S}}^<_{ij} 
 = - i[H_{\rm eff} ,\bar{\cal{S}}^<]_{ij} 
 + \gamma^0 \langle{\cal C}_{ij} + {\cal C}_{ij}^\dagger\rangle\gamma^0\,,
\label{eom_prop}
\end{equation}
where we denote $\langle\;\cdots\;\rangle \equiv \int \frac{{\rm d} k_0}{2\pi}\,(\;\cdots\;)$, and we defined $\langle S \rangle \equiv {\cal S}$ and
\begin{equation}   
(H_{\rm eff})_{ij} \,\equiv\, ({\bf k}\cdot \vec\alpha + \gamma^0 m_i)\delta_{ij} - \Xi'_{ij} \, = \, H_i\delta_{ij} - \Xi'_{ij}\,.
\label{eq:heff}
\end{equation}
Note that the $k_0$-derivatives in $\hat H_i$ have now vanished as total derivatives, reducing $\hat H_i$ to the familiar diagonal form given by Eq.~(\ref{eq:diaghamil}). The explicit form of the integrated Wightman function is:
\begin{equation}
\bar{{\cal S}}^<_{ij} 
= \sum_{h\pm} P_h P_{i\pm} \gamma^0
     \Big( P_{j\pm} f^m_{ijh\pm}  +  P_{j\mp} f^c_{ijh\pm}\Big) 
\equiv \sum_{\pm} 
    \big( \bar{{\cal S}}^{m <}_{ij\pm} + \bar{{\cal S}}^{c <}_{ij\pm} \big)\,,
\label{int_wightman}
\end{equation}
where we have defined positive- and negative-energy projectors onto mass eigenstates:
\begin{equation}
P_{i\pm} \equiv \frac12\Big(1 \pm \frac{H_i}{\omega_i}\Big)\,.
\end{equation}
Using the identity $H_i^2 = \omega_i^2$ it is easy to verify that $P_{i\pm}$'s are projectors obeying:
\begin{equation}
P_{i\pm}^2 = P_{i\pm}, \qquad  P_{i+}P_{i-} = P_{i-}P_{i+} = 0
\quad {\rm and} \quad
H_i P_{i\pm} = \pm\omega_i P_{i\pm}.
\label{proj_identity}
\end{equation}

%
\subsection{Flavoured quantum Boltzmann equations}
%

Equations (\ref{eom_prop}-\ref{eq:heff}) are in many ways the simplest and most compact form of the kinetic equations in the cQPA scheme. However, it is also useful to derive explicit equations for the on-shell functions. The easiest way to do this is to first solve $f^{m,c}_{ijh \pm}$ from Eq.~(\ref{int_wightman}) by taking  projections and tracing over Dirac indices:
\begin{eqnarray}
f^m_{ijh\pm} &=& N^m_{ij}{\rm Tr}\Big[P_{j\pm}\gamma^0 P_{i\pm} P_h \bar{{\cal S}}^<_{ij} \Big]\,,
\nonumber\\[2mm]
f^c_{ijh\pm} &=& N^c_{ij}{\rm Tr}\Big[P_{j\mp} \gamma^0 P_{i\pm} P_h \bar{{\cal S}}^<_{ij} \Big]\,,
\label{on-shell_rel}
\end{eqnarray}
where the normalization constants are
\begin{equation}
N^m_{ij} = \frac{\omega_i\omega_j}{\Omega^2_{mij}} 
\quad {\rm and} \quad
N^c_{ij} = \frac{\omega_i\omega_j}{\Omega^2_{cij}}    \,,
\end{equation}
and we defined
\begin{eqnarray}
\Omega^2_{mij} &\equiv&   
\frac{1}{2}(\omega_i\omega_j - \mathbf{k}^2 + m_im_j) 
= \bar m^2_{ij} - \Delta \omega_{ij}^2 
\,, \\
\Omega^2_{cij} &\equiv&   
\frac{1}{2}(\omega_i\omega_j + \mathbf{k}^2 - m_im_j)
= \bar \omega^2_{ij} - \bar m^2_{ij} 
\,. 
\end{eqnarray}
The equations of motion for $f^{m,c}_{ijh \pm}$ can now be obtained by taking time-derivatives of Eqs.~(\ref{on-shell_rel}), and then using the kinetic equation (\ref{eom_prop}) for $\partial_t \bar{{\cal S}}^<_{ij}$ in the trace, the projection identities (\ref{proj_identity}), as well as the result $\partial_t H_i = \gamma^0 m_i^\prime $, and finally computing the traces of Dirac algebra. 
It is convenient to define the following generalized (anti)commutator operations in flavour space:
\begin{eqnarray}
&\big[O , f_{s}  \big]^m  \equiv O f_{s}  - ( O f_{s } )^{\dagger} \,, 
\qquad &\{O , f_{s} \}^m   \equiv O f_{s} + ( O f_{s} )^{\dagger} \,, 
\nonumber\\
&\big[O , f_{s}  \big]^c  \equiv O f_{s}  - ( O f_{-s} )^{\dagger} \,, 
\qquad &\{O , f_{s} \}^c      \equiv O f_{s}  + (O f_{-s} )^{\dagger} \,,
\label{gen_comm3}
\end{eqnarray}
where $O$ and $f_s$ are matrices in flavour indices with a (possible) dependence on a generic energy index $s = \pm$.\footnote{Note that $\tfrac{1}{2} \{O, f \}^m$ is just the hermitian part of matrix $O f$ and $-\tfrac{i}{2}  [ O, f ]^m$ is the anti-hermitian part of same matrix. The same interpretation does not apply for the $[ O, f ]^c$ and $\{ O, f \}^c$ constructs because of the change of the energy index in the second term.} Using this notation we find the following coupled set of equations of motion for $f^{m,c}_{ijh \pm}$:
\begin{eqnarray}
 \partial_t  f^m_{ijh\pm}  
  &=&  \mp i 2 \Delta \omega_{ij}  f^m_{ijh\pm } 
 - \frac{ \mathbf{k}^2\bar m_{ij} }{\Omega^2_{mij} } 
 \{ \sfrac{m^\prime}{\omega^2}, f^-_{h\pm} \}^m_{ij}   
 + i X^m_{h\pm}[f]_{ij} + {\cal C}^m_{h\pm}[f]_{ij} \,,
\label{eom_on-shell1}
\\[4mm]
\partial_t f^c_{ijh\pm} 
  &=&  \mp i 2 \bar \omega_{ij} f^c_{ijh\pm } 
 - \frac{ \mathbf{k}^2\bar m_{ij} }{\Omega^2_{cij} } 
 \{ \sfrac{m^\prime}{\omega^2} , f^-_{h\mp} \}^c_{ij}   
 + i X^c_{h\pm}[f]_{ij} + {\cal C}^c_{h\pm}[f]_{ij} \,, 
\label{eom_on-shell2}
\end{eqnarray}
where the $X^{m,c}_{h\pm}[f]$-terms involving all flavour-mixing gradients are given by
\begin{eqnarray} 
X^{m}_{h\pm}[f]_{ij} &\equiv& 
       \big[\Xi'^{+} , f^+_{h\pm} \big]^m_{ij} 
\pm h \, \big[ \Xi'^{-}  , v_{{\bf k}}\, f^-_{h\pm}  \big]^m_{ij}   
\nonumber\\
&-&\,  \frac{|\mathbf{k}|\Delta\omega_{ij}}{\Omega^{2}_{m ij}} 
\Big(\{ \Xi'^{+} ,v_{{\bf k}} \, f^-_{h\pm} \}^m_{ij} \pm  h \, \{ \Xi'^{-} ,  f^+_{h\pm} \}^m_{ij}   
\Big) \phantom{mi}
\nonumber \\
&+&\, \frac{\bar \omega_{ij}}{\bar m_{ij}} 
\big[ \Xi'^{+} , \tfrac{m}{\omega} \, f^-_{h\pm}  \big]^m_{ij}
\pm h \, \frac{|\mathbf{k}|\bar m_{ij}}{\Omega^{2}_{m ij}} \big[   \Xi'^{-} , \tfrac{m}{\omega} \, f^-_{h\pm} \big]^m_{ij}
\label{f_m_mixing_gradients}
\end{eqnarray}
and
\begin{eqnarray} 
X^{c}_{h\pm}[f]_{ij} &\equiv&
\big[\Xi'^{+} , f^+_{h\mp} \big]^{c}_{ij} 
\mp h \, \{ \Xi'^{-} , v_{{\bf k}} \, f^-_{h\mp} \}^{c}_{ij}
\nonumber\\
&-&\,  \frac{|\mathbf{k}| \bar\omega_{ij} }{\Omega^{2}_{c ij}} 
\Big( \big[ \Xi'^{+} , v_{{\bf k}} \, f^-_{h\mp} \big]^{c}_{ij}  \mp h \,  \{ \Xi'^{-} , f_{h\mp}^+ \}^{c}_{ij}
\Big) \,,\phantom{i}
\nonumber \\
&-&\, \frac{\Delta \omega_{ij} }{\bar m_{ij} } \{  \Xi'^{+} ,\tfrac{m}{\omega} \, f^-_{h\mp} \}^{c}_{ij} 
\pm h \,  \frac{|\mathbf{k}| \bar m_{ij} }{\Omega^{2}_{c ij}} \{ \Xi'^{-} , \tfrac{m}{\omega} \, f^-_{h\mp} \}^{c}_{ij}\,,
\label{f_c_mixing_gradients}
\end{eqnarray}
the diagonal matrix kernels are respectively
\begin{equation}
({\frac{m^\prime}{\omega^2}})_{ij} = \delta_{ij}\frac{m_i^\prime}{\omega_i^2}  \,, \quad
{v}_{{\bf k}\,ij} = \delta_{ij}\frac{ |\mathbf{k}| }{\omega_i} 
\quad {\rm and} \quad
(\frac{ m}{\omega})_{ij} = \delta_{ij}\frac{m_i}{\omega_i} \,,
\label{diag_kernels}
\end{equation}
and we have used a further shorthand notation:
\begin{equation}
f^+_{h\pm} \equiv \frac{1}{2}( f^m_{h\pm} + f^c_{h\mp}),
\quad {\rm and} \quad
f^-_{h\pm} \equiv \frac{1}{2}( f^m_{h\pm} - f^c_{h\mp}).
\end{equation}
Finally, $\Xi'^{\pm}$ are the hermitian chiral components of the $\Xi'$-matrix appearing in the commutator in Eq.~(\ref{eq:DDerivative}): 
\begin{equation}
\Xi'^{\pm}_{ij} \equiv \frac{i}{2}\big(V\partial_tV^\dagger \pm U\partial_tU^\dagger\big)_{ij}\,,
\label{mixing_grad}
\end{equation}
and the collision integrals are given by:
\begin{eqnarray}
{\cal C}^m_{h\pm}[f]_{ij} &=& N^m_{ij}{\rm Tr}\Big[P_{j\pm}\gamma^0 P_{i\pm} P_h \gamma^0 \langle{\cal C}_{ij} + {\cal C}_{ij}^\dagger\rangle\gamma^0 \Big]\,,
\nonumber\\[2mm]
{\cal C}^c_{h\pm}[f]_{ij} &=& N^c_{ij}{\rm Tr}\Big[P_{j\mp}\gamma^0 P_{i\pm} P_h \gamma^0 \langle{\cal C}_{ij} + {\cal C}_{ij}^\dagger\rangle\gamma^0 \Big]\,.
\label{coll_integrals}
\end{eqnarray}
Eqs.~(\ref{eom_on-shell1}-\ref{eom_on-shell2}) together with the collision integrals (\ref{coll_integrals}) form a closed set of equations of motion for the on-shell functions $f_{\alpha}$, once the self-energy functionals $\Sigma^{<,>}$ appearing in the collision integrals are specified. Note that despite their somewhat involved appearance, they still are of a similar form as the usual kinetic Boltzmann equations, only with a larger number of independent variables. Also, note that most of the complications arise from the explicit rotation to the mass eigenbasis; the starting equation (\ref{eom_prop}) is remarkable simple in the matrix form. More explicit expressions for the collision integrals will be derived in the section \ref{subsec-explicit} below (see equations (\ref{effectiveC3b}-\ref{effectiveS1}) and (\ref{Coll_eff_proj_m}-\ref{Coll_eff_proj_c})).

%
\subsection{Single-flavour and flavour-diagonal limits}
%
%
It is instructive to consider the single-flavour and diagonal limits of the full evolution equations (\ref{eom_on-shell1}-\ref{eom_on-shell2}). It is easy to see that if we drop the mixing gradients and flavour mixing by the collision term, diagonal equations with $i=j$ reduce to the single-flavour equations presented in ref.~\cite{HKR4} with a positive real mass $m = m_R$. However, 
for a more complete comparison we have to account for the complex mass $m= |m|e^{i\theta}$ in the single-field case. In the current approach the complex phase has to be removed by a chiral rotation $U_A = \exp(-i\shalf \gamma^5\theta)$.  After applying this rotation to both constraint and kinetic equations we get the following rotated single-flavour quantum Boltzmann equations:
\begin{eqnarray}
  \partial_t f_{mh\pm} &=& \pm 
      \frac12 \big( \Phi_{{\bf k}h+}^\prime  \hat f_{ch+} 
                   + \Phi^\prime_{{\bf k}h-} \hat f_{ch-} \big)
                   + {\cal C}_{mh\pm}[f]\,,
\label{sf_eom_on-shell1}
\\
   \partial_t\hat f_{ch\pm} &=& \mp i (2 \omega_{\bf k} + v_{{\bf k}} h \theta^\prime )\hat f_{ch\pm} -  
\frac{1}{2}\Phi^\prime_{{\bf k}h\mp} (f_{mh+} - f_{mh-}) + v_{{\bf k}} h {\cal C}_{ch\pm}[f]\,,
\label{sf_eom_on-shell2}
\end{eqnarray}
where $v_{{\bf k}} \equiv |\mathbf{k}|/\omega_{\bf k}$ and all dependence on the mass gradients in mixing terms 
is encoded in 
\begin{equation}
\Phi^\prime_{{\bf k} h\pm} \equiv v_{{\bf k}} h \frac{|m|^\prime}{\omega_{\bf k}} 
              \mp i \frac{|m| }{\omega_{\bf k}}\theta^\prime \,.
\end{equation}
The new coherence functions $\hat f_{ch\pm}$ are related to the ones used in ref.~\cite{HKR4} as follows:
\begin{equation}
\hat f_{ch\pm} = v_{{\bf k}} h \cos\theta \Re f_{ch\pm} \mp \sin\theta \Im f_{ch\pm}   
  + i \big( v_{{\bf k}} h \cos\theta \Im  f_{ch\pm} \pm \sin\theta \Re f_{ch\pm}\big)\,,
\label{sf_eom_on-shell3}
\end{equation}
Note that equations (\ref{sf_eom_on-shell2}) are even somewhat simpler than those introduced in ref.~\cite{HKR4}. Beyond simplicity, they are also convenient because $\hat f_{ch\pm}$'s turn out to be the functions with the (local) canonical normalization described in~\cite{HKR4}.

Let us now move on to the diagonal limit of the mixing case. The above example underlines the significance of the diagonal chiral rotations, which are relevant also in the mixing case. Indeed, it is important to note that the rotation matrices $U$ and $V$ are elements of the full $U(N)$ group. We can write the diagonalizing matrix $Y$ as a product of chiral $U(1)$-rotations and the elements of the $SU(N)$ groups:  
\begin{equation}
Y = P_L \otimes U + P_R \otimes V  = P_L \otimes U^N_LU_VU_A + P_R \otimes U^N_RU_VU^*_A \,, 
\label{mixing_matrix3}
\end{equation}
where $U_V = \exp(i\shalf \theta_V)$ and $U_A = \exp(i\shalf \theta_A)$ are the vector and axial vector $U(1)$ transformations, and the $SU(N)_{L,R}$ transformation matrices can be parametrized as $U^N_{L,R} = \exp(i \alpha_{L,R}^a \tau^a)$. In this parametrization the mixing-gradient matrices (\ref{mixing_grad}) become:
\begin{eqnarray}
  \Xi^\prime_{ij} &=& 
  \frac12 \Big[ \theta_V^\prime \delta_{ij} 
    + \big(\alpha_R^{ \prime a} + \alpha_L^{ \prime a} \big)\tau^a_{ij} \Big] 
\; + \;  \frac12  \Big[ - \theta_A^\prime \delta_{ij}
  + \big(\alpha_R^{ \prime a} - \alpha_L^{ \prime a} \big)\tau^a_{ij} \Big] \gamma^5  
  \nonumber \\
  &=& \Xi'^+_{ij} + \gamma^5\Xi'^-_{ij}\,,
\label{sf_mixing_gradient}
\end{eqnarray}
The pure $U_V$-part of $\Xi^\prime$ is proportional to identity operator and therefore it does not contribute to kinetic equations. This is as expected, since $U_V$ is
actually a pure gauge transformation. In single-flavour limit $U^{N= 1}_{L,R} = 1$ and only axial vector transformation $U_A$ contributes to mixing 
gradients in this limit, as shown above.    

It is interesting to observe that an effective (varying) complex phase is generated in the non-hermitian flavour-mixing scenario by the flavour-diagonal mixing gradients. Indeed, one can show after some algebra that the diagonal mixing-gradient terms (\ref{f_m_mixing_gradients}-\ref{f_c_mixing_gradients}) reduce to
\begin{eqnarray}
X^{m}_{h\pm}[f]_{ii} &=& i \,2 \Xi'^-_{ii}  \frac{h |\mathbf{k}|} {2 \omega_i} \big( f^c_{iih+} - f^c_{iih-} \big) + \ldots\,,
\label{eff_mI1}\\[2mm]
X^{c}_{h\pm}[f]_{ii} &=& \pm i \,2 \Xi'^-_{ii}  \frac{h |\mathbf{k}| }{\omega_i}  f^c_{iih\pm}  
  \pm i \,2 \Xi'^-_{ii}  \frac{h \omega_i  }{|\mathbf{k}|}  \frac{ m^2_i  }{2\omega_i^2  }\big(  f^m_{iih+} +  f^m_{iih-} \big) + \ldots\,,
\label{eff_mI2}
\end{eqnarray}
respectively, where we have written down only the flavour-diagonal self-coupling terms. By identifying:  $ -2 \Xi'^-_{ii} =  \theta_A^\prime + ( \alpha^{ \prime a}_L - \alpha^{ \prime a}_R )\tau^a_{ii} \sim \theta^\prime $, and accounting for the normalization differences: $v_{{\bf k}} h f^c_{ii} =  \hat f_c$ and $ \pm \tfrac{m_i}{\omega_i} f^m_{ii \pm} = f_{m\pm}$, we see that the terms in Eqs.~(\ref{eff_mI1} - \ref{eff_mI2}) reduce to the nondiagonal (in energy) mixing terms appearing in the single flavour equations (\ref{sf_eom_on-shell1} - \ref{sf_eom_on-shell2}).
Since scalar-like mixing gradients $\Xi'^+_{ii}$ do not contribute to  flavour diagonals, this means  that neglecting all off-diagonal flavour-mixing terms in  
Eqs.~(\ref{eom_on-shell1} - \ref{eom_on-shell2}) reduces the system to just $N$ copies of single-flavour equations, as expected. Note that taking the diagonal flavour-mixing gradients in consideration is necessary to get independent 
mass-function phase changes for all diagonal components.

%
\section{Resummed fermion collision term}
\label{sec:fermicollision}
%

The basic quantity appearing in the equations of motion (\ref{eom_on-shell1}-\ref{eom_on-shell2}) is the zeroth 
moment-integral of the mass-basis collision term $\langle {\cal C}_{\rm d} \rangle = \langle X{\cal C}_{\rm coll}X^\dagger \rangle$, or more precisely, its projections defined in 
Eq.~(\ref{coll_integrals}). 
In this section we present a general method of expanding this quantity to the first (leading) order in the 
interaction width and the gradient of the mass function, denoted by 
${\cal O}^1 \equiv {\cal O}(\Gamma, \partial_t m)$. We follow closely the analysis presented in 
ref.~\cite{HKR4} for the single-flavour case. 

As was observed in ref.~\cite{HKR4}, the rapidly oscillating coherence functions $f_\alpha$ give rise to a leading order ${\cal O}^1$ term in each order of the naive ($\Diamond$-)gradient expansion of the collision integral:
\beq
\langle {\cal C}_{\rm d} \rangle = \int \frac{{\rm d} k_0}{2\pi}\,X(t)\frac{1}{2} e^{-i\Diamond}
                             \big( \{\Sigma^>(k,t)\}\{S^<(k,t)\} -
                                    \{\Sigma^<(k,t)\}\{S^>(k,t)\}\big)X^\dagger(t)\,.
\label{coll_term}
\eeq
If the coherence contributions are present only in the external propagators $S^{<,>}$, the oscillatory gradients can be directly resummed in the Wigner representation~\cite{HKR4}. In the general case the $\Diamond$-expansion becomes very complicated however, and it is much more convenient to rewrite the collision term in the {\em two-time representation}:
\begin{equation}
\langle {\cal C}_{\rm d} \rangle = \frac{1}{2}\int {\rm d} w_0\,X(t) \big(\Sigma^>(t,w_0,\mathbf{k})S^<(w_0,t,\mathbf{k}) - \Sigma^<(t,w_0,\mathbf{k})S^>(w_0,t,\mathbf{k})\big) X^\dagger(t)\,,
\label{coll_two-time}
\end{equation}
where the propagators are defined as
\begin{eqnarray}
S(w_0,w_0',\mathbf{k}) &\equiv& \int {\rm d}^{\,3}(\mathbf{w} - \mathbf{w'}) \, e^{-i\mathbf{k}\cdot(\mathbf{w} - \mathbf{w'})} S(w,w')
\nonumber\\ 
&=& \int \frac{{\rm d} k_0}{2\pi}\, e^{-i k_0(w_0 - w_0')}S\big(k, \frac{w_0+w_0'}{2}\big)
\label{prop_two-time}
\end{eqnarray}
and similarly for self-energies.
We emphasize that we employ the two-time representation only temporarily to perform the resummation of the oscillatory gradients. At the end we will recover a Boltzmannian-type collision term involving momentum loop-integrals and the distribution functions $f^{<,>}_{\alpha}(\mathbf{k},t)$ in the Wigner representation. Next, we want to express the flavour-basis propagators $S^{<,>}$ and the self-energies $\Sigma^{<,>}$ in Eq.~(\ref{coll_two-time}) in terms of the mass-basis propagators, in order to use the spectral solutions of section \ref{sec:shell}. According to the definition in section \ref{sec:shell}, the propagators transform locally in time in the $(k,t)$-representation, and we now define a similar transformation law for the self energies:
\beqa
S_{\rm d}(k,t) &=& Y(t) S(k,t) X^\dagger(t)\,,
\nonumber\\
\Sigma_{\rm d}(k,t) &=& X(t) \Sigma(k,t) Y^\dagger(t)\,.
\eeqa
It then follows that in two-time representation the transformation laws are still local in the average time-coordinate, and not following different end-point times $w_0$ and $w_0'$ separately:
\begin{equation}
S_{\rm d}(w_0,w_0',\mathbf{k})= Y\big(\frac{w_0+w_0'}{2}\big) S(w_0,w_0',\mathbf{k}) X^\dagger\big(\frac{w_0+w_0'}{2}\big)\,,
\label{prop_trans_two-time}
\end{equation}
with a similar expression for $\Sigma^{<,>}_{\rm d}(w_0,w_0',\mathbf{k})$. Then the collision term (\ref{coll_two-time}) becomes 
\begin{eqnarray}
\langle {\cal C}_{\rm d} \rangle &=& \frac{1}{2}\int {\rm d} w_0\,X(t)X^\dagger\big(\frac{t+w_0}{2}\big) \big(\Sigma^>_{\rm d}(t,w_0,\mathbf{k})S^<_{\rm d}(w_0,t,\mathbf{k}) 
\nonumber\\
&& \qquad\qquad\qquad\qquad\qquad\quad - \Sigma^<_{\rm d}(t,w_0,\mathbf{k})S^>_{\rm d}(w_0,t,\mathbf{k})\big) X\big(\frac{t+w_0}{2}\big) X^\dagger(t)
\nonumber\\[4mm]
&=& \frac{1}{2}\int {\rm d} w_0 \big(\Sigma^>_{\rm d}(t,w_0,\mathbf{k})S^<_{\rm d}(w_0,t,\mathbf{k}) - \Sigma^<_{\rm d}(t,w_0,\mathbf{k})S^>_{\rm d}(w_0,t,\mathbf{k})\big) + {\cal O}^2\,,
\label{coll_mass_basis}
\end{eqnarray}
where ${\cal O}^2 \equiv {\cal O}(\Gamma^2,\Gamma \partial_t m, \Gamma \partial_t X)$.\footnote{
The second order correction to the equation (\ref{coll_mass_basis}) from the explicit mixing gradients can be expressed using 
the notation of the section \ref{sec:exp} as
 $\delta \langle {\cal C} \rangle = \frac{1}{4} \sum_{ \pm} \big[ \gamma^0\Xi^{\prime}\gamma^0\, , \, 
(\partial_{k_0}\Sigma^>_{{\rm eff}}(k_0,\mathbf{k},t) {\cal S}^<_{\pm}(\mathbf{k},t) -
 \partial_{k_0}\Sigma^<_{{\rm eff}}(k_0,\mathbf{k},t) {\cal S}^>_{\pm}(\mathbf{k},t)) \big]|_{k_0=\pm \omega_k} + ...$ , 
where other second order terms arise from the equations of motion of the correlators $S^{<,>}$ or explicitly expanding $\Sigma^>_{{\rm eff}}$
in gradients of $ \partial_t m , \partial_t X$. }
In the last row we have used the zeroth order term of the mixing matrix $X$ expanded in mixing gradients: 
\beq
X\big(\frac{t+w_0}{2}\big) = X(t) + \partial_t X(t)(\frac{w_0-t}{2}) + {\cal O}(\partial^2_t X)\,.
\label{mixing_matrix_zeroth}
\eeq
We can now see that neglecting ${\cal O}^2$ corrections in Eq.~(\ref{coll_mass_basis}) is a good 
approximation if the gradients associated
with the varying background, $ \partial_t \ln m \sim \partial_t X \sim \tau^{-1}$ and energy gradients  
$\partial_{k_0}\ln \Sigma_{} \sim \omega^{-1}$ parametrically obey $\tau \omega \gtrsim 1$.
This is consistent with our approximation that all the elements of the original flavour-basis mass matrix 
are slowly varying in time. 

%
\subsection{Expansion of the propagators $S^{<,>}$}
\label{sec:exp}
%

Next, we want to write the collision term $\langle {\cal C}_{\rm d} \rangle$ in terms of matrix functions ${\cal S}^{m,c <,>}_{ij}(\mathbf{k},t)$ defined in Eq.~(\ref{int_wightman}), locally in the external time-variable $t$. By writing the Wightman function (\ref{full_wightman_nondeg}) as (we suppress the flavour indices $ij$ for the remaining of this section for convenience)
\begin{eqnarray}
S^<_{\rm d}(k,t) &\equiv& 2\pi \sum_{\pm} \big[ {\cal S}^{m <}_{\pm}(\mathbf{k},t) \delta(k_0 \mp \bar\omega) + {\cal S}^{c <}_{\pm}(\mathbf{k},t)\delta(k_0 \mp \Delta\omega) \big] 
\nonumber\\
&\equiv& \sum_\pm \big( S_{\pm}^{m <}(k,t) + S_{\pm}^{c <}(k,t) \big)\,,
\label{wightman_proj}
\end{eqnarray}
where ${\cal S}^{m <}_{\pm}$ and ${\cal S}^{c <}_{\pm}$ can be read off from Eqs.~(\ref{full_wightman_nondeg}) or (\ref{int_wightman}), and inserting it to the definition (\ref{prop_two-time}) we get the two-time representation of the cQPA-Wightman functions as ($t=(u_0+v_0)/2$):
\begin{equation}
S^{<,>}_{\rm d}(u_0,v_0,\mathbf{k}) = \sum_{\pm} \big[ e^{\mp i \bar\omega(u_0 - v_0)} {\cal S}^{m <,>}_{\pm}(\mathbf{k},t) + e^{\mp i \Delta\omega(u_0 - v_0)}{\cal S}^{c <,>}_{\pm}(\mathbf{k},t)\big]\,.
\label{uv_prop}
\end{equation}
We obviously need to expand the functions $S^{<,>}_{\rm d}(w_0,t,\mathbf{k})$
appearing explicitly in Eq.~(\ref{coll_mass_basis}). However, the perturbative expressions for the self-energy functions $\Sigma^{<,>}_{\rm d}(t,w_0,\mathbf{k})$ in general involve further integrations over internal vertices $w_0$ and $w_0'$,
and so we need to expand correlators $S_{\rm d}(w_0,w_0',\mathbf{k})$ with arbitrary time-coordinates $w_0$ and $w_0'$ with respect to the external time $t=(u_0+v_0)/2$ to zeroth order. This can be done by a simple Taylor expansion:
\begin{equation} 
S_{\rm d}^{<,>}\big(k,\frac{w_0+w_0'}{2}\big) = \sum_{n=0}^\infty \frac{1}{n!}\Big(\frac{w_0+w_0'}{2}-t\Big)^n \partial_t^n S_{\rm d}^{<,>}(k, t\big)\,.
\end{equation}
The time-derivatives $\partial_t^n S^<_{\alpha}(\mathbf{k},t)$ can be found from the equations of motion (\ref{eom_on-shell1}-\ref{eom_on-shell2}). To zeroth order these reduce to $\partial_t f^m_{h\pm} = \mp i2\Delta\omega f^m_{h\pm}$ and  $\partial_t f^c_{h\pm} = \mp i 2\bar\omega f^c_{h\pm}$, indicating that the on-shell functions $f^m_{h\pm}$ and $f^c_{h\pm}$ are oscillating in quantum scales with frequencies $2\Delta\omega$ and $2\bar\omega$, respectively. This holds true in the distribution sense also for the functions $S_{\pm}^{m <,>}(\mathbf{k},t)$ and $S_{\pm}^{c <,>}(\mathbf{k},t)$ and we get:
\begin{equation}
\partial_t^n S_{\pm}^{m <,>} = (\mp i2\Delta\omega)^n S_{\pm}^{m <,>} + {\cal O}^1\,,\qquad \partial_t^n S_{\pm}^{c <,>} = (\mp i2\bar\omega)^n S_{\pm}^{c <,>} + {\cal O}^1 \,.
\label{zeroth_eq_higher}
\end{equation}
Using these equations recursively in the Taylor expansions, we find that distributions $S_{\pm}^{m <,>}$ and $S_{\pm}^{m <,>}$ depending on the average time $(w_0 + w_0')/2$ reduce to quantities evaluated at time $t$, multiplied by the phase factors:
\begin{eqnarray}
\sum_{n=0}^\infty \frac{1}{n!}\Big(\frac{w_0+w_0'}{2}-t\Big)^n \partial_t^n S_{\pm}^{m <,>}(k,t) &=&  e^{\mp i \Delta \omega (w_0 + w_0' - 2t)}S_{\pm}^{m <,>}(k,t) + {\cal O}^1\,,
\nonumber\\
\sum_{n=0}^\infty \frac{1}{n!}\Big(\frac{w_0+w_0'}{2}-t\Big)^n \partial_t^n S_{\pm}^{c <,>}(k,t) &=& e^{\mp i \bar \omega (w_0 + w_0' - 2t)}S_{\pm}^{c <,>}(k,t) + {\cal O}^1\,.
\end{eqnarray}
Inserting these results to Eq.~(\ref{prop_two-time}) one eventually finds:
\begin{eqnarray}
S^{<,>}_{\rm d}(w_0,w_0',\mathbf{k}) &=& \sum_\pm\Big[e^{\mp i \bar\omega(w_0 - w_0') \mp i \Delta\omega(w_0 + w_0' - 2t)} {\cal S}_{\pm}^{m <,>}(\mathbf{k},t)
\nonumber\\
&&\quad\;+\, e^{\mp i \Delta\omega(w_0 - w_0') \mp i \bar\omega(w_0 + w_0' - 2t)}{\cal S}_{\pm}^{c <,>}(\mathbf{k},t) \Big] \,+\, {\cal O}^1\,.
\label{eff_prop} 
\end{eqnarray}
As a trivial consistency check we see that this expansion reduces to Eq.~(\ref{uv_prop}) as $w_0 \to u_0$ and $w_0' \to v_0$. Physically the propagator (\ref{eff_prop}) is an approximation which takes into account the rapid temporal variations due to oscillations of the coherence
functions, but neglects the corrections due to temporal variations in the background fields as well as the corrections from the full memory effects of the dynamical interactions.
When applied to the case $w_0'=t$, Eq.~(\ref{eff_prop}) implies a particularly simple phase dependence: 
\beq
S^{<,>}_{\rm d}(w_0,t,\mathbf{k}) = \sum_\pm e^{\mp i \omega_i(w_0-t)} {\cal S}^{<,>}_{\pm}(\mathbf{k},t) + {\cal O}^1 \,,
\label{eq:simple}
\eeq
where we denote ${\cal S}^{<,>}_{\pm} = {\cal S}^{m <,>}_{\pm} + {\cal S}^{c <,>}_{\pm}$. Applying equation (\ref{eq:simple}) to the propagators in the collision term (\ref{coll_mass_basis}), we find
\begin{equation}
\langle {\cal C}_{ij} \rangle = \frac{1}{2} \sum_{k \pm} \Big(\Sigma^>_{{\rm eff},ik}(\pm \omega_k,\mathbf{k},t){\cal S}^<_{kj\pm}(\mathbf{k},t) - \Sigma^<_{{\rm eff},ik}(\pm \omega_k,\mathbf{k},t){\cal S}^>_{kj\pm}(\mathbf{k},t)\Big) + {\cal O}^2\,,
\label{coll_final}
\end{equation}
where the effective self-energy function is defined by:
\beq
\Sigma^{<,>}_{{\rm eff},ij}(k,t) \equiv \int {\rm d} w_0 e^{i k_0(t-w_0)} \Sigma^{<,>}_{{\rm d}, ij}(t,w_0,\mathbf{k})\,.
\label{eff_sigma}
\eeq
This is formally the most important result of this section. Note that the effective self-energy functions $\Sigma_{{\rm eff},ij}^{<,>}(\pm \omega_j,\mathbf{k},t)$ are {\em not} in general equivalent to the projections $\Sigma^{<,>}_{ij}(\pm \omega_j,\mathbf{k},t)$ of the original Wigner representation self-energy. However, both reduce to the same expression in the limit $\partial_t \Sigma^{<,>}(k,t) = {\cal O}^1$, \ie in the case that the oscillatory coherence solutions do not contribute to the self-energy. Perturbation theory rules for evaluating  $\Sigma^{<,>}_{{\rm eff},ij}(k,t)$ will be derived in section \ref{sect:effective-feynman}.  

It is interesting to see that in the end, after the resummation has been carried out, Eq.~(\ref{coll_final}) can be written in an integral form {\em formally} equivalent to the naive lowest order gradient expansion of the equation~(\ref{coll_term}):
\begin{equation}
\langle {\cal C}_{ij} \rangle
\equiv \frac{1}{2} \int \frac{{\rm d}k_0}{2\pi }\big( 
     \Sigma^>_{{\rm eff},ik}(k,t) S^<_{{\rm eff , in},kj}(k,t)
   - \Sigma^<_{{\rm eff},ik}(k,t) S^>_{{\rm eff , in},kj}(k,t)
   \big) \,.
\label{effectiveC3b}
\end{equation}
The crucial difference is the replacement of the original self-energy functions and the cQPA Wightman propagators by the effective self-energy functions $\Sigma^{<,>}_{\rm eff}$ and the effective in-propagators:
\begin{align}
iS^{<,>}_{{\rm eff , in},ij}(k,t) \equiv& 2\pi i \sum_{\pm} 
 \big( {\cal S}^{m <,>}_{ij\pm}({\bf k},t) + {\cal S}^{c <,>}_{ij\pm}({\bf k},t)\big) \,\delta(k_0 \mp \omega_i)
\nonumber\\
 \equiv&  iS^{m <,>}_{{\rm eff , in}, ij}(k,t)  + iS^{c <,>}_{{\rm eff , in}, ij}(k,t) \,. 
\label{effectiveS1}
\end{align}
We conclude this section by finding the most general explicit expressions for the collision integrals Eq.~(\ref{coll_integrals}), consistent with the symmetries (spatial homogeneity and isotropy) of the problem.

%
\subsection{Explicit collision integrals}
\label{subsec-explicit}
%

By using covariant formulation for the Dirac structures the collision integrals (\ref{coll_integrals}) can be rewritten as
\begin{eqnarray}
{\cal C}^m_{h\pm}[f]_{ij} &=& \frac{1}{4 \Omega^2_{mij} }
{\rm Tr}\Big[\langle{\cal C}_{ij} + {\cal C}_{ij}^\dagger\rangle\gamma^0 
 P_h (\kdag_{j\pm} + m_j)(\kdag_{i\pm} + m_i) \Big] \, ,
\nonumber\\[2mm]
{\cal C}^c_{h\pm}[f]_{ij} &=& -\frac{1}{4 \Omega^2_{cij} } 
{\rm Tr}\Big[\langle{\cal C}_{ij} + {\cal C}_{ij}^\dagger\rangle\gamma^0  
 P_h (\kdag_{j\mp} + m_j)(\kdag_{i\pm} + m_i) \Big] \, .
\label{coll_integrals2}
\end{eqnarray}
Similarly to Eq.~(\ref{bloch_prop}) we can write the helicity projections of the most general spatially homogeneous and isotropic effective self-energy $ \Sigma_{{\rm eff}}^{<,>} $ in the Dirac notation as 
\begin{equation}
  P_h i\Sigma_{{\rm eff}}^{<,>}(k,t)  =  P_h \big[
 \gamma^0 \, A^{<,>}_{h} +  \hat {\bf k} \cdot \gamma \,B^{<,>}_{h} +  C^{<,>}_{h} + i h \gamma^5\,D^{<,>}_{h} \big]\,.
\label{dirac_sigma}
\end{equation}
Now, using Eq.~(\ref{coll_final}) for $\langle {\cal C}_{ij} \rangle$ and the projector properties of the propagator matrices ${\cal S}^{m,c <,>}_{ij\pm}$ on the self-energies (\ref{dirac_sigma}) it is relatively straightforward to show that the collision integrals of the Boltzmann equations (\ref{eom_on-shell1}-\ref{eom_on-shell2}) take the following form:
\begin{eqnarray} 
{\cal C}^m_{h\pm}[f]_{ij} &=& - \frac{1}{2}\bigg( \,
    \{ A^>_{h} ,  f^{<+}_{h\pm} \}^{m}_{ij} 
\pm \{ B^>_{h} , {v}_{{\bf k}} \, f^{<-}_{h\pm}  \}^{m}_{ij}  
\pm  \{ C^>_{h} , \tfrac{m}{\omega} \, f^{<-}_{h\pm} \}^{m}_{ij} 
\nonumber\\
&+&\,  \frac{i |\mathbf{k}| \Delta\omega_{ij} }{\Omega^{2}_{m ij}} \Big( 
  i\, \big[ A^>_{h}  , {v}_{{\bf k}} \, f^{<-}_{h\pm} \big]^{m}_{ij} 
\pm i \, \big[  B^>_{h} , f^{<+}_{h\pm} \big]^{m}_{ij}   
+ \{ D^>_{h}  ,  \tfrac{m}{\omega} \, f^{<-}_{h\pm} \}^{m}_{ij}\Big) 
\nonumber\\
&+&\, \frac{\bar \omega_{ij} }{ \bar m_{ij} } \Big( 
\{ A^>_{h} ,  \tfrac{m}{\omega} \, f^{<-}_{h\pm} \}^{m}_{ij} 
\pm \{ C^>_{h} , f^{<+}_{h\pm} \}^{m}_{ij} 
+ i\, \big[ D^>_{h} , {v}_{{\bf k}} \,  f^{<-}_{h\pm} \big]^{m}_{ij} \Big) 
\nonumber\\
&\pm& \, \frac{|\mathbf{k}| \bar m_{ij} }{ \Omega^{2}_{m ij} } \Big(
\{ B^>_{h} ,  \tfrac{m}{\omega} \,  f^{<-}_{h\pm} \}^{m}_{ij} 
- \{ C^>_{h} ,  {v}_{{\bf k}} \,  f^{<-}_{h\pm} \}^{m}_{ij} 
\mp i \, \big[ D^>_{h} , f^{<+}_{h\pm} \big]^{m}_{ij} \Big)
\nonumber\\
&-& \Big[ > \leftrightarrow < \Big] \bigg) \,,
\label{Coll_eff_proj_m}
\end{eqnarray}
\begin{eqnarray} 
{\cal C}^c_{h\pm}[f]_{ij} &=& - \frac{1}{2}\bigg( \,
    \{ A^>_{h} , f^{<+}_{h\mp} \}^{c}_{ij} 
\mp \big[ B^>_{h}  , {v}_{{\bf k}} \, f^{<-}_{h\mp} \big]^{c}_{ij}  
\mp \big[ C^>_{h} , \tfrac{m}{\omega} \, f^{<-}_{h\mp} \big]^{c}_{ij}
\nonumber\\
&+&\,  \frac{|\mathbf{k}| \bar \omega_{ij} }{\Omega^{2}_{c ij}} \Big( 
- \{ A^>_{h} , {v}_{{\bf k}} \, f^{<-}_{h\mp}  \}^{c}_{ij} 
\pm \big[ B^>_{h} , f^{<+}_{h\mp} \big]^{c}_{ij} 
+ i\, \big[ D^>_{h}  ,  \tfrac{m}{\omega} \, f^{<-}_{h\mp} \big]^{c}_{ij} \Big)  
\nonumber\\
&+&\, \frac{i \Delta \omega_{ij} }{ \bar m_{ij} } \Big( 
i\,\big[ A^>_{h} , \tfrac{m}{\omega} \, f^{<-}_{h\mp} \big]^{c}_{ij} 
\mp i\, \{ C^>_{h} , f^{<+}_{h\mp} \}^{c}_{ij}
- \{ D^>_{h} , {v}_{{\bf k}} \,  f^{<-}_{h\mp} \}^{c}_{ij} \Big)
\nonumber\\
&\pm& \, \frac{|\mathbf{k}| \bar m_{ij} }{ \Omega^{2}_{c ij} } \Big(
\big[ B^>_{h} , \tfrac{m}{\omega} \, f^{<-}_{h\mp} \big]^{c}_{ij} 
- \big[ C^>_{h} ,  {v}_{{\bf k}} \,  f^{<-}_{h\mp} \big]^{c}_{ij} 
\pm i \, \big[ D^>_{h} , f^{<+}_{h\mp} \big]^{c}_{ij}\Big)
\nonumber\\
&-& \Big[ > \leftrightarrow < \Big] \bigg) \,.
\label{Coll_eff_proj_c}
\end{eqnarray}
Keeping in mind that here $i \Sigma^{<,>}_{{\rm eff},ij\pm} \equiv i \Sigma^{<,>}_{{\rm eff},ij}(\pm \omega_j)$ 
depends on the energy index $\pm$, we have enlarged the notation (\ref{gen_comm3}) to accommodate this:
\begin{eqnarray}
 &\big[\Sigma_s , f_{s}  \big]^m  \equiv \Sigma_s f_{s}  - ( \Sigma_s f_{s } )^{\dagger} \,, 
\qquad &\{\Sigma_s , f_{s} \}^m   \equiv \Sigma_s f_{s} + ( \Sigma_s f_{s} )^{\dagger} \,, 
\nonumber\\
 &\big[\Sigma_s , f_{s}  \big]^c  \equiv \Sigma_s f_{s}  - ( \Sigma_{-s} f_{-s} )^{\dagger} \,, 
\quad &\{\Sigma_s , f_{s} \}^c   \equiv \Sigma_s f_{s}  + (\Sigma_{-s} f_{-s} )^{\dagger} \,,
\label{Coll_gen_comm3}
\end{eqnarray}
and the generalized anticommutators of $f^{\pm}_{\alpha}$:s with the self-energies $\Sigma$ without $\pm$ indices are reduced as follows (with obvious generalization to commutators):
\begin{eqnarray} 
    \{ \Sigma , f^{<\pm'}_{h\pm} \}^{m} &\equiv& \frac{1}{2} \{ \Sigma_{\pm} , f^{m <}_{h\pm} \}^{m} \, \pm' \,
    \frac{1}{2} \{ \Sigma_{\mp} , f^{c <}_{h\mp} \}^{m} \,,
\nonumber\\
    \{ \Sigma , f^{<\pm'}_{h\pm} \}^{c} &\equiv& \frac{1}{2} \{ \Sigma_{\pm} , f^{m <}_{h\pm} \}^{c} \, \pm' \,
    \frac{1}{2} \{ \Sigma_{\mp} , f^{c <}_{h\mp} \}^{c} \,.
\label{Coll_eff_proj_mA}
\end{eqnarray}
For example the first term of Eq.~(\ref{Coll_eff_proj_m}) is explicitly written as follows:
\begin{eqnarray} 
    \{ A^{>}_{h} , f^{<+}_{h\pm} \}^{m}_{ij} = 
    \frac{1}{2} \sum_k &\Big(& A^{>}_{ikh}(\pm \omega_k) f^{m <}_{kjh\pm}  
           +  f^{m <}_{ikh\pm} A^{> \dagger}_{kjh}(\pm \omega_k)  
\nonumber\\
                           &+& A^{>}_{ikh}(\mp \omega_k) f^{c <}_{kjh\mp} 
           +  f^{c <}_{ikh\pm} A^{> \dagger}_{kjh}(\mp \omega_k) \Big) \,,
\label{Coll_eff_proj_mA1}
\end{eqnarray}
and respectively the first term of Eq.~(\ref{Coll_eff_proj_c}) is 
\begin{eqnarray} 
    \{ A^{>}_{h} , f^{<+}_{h\mp} \}^{c}_{ij} = 
    \frac{1}{2} \sum_k &\Big(& A^{>}_{ikh}(\pm \omega_k) f^{c <}_{kjh\pm} 
           +  f^{c <}_{ikh\pm} A^{> \dagger}_{kjh}(\mp \omega_k)  
\nonumber\\
                           &+& A^{>}_{ikh}(\mp \omega_k) f^{m <}_{kjh\mp}  
           +  f^{m <}_{ikh\pm}  A^{> \dagger}_{kjh}(\pm \omega_k)  \Big) \,.
\label{Coll_eff_proj_cA1}
\end{eqnarray}
Note that the flavour index of the matrix multiplication dictates the energy dispersion $k_0 = \pm\omega_k$ of the effective self-energy, and observe in particular the delicate inversion of signs in the dispersion for the second $f^c$-terms, in accordance with Eqs.~(\ref{coll_final}) and (\ref{eq:hermiticitycond}). 

Equations (\ref{Coll_eff_proj_m}-\ref{Coll_eff_proj_c}) present the most general form of the fermionic collision integrals in the cQPA approximation, consistent with the spatial homogeneity and isotropy. Further reduction is possible only after an explicit expression for the self-energy functions is specified.\footnote{Let us again emphasize that if there is no coherence contributions {\em inside} the self energies, \eg in the case of interaction with thermal bath, then to leading order $\Sigma^{<,>}_{\rm eff}(k,t)$ just reduce to standard Wigner representation self-energies $\Sigma^{<,>}(k,t)$, which can be computed using the standard techniques of CTP-formalism or real-time formalism of thermal field theory. This type of simple example is considered in section \ref{sec:applications}, where we apply our formalism to $CP$-violating flavour mixing in the presence of collisions with thermal background.}  A perturbative expansion for $\Sigma^{<,>}_{\rm eff}$'s, which automatically accounts for the resummation of the oscillations of the coherence propagators, could easily be written down in the two-time representation using the zeroth order expanded propagators Eq.~(\ref{eff_prop}). However, in section \ref{sect:effective-feynman} we will show how this expansion can be rewritten in terms of effective momentum space Feynman rules, which make the computation of $\Sigma^{<,>}_{\rm eff}$ straightforward up to a desired order in perturbation expansion including an arbitrary number of coherent propagators, without leaving the Wigner representation.  Before this, we shall extend our analysis to the case of scalar fields, however.

%
\section{Scalar fields}
\label{sec:scalar}
%

The cQPA approximation can be employed also for other types of quantum fields than fermions, although the analysis is qualitatively different since the spinor structure and corresponding projection operators are not present. In this section we formulate the cQPA quantum transport equations for flavour-mixing complex scalar fields following the same approximation scheme we used for fermions, described in section \ref{sec:approx}. As a result, we will find the same singular phase space structure with the shells: $k_0 = \pm\bar\omega = \pm(\omega_i + \omega_j)/2$ and $k_0 = \pm\Delta\omega = \pm(\omega_i - \omega_j)/2$, as for fermions.\footnote{This complete shell structure for mixing scalar fields was also observed in ref.~\cite{CLR-MT10}. However, in this work the rapidly oscillating $k_0 = \pm\Delta\omega$ coherence-solutions were subsequently neglected in the derivation of the transport equations.} However, because of the structural difference between scalar and fermionic KB-equations, the derivation of the transport equations is now conceptually different to the method for fermions; analogously to our earlier works for single flavour scalar field~\cite{HKR3,HKR4}, we need to introduce multiple $k_0$-moment integrals to obtain closure.

As with fermions, we start by writing down the Kadanoff-Baym equations for the Wightman functions $i\Delta^<(u,v) = \langle \phi^\dagger(v)\phi(u) \rangle$ and $i\Delta^>(u,v) = \langle \phi(u) \phi^\dagger(v)\rangle$ in the Wigner representation (see \eg ref.~\cite{PSW}):
\begin{equation}
\Big( k^2-\frac{1}{4}\partial_x^2 + i k \cdot \partial_x - m^2e^{-\frac{i}{2}\overleftarrow{\partial_x}\partial_k} \Big) \Delta^{<,>}  -  e^{-i\Diamond}\{ \Pi^h \}\{ \Delta^{<,>} \}
  -  e^{-i\Diamond}\{ \Pi^{<,>} \}\{ \Delta^h \}
= \mathcal{C}^\phi_{\rm coll}\,,
\label{KB_scalar}
\end{equation}
where $\Delta^h = \Delta^t - (\Delta^> + \Delta^<)/2$ and $\Pi^h = \Pi^t - (\Pi^> + \Pi^<)/2$, while $\Delta^t$ and $\Pi^t$ denote the time-ordered Green's function and the corresponding self-energy. The form of the collision term $\mathcal{C}^\phi_{\rm coll}$ can be obtained from Eq.~(\ref{collintegral}) by replacing $S \to \Delta$ and $\Sigma \to \Pi$. The mass $m^2$ as well as $i\Delta^{<,>}(k,x)$ and $i\Pi^{<,>}(k,x)$ are hermitian $N \times N$ matrices in flavour indices. Proceeding through the steps 1-3 in the approximation scheme described in section \ref{sec:approx}, we drop the terms proportional to $\Delta^h$ and $\Pi^h$ in KB-equations (\ref{KB_scalar}) and (by spatial homogeneity) set $\vec\partial_x i\Delta^{<,>}(k,x) = \vec\partial_x m^2(x) = 0$. After these approximations we break the KB-equation into hermitian (H) and antihermitian (AH) parts, finding eventually:
\begin{eqnarray}
 \left(k^2 -\frac{1}{4}\partial_t^2 -\cos\big(\frac{1}{2}\partial^m_t\partial^\Delta_{k_0}\big) \frac{1}{2} \left\{m^2,\circ\right\} + \sin\big(\frac{1}{2}\partial^m_t\partial^\Delta_{k_0}\big) \frac{i}{2} \left[ m^2,\circ\right]\right)i\Delta^{<,>} &=& -\mathcal{C}_{A\phantom{}}\,,
\label{ConstraintEq}
\\ 
\left( k_0 \partial_t + \cos\big(\frac{1}{2}\partial^m_t\partial^\Delta_{k_0}\big) \frac{i}{2} \left[m^2,\circ\right] + \sin\big(\frac{1}{2}\partial^m_t\partial^\Delta_{k_0}\big)\frac{1}{2}\left\{ m^2,\circ\right\}\right)i\Delta^{<,>}  &=&  \mathcal{C}_{H} \,,
\label{EvoEq}
\end{eqnarray}
where $[X,\circ]Y \equiv [X,Y]$, $\mathcal{C}_{H} \equiv (\mathcal{C}^\phi_{\rm coll} + \mathcal{C}_{\rm coll}^{\phi\dagger})/2$ and $\mathcal{C}_{A} \equiv (\mathcal{C}^\phi_{\rm coll} - \mathcal{C}_{\rm coll}^{\phi\dagger})/(2i)$, and the superscripts on derivatives denote explicitly that $t$-derivatives always act on the mass function $m^2(t)$, while $k_0$-derivatives act on the correlators $i\Delta^{<,>}(k,t)$. 

%
\subsection{Phase space structure}
%

We now proceed to steps 4-5 in the cQPA approximation scheme, and analyze the KB-equations (\ref{ConstraintEq}-\ref{EvoEq}) in the zeroth order in $\Gamma$ and $\partial_t m$ to find out the singular phase-space structure. Unlike in the case with fermions, for scalars both H- and AH-equations contain explicit time-derivatives even in zeroth order in mass gradients, and we need to use both equations to derive the algebraic constraint: first the collisionless AH-equation (\ref{EvoEq}) is solved to the lowest order to give $\partial_t i\Delta^{<,>} = -\frac{i}{2k^0}[m^2,i\Delta^{<,>}]$, and then this relation is used recursively to compute the second derivative term $\partial_t^2 i\Delta^{<,>}$ in Eq.~(\ref{ConstraintEq}), again to the lowest order in mass gradients. This procedure leads to the zeroth-order algebraic constraint equation
\begin{equation}
\Big(k_0^2-\mathbf{k}^2-\frac{1}{2}\{m^2,\circ \}+\frac{1}{16 k_0^2}\left[m^2,[m^2,\circ]\right]\Big) i\Delta^{<,>}(k,t) = 0\,.
\label{shell_constraint}
\end{equation}
Next, we go to the mass basis by diagonalizing the hermitian mass matrix: $m^2 \to m^2_{\rm d} = U m^2 U^\dagger$, where $U$ is a unitary mixing matrix. The correlators transform similarly: $\Delta \to \Delta_{\rm d} = U \Delta U^\dagger$. In the mass basis Eq.~(\ref{shell_constraint}) becomes
\begin{equation}
\Big((k_0^2 - \mathbf{k}^2 - M^2_{ij})k_0^2 + (\Delta m_{ij} \bar m_{ij})^2\Big)i\Delta^{<,>}_{ij}(k,t) = 0\,.
\label{shell_constraint_mass}
\end{equation}
where $M_{ij}^2 \equiv (m_i^2 + m_j^2)/2$. We see that the multiplying factor in parenthesis is equal to the fermionic constraint determinant in Eq.~(\ref{constraint_det}), and thus the nonzero solutions of $\Delta^{<,>}_{ij}\equiv (\Delta^{<,>}_{\rm d})_{ij}$ must be proportional to either $\delta(k_0 \mp \bar\omega_{ij})$ or $\delta(k_0 \mp \Delta\omega_{ij})$, as before. There are no further algebraic constraints and we can parametrize the cQPA-propagators $\Delta^{<,>}_{ij}$ as 
\begin{equation} 
 i\Delta^{<,>}_{ij}(k,t) =\frac{\pi \, \bar \omega_{ij}}{{\omega_i\omega_j}} \sum \limits_{\pm} \Big( \pm f_{ij\pm}^{m <,>}\delta\left(k_0 \mp \bar{\omega}_{ij}\right) + f_{ij\pm}^{c <,>}\delta\left(k_0 \mp \Delta\omega_{ij}\right)\Big)\,.
 \label{Dist}
\end{equation}

To zeroth order the spectral function ${\cal A}^\phi = \frac{i}{2}(\Delta^> - \Delta^<)$ obeys identical KB-equations to Eqs.~(\ref{ConstraintEq}-\ref{EvoEq}), and consequently the solution is of the same form as Eq.~(\ref{Dist}). In addition however, the spectral function must obey the sum rule
\beq
\int \frac{{\rm d} k_0}{\pi} \big(k_0 + \frac{i}{2}\partial_t \big) {\cal A}^\phi_{ij}(k,t) = \delta_{ij}\,,
\eeq
following from the equal-time commutation relations of the scalar fields $\phi_i$. It turns out that this relation completely fixes the values of the corresponding on-shell functions (see ref.~\cite{HKR3}) to $f^{m \cal A}_{ij\pm} = \frac12 \delta_{ij}$ and $f^{c \cal A}_{ij\pm} = 0$, reducing the spectral function to its standard form
\begin{equation}
 {\cal A}^\phi_{ij}(k,t) = \pi\,{\rm sgn}(k_0) \delta(k^2 - m_i^2)\delta_{ij} \,.
\label{full_spectral_sca}
\end{equation}
Furthermore, using this result with the defining relation $2i{\cal A}^\phi = \Delta^< - \Delta^>$, one can easily show that the dynamic functions $f^{>}(\mathbf{k},t)$ and $f^{<}(\mathbf{k},t)$ are related:
\begin{equation}
f_{ij\pm}^{m >} = \delta_{ij} + f_{ij\pm}^{m <} \,,
\qquad\quad 
f_{ij\pm}^{c >} = f_{ij\pm}^{c <}\,.
\label{gtr_less_sca}
\end{equation}
That is, only half of the on-shell functions appearing in $\Delta^{<,>}$ are free variables. In what follows, we derive equations of motion for the on-shell functions $f^{m,c <}_{ij\pm} \equiv f^{m,c}_{ij\pm} \equiv f_\alpha$.

%
\subsection{Equations of motion}
%

We derive the transport equations for the on-shell functions $f_{\alpha}$ by inserting the singular cQPA-propagator (\ref{Dist}) as an ansatz into the full KB-equations (\ref{ConstraintEq}-\ref{EvoEq}). The unknown on-shell functions can be extracted from $\Delta^{<}_{ij}(k,t)$ by integration with different weight functions. To be specific, we use moment functions:
\begin{equation}
\rho_n(k,t)= \int \frac{{\rm d} k_0}{2\pi}\,k_0^n\,i\Delta^<_{\rm d}(k,t)\,.
\label{n-moment}
\end{equation}
To get the closure we need at least four different moments, or alternatively, we could use three moments and one time-derivative of a moment since the constraint equation (\ref{ConstraintEq}) is a second order differential equation w.r.t. $t$. It appears that the most natural choice is to use the three lowest moments $\rho_{0,1,2}$ and the derivative $\partial_t \rho_0$. In order to relate these moments, and in particular the derivative $\partial_t \rho_0$, to the $f$-functions of Eq.~(\ref{Dist}), one needs to solve the derivatives $\partial_t f_\alpha$ in terms of $f_\alpha$ to zeroth order. The required zeroth order equations of motion can be obtained by solving Eq.~(\ref{EvoEq}) for any four moments, for example $\rho_{1,2,3,4}$, using the singular cQPA-propagator (\ref{Dist}). In this way one finds the equations
\begin{equation}
\label{Firtsfscal}
\partial_t f^m_{ij\pm} = \mp 2i \Delta\omega_{ij} f^m_{ij\pm} + {\cal O}^1\,, \qquad
\partial_t f^c_{ij\pm} = \mp 2i \bar\omega_{ij} f^c_{ij\pm} + {\cal O}^1\,,
\end{equation}
which are identical to fermionic equations of motion (\ref{eom_on-shell1}-\ref{eom_on-shell2}) to zeroth order. By computing the moments (\ref{n-moment}) of the cQPA-propagator (\ref{Dist}), and using Eqs.~(\ref{Firtsfscal}) for the derivatives $\partial_t f_\alpha$ appearing in $\partial_t \rho_0$, we then obtain a set of invertible relations between $f_\alpha$ and the moment functions:
\begin{equation}
\left(\begin{array}{c} 
\rho_0            \\ 
\partial_t \rho_0 \\ 
\rho_1            \\ 
\rho_2 
\end{array}\right)_{ij}
= \frac{\bar \omega_{ij} }{2 \omega_i \omega_j}
\left(\begin{array}{cccc} 
      1               &   -1            &   1            & 1              \\
-2 i \Delta\omega     & -2i\Delta\omega & -2i\bar\omega  & 2i\bar\omega   \\
   \bar\omega         &  \bar\omega     & \Delta\omega   & -\Delta\omega  \\
\bar\omega^2          & -\bar\omega^2   & \Delta\omega^2 &\Delta\omega^2
\end{array}\right)_{ij}
\left(\begin{array}{c} 
  f^m_{+} \\ 
  f^m_{-} \\ 
  f^c_{+} \\ 
  f^c_{-} 
\end{array}\right)_{ij} + {\cal O}^1\,,
\label{RelationEq}
\end{equation}
where an element wise rule of matrix multiplication is applied to the flavour indices which are suppressed to generic subscripts for convenience.
 
The evolution equations for the moments $\rho_{0,1,2}$ and $\partial_t \rho_0$ are found by taking the zeroth moment of Eq.~(\ref{ConstraintEq}) and the zeroth and the first moments of Eq.~(\ref{EvoEq}). Transforming these equations to the mass-diagonal basis amounts to replacement: 
\beq
\partial_t \rho_n \to D_t \rho_n \equiv \partial_t \rho_n - i[\Theta^\prime\,,\rho_n]\,,\qquad \Theta^\prime \equiv i U\partial_t U^\dagger\,,
\eeq
analogous to fermions (see Eqs.~(\ref{eq:DDerivative}-\ref{eq:ekamultikinetic})). The resulting equations of motions for the moments are given by  
\begin{eqnarray}
\nonumber 
D_t^2 \rho_0 - 4\rho_2 + 4(\bar\omega^2+\Delta\omega^2)\rho_0 &=& 4 \langle\mathcal{C}_{A{\rm d}}\rangle \,, 
\\ \nonumber
D_t \rho_1 + 2 i \Delta\omega \bar\omega \rho_0 &=& \langle\mathcal{C}_{H{\rm d}}\rangle \,, 
\\
D_t \rho_2 + 2 i \Delta\omega \bar\omega \rho_1 - (\bar\omega \bar\omega' + \Delta\omega \Delta\omega')\rho_0 &=& \langle k_0 \mathcal{C}_{H{\rm d}}\rangle\,,
\label{EvoEqMoments}
\end{eqnarray}
where $\mathcal{C}_{(H,A){\rm d}} \equiv U \mathcal{C}_{H,A} U^\dagger$, and we have suppressed the flavour indices, for which the multiplication rule is elementwise as in Eq.~(\ref{RelationEq}).

Unlike in the case of fermions, we do not attempt to rewrite the evolution equations in terms of $f$-functions, although it could be done \eg by differentiating the inverted equations (\ref{RelationEq}) and using the evolution equation (\ref{EvoEqMoments}). This change of variables carries a subtle issue related to the gradient expansion however, as the moment equations are exact in the sense of gradient expansion but the inversion relations are only valid to zeroth order.  To avoid this unnecessary loss of accuracy, and also to keep the evolution equations simpler, we prefer to write the qBE's in terms of the moments rather than the $f$-functions in the scalar case. In practice, this means that the $f$-functions are only used in the collision integrals, and only in this context their relation to moments needs to be defined through the inverse of equations (\ref{RelationEq}).

The subtlety with the inversion is related to an inherent ``ambiguity" in choosing the particular set of moments (or choice of moment weight functions) to define the integrated equations of motion. Indeed, transport equations based on higher moments (\eg for $\rho_{1,2,3,4}$) would be equivalent to Eqs.~(\ref{EvoEqMoments}) only up to zeroth order in gradients and scattering width, because the singular correlator (\ref{Dist}) and consequently the relations (\ref{RelationEq}) between $f$-functions and the moments are valid only to zeroth order. Our choice of moments is the most natural one in the following sense however: using the flat weight (lowest moment) to integrate the KB-equations (\ref{ConstraintEq}-\ref{EvoEq}) corresponds to encoding no exterior information on the dynamical equations (see the discussion in section 5 in ref.~\cite{HKR1}). This creates the two first equations of Eqs.~(\ref{EvoEqMoments}). However, the flat weight integration generates an extra moment $\rho_2$ into the integrated H-equation, and the {\em only way} to reach closure without adding new moments is to take the first moment of AH-equation. This we have done to obtain the last equation in Eqs.~(\ref{EvoEqMoments}).\footnote{For fermions this issue did not arise because there the flow term of the ($k_0$-integrated) dynamical KB-equation does not couple zeroth moment of the correlator $\langle S^<(k,t) \rangle$ to higher moments. For scalars the issue is more delicate because we are forced to use a mixture of different moment functions.}

The moment choice problem is also related to which order in gradients one is working. For example, going to higher order in gradients would involve more independent moments, but a consistent treatment would then require using more complicated phase-space structure for the correlator $\Delta^<(k,t)$, which can be obtained as a generalized distributive expansion in gradients. We explore this issue to some detail in Appendix \ref{sec:appendix}.  

%
\subsection{Resummed scalar collision term}
%

We need to express the collision integrals appearing in Eq.~(\ref{EvoEqMoments})
in terms of the distribution functions $f^m_\pm$ and $f^c_{\pm}$ (and eventually in terms of the moments using the connection Eqs.~(\ref{RelationEq})). Likewise with fermions, the basic quantity we encounter is the $\alpha$-moment integral of the collision term in the mass basis: 
\beq
\langle{\mathcal{C}}_\alpha\rangle = 
 \int \frac{{\rm d} k_0}{2\pi} k_0^{\alpha}\,
   U(t)\frac{1}{2} e^{-i\Diamond} 
 \big(\{\Pi^>(k,t)\}\{\Delta^<(k,t)\}
     -\{\Pi^<(k,t)\}\{\Delta^>(k,t)\}
 \big)U^\dagger(t)\,, 
\eeq
where we now need $\alpha=0,1$. Again, to resum the oscillatory gradients of the distribution functions in the $\Diamond$-expansion, we write the collision term in the two-time representation:
\beqa
\langle\mathcal{C}_\alpha\rangle &=& \frac12 \int {\rm d} w_0 (i\partial_{r_0})^{\alpha} \Big[ \Pi_{\rm d}^>\big(t+\frac{r_0}{2},w_0\big) \Delta_{\rm d}^<\big(w_0,t-\frac{r_0}{2}\big) 
\nonumber\\
&& \qquad\qquad\qquad\;\;\; - \Pi_{\rm d}^<\big(t+\frac{r_0}{2},w_0\big) \Delta_{\rm d}^>\big(w_0,t-\frac{r_0}{2}\big)\Big]_{r_0 = 0}\,, 
\label{coll_int_sca}
\eeqa
where we have dropped the time-gradients of the mixing matrix $U(t)$, as before. For $\alpha=0$ this immediately reduces to a formula analogous to Eq.~(\ref{coll_mass_basis}) for fermions. For $\alpha=1$ the extra $\partial_{r_0}$-derivative gives rise to an additional complication, but we still continue to search for a consistent expansion around the external time $t$ as before. We begin by writing the spectral propagator (\ref{Dist}) in two-time representation as 
\begin{eqnarray}\label{two-time_sca}
i \Delta_{ij}^<(w_0,w_0',\mathbf{k}) &=& \int \frac{{\rm d} k_0}{2\pi} e^{-ik_0(w_0-w_0')} i \Delta_{ij}^< \big(k_0,\mathbf{k},\frac{w_0+w_0'}{2}\big)
\nonumber\\[2mm]
&=&  \frac{\bar \omega_{ij}}{2 {\omega_i\omega_j}} \sum \limits_{\pm}\Big[ \pm e^{\mp i \bar\omega (w_0-w_0') \mp i\Delta\omega (w_0+w_0'-2t)} f_{\pm}^{m <,>}(\mathbf{k},t) 
\nonumber\\
&& \qquad\qquad\;\;\;\;
+ e^{\mp i \Delta\omega(w_0-w_0') \mp i \bar\omega (w_0+w_0'-2t)} f_{\pm}^{c <,>}(\mathbf{k},t)\Big]_{ij} + {\cal O}^1\,,
\label{eff_prop_sca}
\end{eqnarray}
where we Taylor-expanded $f_{\alpha}\big(\mathbf{k},\frac{w_0+w_0'}{2}\big)$ around the external time variable $t$, and used the recursive zeroth order equations of motion: $\partial_t^n f^m_{\pm} = (\mp i2\Delta\omega)^n f^m_{\pm} + {\cal O}^1$ and $\partial_t^n f^c_{\pm} = (\mp i 2\bar\omega)^n f^c_{\pm} + {\cal O}^1$,
exactly as in the fermionic case. Now, using the expanded propagator (\ref{eff_prop_sca}) in the collision integrals (\ref{coll_int_sca}) we get:
\begin{align}
\langle{\cal C}_\alpha\rangle_{ij} = -\frac{1}{2}\sum \limits_{\pm, k} &\frac{\bar \omega_{kj}}{2 \omega_k \omega_j} \bigg( 
\pm \big( \pm \bar\omega_{kj} + \frac{i}{2} \partial_t \big)^\alpha i \Pi^{>}_{{\rm eff},ik}(\pm \omega_k,\mathbf{k},t)\,f^{m <}_{kj\pm}(\mathbf{k},t) 
\nonumber\\
 +& \,\big(\pm \Delta\omega_{kj} + \frac{i}{2} \partial_t \big)^\alpha i \Pi^{>}_{{\rm eff},ik}(\pm \omega_k,\mathbf{k},t)\,f^{c <}_{kj\pm}(\mathbf{k},t) 
\bigg) \,-\, \Big[ > \leftrightarrow < \Big] \,,\quad
\label{coll_final_sca}
\end{align}
where the time-derivative operator acts only on the effective self-energies defined as
\beq
\Pi^{<,>}_{{\rm eff},ij}(k,t) \equiv \int {\rm d} w_0 e^{ik_0(t-w_0)}\Pi^{<,>}_{ij}(t,w_0,{\bf k}) \, .
\label{Sigma_eff_proj}
\eeq
Note again the simplicity of these results; to get the collision integral for an arbitrary moment equation, we only need to evaluate a single generic function $\Pi^{<,>}_{{\rm eff},ik}(\pm \omega_{k},\mathbf{k},t)$, which is of the same form as the fermionic effective self-energy function (\ref{eff_sigma}).

The collision integrals appearing in the equations of motion (\ref{EvoEqMoments}) are just the hermitian and the antihermitian parts of Eqs.~(\ref{coll_final_sca}): $\langle C_{H {\rm d}} \rangle = 
\frac{1}{2} ( \langle C_0 \rangle + \langle C_0^\dagger \rangle)$,
$\langle C_{A {\rm d}} \rangle 
= \frac{1}{2i} (\langle C_0 \rangle -\langle C_0^\dagger \rangle )$  and 
$\langle k_0 C_{H {\rm d}} \rangle = \frac{1}{2}( \langle C_1 \rangle + \langle C_1^\dagger \rangle)$.
Using the conventions (\ref{Coll_gen_comm3}) for the generalized (anti)commutators in flavour space, we finally obtain the following explicit expressions:
\begin{eqnarray}
\langle C_{A {\rm d}} \rangle_{ij} &=& - \frac{1}{8}\sum_\pm \mp \bigg( 
i\, \big[ i\Pi^{>}_{{\rm eff}\pm} , \tfrac{1}{\omega} \, f^{<+}_{\pm}\big]^{m}_{ij} 
+ \frac{\bar \omega_{ij} }{\omega_{i}\omega_{j}}\, 
   i\, \big[ i\Pi^{>}_{{\rm eff}\pm} , f^{<+}_{\pm} \big]^{m}_{ij}
\nonumber\\
&& \hskip2cm
+ \frac{i \Delta \omega_{ij} }{\omega_{i}\omega_{j}} \{ i\Pi^{>}_{{\rm eff}\pm} , f^{< +}_{\pm} \}^{m}_{ij} 
- \Big[ > \leftrightarrow < \Big] \bigg) \,,
\label{Scalar_real_CT1} \\
\langle C_{H {\rm d}} \rangle_{ij}  &=& - \frac{1}{8}\sum_\pm \pm \bigg( 
\{ i\Pi^{>}_{{\rm eff}\pm} , \tfrac{1}{\omega} \, f^{< +}_{\pm} \}^{m}_{ij} 
+ \frac{\bar \omega_{ij} }{\omega_{i}\omega_{j}}  \{ i\Pi^{>}_{{\rm eff}\pm} , f^{< +}_{\pm} \}^{m}_{ij}
\nonumber\\
&& \hskip2cm
- \frac{i \Delta \omega_{ij} }{\omega_{i}\omega_{j}}\,i\, \big[ i\Pi^{>}_{{\rm eff}\pm} , f^{< +}_{\pm} \big]^{m}_{ij}
- \Big[ > \leftrightarrow < \Big] \bigg) 
\label{Scalar_real_CT2}
\end{eqnarray}
and
\begin{eqnarray}
\langle k_0 C_{H {\rm d}} \rangle_{ij} &=&  - \frac{1}{16}\sum_\pm \bigg( 
\{ i\Pi^{>}_{{\rm eff}\pm} , (f^{< +}_\pm + f^{< -}_\pm) \}^{m}_{ij} 
\pm \{ i \partial_t( i\Pi^{>}_{{\rm eff}\pm} ) , \tfrac{1}{\omega} \,  f^{< +}_{\pm} \}^{m}_{ij}
\nonumber\\
&+&  \frac{\bar \omega_{ij} }{\omega_{i}\omega_{j}} \Big(
\{ i\Pi^{>}_{{\rm eff}\pm} , \omega \, f^{< +}_{\pm} \}^{m}_{ij}
+ \omega_i \omega_j \{  i\Pi^{>}_{{\rm eff}\pm} , \tfrac{1}{\omega} \, f^{< -}_{\pm} \}^{m}_{ij} 
\pm \{ i \partial_t( i\Pi^{>}_{{\rm eff}\pm} ) , f^{< +}_{\pm} \}^{m}_{ij} \Big)
\nonumber\\
&+& \frac{\Delta \omega_{ij} }{\omega_{i}\omega_{j}} \Big(
\big[ i\Pi^{>}_{{\rm eff}\pm} , \omega \, f^{< +}_{\pm} \big]^{m}_{ij}
- \omega_i \omega_j \big[ i\Pi^{>}_{{\rm eff}\pm} , \tfrac{1}{\omega} \, f^{< -}_{\pm} \big]^{m}_{ij}    
\pm \big[ i \partial_t( i\Pi^{>}_{{\rm eff}\pm} ) , f^{< +}_{\pm} \big]^{m}_{ij} \Big)
\nonumber\\
&-& \Big[ > \leftrightarrow < \Big] \bigg) \,,
\label{Scalar_real_CT3}
\end{eqnarray}
where the diagonal matrix kernels are defined in analogy to Eq.~(\ref{diag_kernels}):
\begin{equation}
(\omega)_{ij} = \delta_{ij}{\omega_i}
\quad {\rm and} \quad
(\frac{ 1}{\omega})_{ij} = \delta_{ij}\frac{1}{\omega_i} \,,
\end{equation}
and we have again used shorthand notations:
\begin{equation}
f^{< +}_{\pm} \equiv \frac{1}{2}(f^{m <}_{\pm} \pm f^{c <}_{\pm})
\, , \quad
f^{< -}_{\pm} \equiv \frac{1}{2}(f^{m <}_{\pm} \mp f^{c <}_{\pm})
\quad {\rm and} \quad
\Pi^{<,>}_{{\rm eff}, ij \pm} \equiv   \Pi^{<,>}_{{\rm eff},ij}(\pm \omega_j) \,.
\end{equation}
As emphasized before, in order to perform practical calculations using Eqs.~(\ref{EvoEqMoments}) with the moment integrals as dynamical variables, one has to use the inverse relations of Eqs.~(\ref{RelationEq}) to express the collision integrals (\ref{Scalar_real_CT1}-\ref{Scalar_real_CT3}) in terms of the moment functions $\rho_{0,1,2}$ and $\partial_t \rho_0$.  In practice, this inversion can be left to numerical routines, since the expressions of the collision terms take the simplest form in terms of $f$-functions.

%
\section{Momentum space Feynman rules}
\label{sect:effective-feynman}
%

In this section we derive generalized Feynman rules for computing the effective self-energy functions $i \Sigma_{\rm eff}$ and $i \Pi_{\rm eff}$ through perturbative techniques, including the coherence effects. Standard methods, such as the 2PI formalism~\cite{2PI}, exist for diagrammatic expansion of the two-time self-energies $\Sigma^{ab}(u,v)$ and $\Pi^{ab}(u,v)$ appearing in equations (\ref{eff_sigma}) and (\ref{Sigma_eff_proj}), and our task is to reduce the computation of the diagrams into a set of momentum space Feynman rules.
We derive these rules by replacing all propagators in a given diagram by our resummed propagators (\ref{eff_prop}) and (\ref{two-time_sca}), transforming these propagators to momentum space and performing all time integrations related to internal vertices.  To be specific we shall consider a generic Yukawa type interaction Lagrangian between the scalar and fermion fields:
\begin{equation} 
{\cal L}_{\rm int} = - y_{ij}^l\; \phi_l \,\bar \psi_i\, \psi_j + h.c. \,.
\label{eq:yukawaint}
\end{equation}
However, from our final results it is obvious how these Feynman rules generalize to arbitrary interactions. 

The main part of the derivation consists of completing the time-integrals at vertices, accounting for the nontrivial phase factors in propagators.  To begin with, it is convenient to rewrite the propagators (\ref{eff_prop}) and (\ref{two-time_sca}) in a generic 4-dimensional representation:
\begin{align}
S^{<,>}_{ij}(w_0,w_0',{\bf k}) =& \int \frac{{\rm d}k_0}{2\pi} e^{-ik_0(w_0-c_{ij} w_0') 
+ ik_0(1-c_{ij} )t} S^{<,>}_{{\rm eff, in}, ij}(k,t)\,,
\nonumber\\
\Delta^{<,>}_{kl}(w_0,w_0',{\bf p}) =& \int \frac{{\rm d}p_0}{2\pi} e^{-ip_0(w_0-c_{kl} w_0') 
+ ip_0(1-c_{kl} )t} \Delta^{<,>}_{{\rm eff, in}, kl}(p,t)\,.
\label{eq:effphase}
\end{align}
where $S^{<,>}_{{\rm eff, in}, ij}(k,t)$ are the (effective) in-propagators, given by Eq.~(\ref{effectiveS1}), with a similar definition for the scalar propagators $\Delta^{<,>}_{{\rm eff, in}, kl}(p,t)$.\footnote{In this section we use $k,l$ instead of $i,j$ to denote scalar flavour-indices.} The phase factors are given by $c_{ij} = + \omega_j / \omega_i$ for $G^{m <,>}_{{\rm eff, in}, ij}$, and  $c_{ij} = - \omega_j / \omega_i$ for $G^{c <,>}_{{\rm eff, in}, ij}$, where we use generic notation $G=\{S,\Delta\}$ for fermion and scalar propagators. Because the phase factors are different for $m$- and $c$-parts of the full $G^{<,>}_{{\rm eff, in}, ij}$, Eq.~(\ref{eq:effphase}) is understood to hold for the corresponding parts separately. For the flavour-diagonal pole-propagators we use the standard expressions, Eq.~(\ref{pole-props}) in two-time representation and similarly for scalars, which are directly of the form (\ref{eq:effphase}) with ``normal'' phase factors $c_{ii}=+1$. We note that the normal phase factor also applies to the flavour-diagonal mass-shell propagators $G^{m <,>}_{ii}$. For all other (component) propagators, which encode the flavour or particle-antiparticle coherence, the overall phase factors are ``abnormal", except in the particular case of $w_0'=t$, where the nontrivial $c_{ij}$-terms cancel. 

To proceed, we need to show that the total $t$-phase arising from $\exp(ik_0(1-c_{ij} )t)$-factors in the decompositions (\ref{eq:effphase}) vanishes for an arbitrary self-energy $\Sigma_{\rm eff}(k,t)$ or $\Pi_{\rm eff}(k,t)$ diagram. This can be shown by a direct computation using Eqs.~(\ref{eq:effphase}) and following the identical steps as in~\cite{HKR4} for single-flavour case. Alternatively, by the flavour-covariant formulation of \cite{HKR5} the cancellation of the total $t$-phase follows trivially by the shift of vertex time-integration variables.
After the cancellation of the $t$-phase, the derivation of the Feynman rules is straightforward. Based on Eq.~(\ref{eq:effphase}) one would obtain rules with (effective) propagators $iS^{<,>}_{{\rm eff, in}, ij}(k,t)$ and $i\Delta^{<,>}_{{\rm eff, in}, kl}(p,t)$ for fermion and scalar lines, and the vertex rule with modified energy-conservation delta-functions involving $c_{ij}$-factors. These rules would provide a generalization of the Feynman rules derived in ref.~\cite{HKR4} to the case of flavour-mixing fields. However, these Feynman rules are somewhat complicated to use in practice, because of the nontrivial $c_{ij}$-factors in the vertex rule with a delicate dependence on the flavour and particle-antiparticle coherence components and the direction signatures associated with the propagators. We shall therefore follow a different approach here.

%
\subsection{Extended Kadanoff-Baym ansatz}
\label{sect:extendedKB}
%

An equivalent, but more intuitive and transparent set of Feynman rules can be obtained by rewriting the two-time Wightman functions (\ref{eff_prop}) and (\ref{two-time_sca}) in the {\em doubly Fourier transformed} form (this step is not necessary for the pole propagators):
\beqa 
S^{<,>}_{ij}(w_0,w_0',\mathbf{k}) &=& \int \frac{{\rm d}k_0}{(2\pi)} \frac{{\rm d}^4k'}{(2\pi)^4}\, 
e^{-i(k_0w_0-k_0'w_0')} S^{<,>}_{{\rm eff}\,ij}(k,k';t) \,,
\nonumber\\
\Delta_{kl}^{<,>}(w_0,w_0',\mathbf{p}) &=& \int \frac{{\rm d}p_0}{(2\pi)} \frac{{\rm d}^4p'}{(2\pi)^4}\, 
e^{-i(p_0w_0-p_0'w_0')} \Delta^{<,>}_{{\rm eff}\,kl}(p,p';t) \,,  
\label{FeynmanRules0}
\eeqa
where we have dropped the $t$-phase factor which cancels at the end, as discussed above. The idea here is that after performing time integrations in internal vertices $w_0$ and $w_0'$ we are left with normal energy conservation delta functions in momentum space at the price of having different 4-momenta $k$ and $k'$ at the ``in" and ``out" ends of each Wightman function. This structure is nicely reflected in the explicit forms of the doubly transformed propagators:
\beqa 
i S^{<,>}_{{\rm eff}\,ij}(k,k';t) &=&  
{\cal A}^{\psi}_{ii}(k)\, F^{<,>}_{\psi\, ij}(k,k';t)\, {\cal A}^{\psi}_{jj}(k') \,, \nonumber\\
i \Delta^{<,>}_{{\rm eff}\,kl}(p,p';t) &=& {\cal A}^{\phi}_{kk}(p) \,  F^{<,>}_{\phi \, kl}(p,p';t) \, {\cal A}^{\phi}_{ll}(p') \,,  
\label{FeynmanRules1}
\eeqa
where the spectral functions ${\cal A}^{\psi}_{ii}(k)$ and ${\cal A}^{\phi}_{kk}(p)$ are defined in Eqs.~(\ref{full_spectral}) and (\ref{full_spectral_sca}), and
\begin{align}
F^{<,>}_{\psi \, ij}(k,k';t) \equiv& 4 (2 \pi)^3 \delta^3({\bf k} - {\bf k'})
\sum_{h,\pm} P_h({\bf \hat k}) \theta_\pm^k 
\Big( \theta_\pm^{k'}f^{m <,>}_{ij h \pm}  
    + \theta_\mp^{k'} f^{c <,>}_{ij h \pm}  \Big) \,,
\nonumber\\
F^{<,>}_{\phi\,kl}(p,p';t) \equiv& 4 (2 \pi)^3 \delta^3({\bf p} - {\bf p'})
\sum_{\pm} \pm 2 \bar\omega_{kl} \theta_\pm^p  
\Big(  \theta_\pm^{p'} f^{m <,>}_{kl \pm}  
   \mp \theta_\mp^{p'} f^{c <,>}_{kl \pm}  \Big) \,,
\label{FeynmanRules2}
\end{align}
where \eg $\theta_\pm^k \equiv \theta(\pm k_0)$ and $f^{m <,>}_{ij h \pm} \equiv f^{m <,>}_{ij h \pm}({\bf k},t)$. The $F^{<,>}_{\psi,\phi \, ij}$-functions can be understood as {\em effective 2-point vertex functions}, which encode all quantum statistical information of the Wightman functions, including flavour and particle-antiparticle coherence, as well as the helicity structure for fermions. Each Wightman function can then be viewed as a composite operator consisting of two standard on-shell spectral functions coupled by effective 2-point interactions; the amount of coherence in the system is directly manifested as the strength of these couplings.

Now, using the following properties of the spectral functions:
\beqa 
\int \frac{{\rm d}^4k'}{(2\pi)^4}  
{\cal A}^{\psi}_{ii}(k) \, 
\Big[ \pm \frac{\omega_i}{m_i} \theta_\pm^k \theta_\pm^{k'} \delta_{ij} 
     (2 \pi)^3 \delta^3({\bf k} - {\bf k'}) 
\Big] \, 
{\cal A}^{\psi}_{jj}(k')  
&=& \frac{1}{2} {\cal A}^{\psi}_{ii}(k) \theta_\pm^k \,,
\nonumber\\
\int \frac{{\rm d}^4p'}{(2\pi)^4} 
{\cal A}^{\phi}_{kk}(p) \, 
\Big[ \pm 2 \omega_{k} \theta_\pm^p \theta_\pm^{p'} \delta_{kl} 
    (2 \pi)^3 \delta^3({\bf p} - {\bf p'}) \Big] \,  
{\cal A}^{\phi}_{ll}(p')  
&=& \frac{1}{2} {\cal A}^{\phi}_{kk}(p) \theta_\pm^p \,,  
\label{FeynmanRules3}
\eeqa
we can connect our fully diagonal (both in flavour and in energy) Wightman functions to their usual kinetic theory counterparts:
\beqa 
\int \frac{{\rm d}^4k'}{(2\pi)^4} \, i S^{m <,>}_{ii \, \rm eff}(k,k';t) 
&=& 2 {\cal A}^{\psi}_{ii}(k)\, \sum_{h,\pm}\pm P_h({\bf \hat k})  \theta_\pm^k \frac{m_i}{\omega_i} f^{m <,>}_{ii h \pm} \, , \nonumber\\
\int \frac{{\rm d}^4p'}{(2\pi)^4} \, i \Delta^{m <,>}_{kk \, \rm eff}(p,p';t) 
&=& 2 {\cal A}^{\phi}_{kk}(p) \, \sum_{\pm} \theta_\pm^p f^{m <,>}_{kk \pm} \,.  
\label{FeynmanRules4}
\eeqa
One readily recognizes the form of the {\em Kadanoff-Baym} ansatz here. Thus, our effective Wightman functions can be understood as an {\em enlarged} Kadanoff-Baym ansatz, which allows to encode the coherence information as well. Conversely, it is now clear why making the KB-ansatz for the propagator from the outset completely misses all coherence information.

%
\subsection{cQPA Feynman rules}
\label{sect:practical-rules}
%

We will now present the complete set of momentum space Feynman rules to compute the effective self-energies $i\Sigma_{\rm eff}(k,t)$ and $i\Pi_{\rm eff}(k,t)$ involving the flavour-coherent propagators in the Yukawa theory with interaction (\ref{eq:yukawaint}). Let us first write down the general relations between the primary CTP-propagators $G^{ab}$, $a,b=\pm$, appearing in generic loop-diagrams in the CTP-formalism, and the pole- and the Wightman-propagators:
\begin{align}
G^{++}&\equiv G^t = G^r \mp G^< \,,\quad && G^{+-} \equiv \mp G^<\,,
\nonumber\\ 
G^{--}&\equiv G^{\bar t} = - G^a \mp G^< \,,\quad && G^{-+} \equiv G^>\,,
\label{CTP-props}
\end{align}
where the upper $-$ sign refers to fermionic correlator $S^<$ and lower $+$ sign to scalar correlator $\Delta^<$. The cQPA Feynman rules can now be stated as follows:

\begin{itemize}

\item{} Draw all perturbative self-energy diagram(s) according to standard CTP-formalism, associate the overall signs and symmetry factors for the diagrams and the CTP-indices for the propagators.

\item{} For each vertex, except the out-vertex of $i\Sigma_{\rm eff}(k,t)$ or $i\Pi_{\rm eff}(k,t)$, associate a normal vertex Feynman rule. For example, in the Yukawa theory (\ref{eq:yukawaint}) the vertex rule for the incoming scalar line is (here $a$ is the CTP branch index of the vertex): 
\beq
 y^l_{ij} \phi_l\bar\psi_i\psi_j : -i \, y_{ij}^l \, 
 (2\pi)^4 \delta^4(k' - k + p)\,a\,.
\label{eq:yfrule1}
\eeq
The rule for the scalar out-vertex is identical to Eq.~(\ref{eq:yfrule1}) with complex conjugated coupling constant $y \rightarrow y^{*} $.
The out-vertex of the self-energy has the same rule but {\em without} 4-momentum conservation delta
 function.

\item{}
Use the relations (\ref{CTP-props}) to write the CTP-propagators in terms of $iG^{r,a}$ and $iG^{<,>}$. 

\item{} For all pole-type propagators $iG^{r,a,t_0,\bar t_0}(k)$ substitute the standard (vacuum) propagator (\eg Eq.~(\ref{pole-props})) and integrate over $k : \int \frac{{\rm d}^4k}{(2\pi)^4}\,$ as usual. 
(Here $iG^{t_0,\bar t_0}$ refer to the pole-parts of the full Feynman and anti-Feynman propagators.)\footnote{Sometimes it may be useful to include the vacuum parts of the Wightman functions $iG^{<,>}$ to the Feynman and anti-Feynman propagators, and use $iG^{t_0,\bar t_0}$ instead of $iG^{r,a}$ as the basic pole-propagators.}

\item{} For each Wightman function substitute a composite propagator $iG^{<,>}_{\rm eff}(k,k')$ 
given by Eqs.~(\ref{FeynmanRules1}) and integrate over both 4-momentum variables $k$ and $k'$: 
$\int \frac{{\rm d}^4k}{(2\pi)^4} \frac{{\rm d}^4k'}{(2\pi)^4}\,$.
\end{itemize}
\begin{figure}[t]
\centering
\includegraphics[width=1.0 \textwidth]{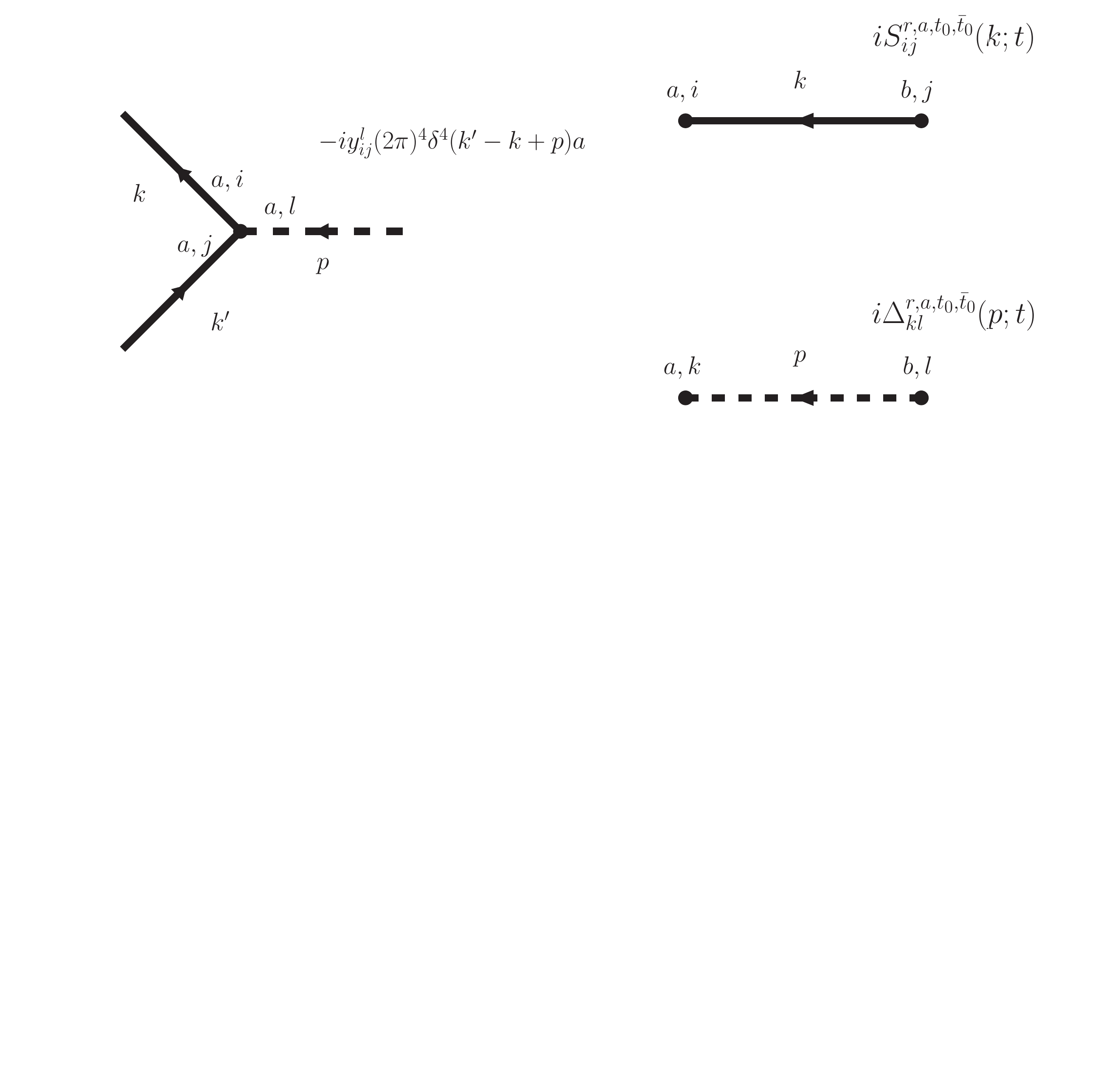}
\vskip -8cm
\caption{The Feynman rules for the pole-type propagators $iG^{r,a,t_0,\bar t_0}$ and the scalar in-vertices.}
\label{fig:FeynmanRules1}
\end{figure}
\begin{figure}[t]
\centering
\vskip 1cm
\includegraphics[width=1.0 \textwidth]{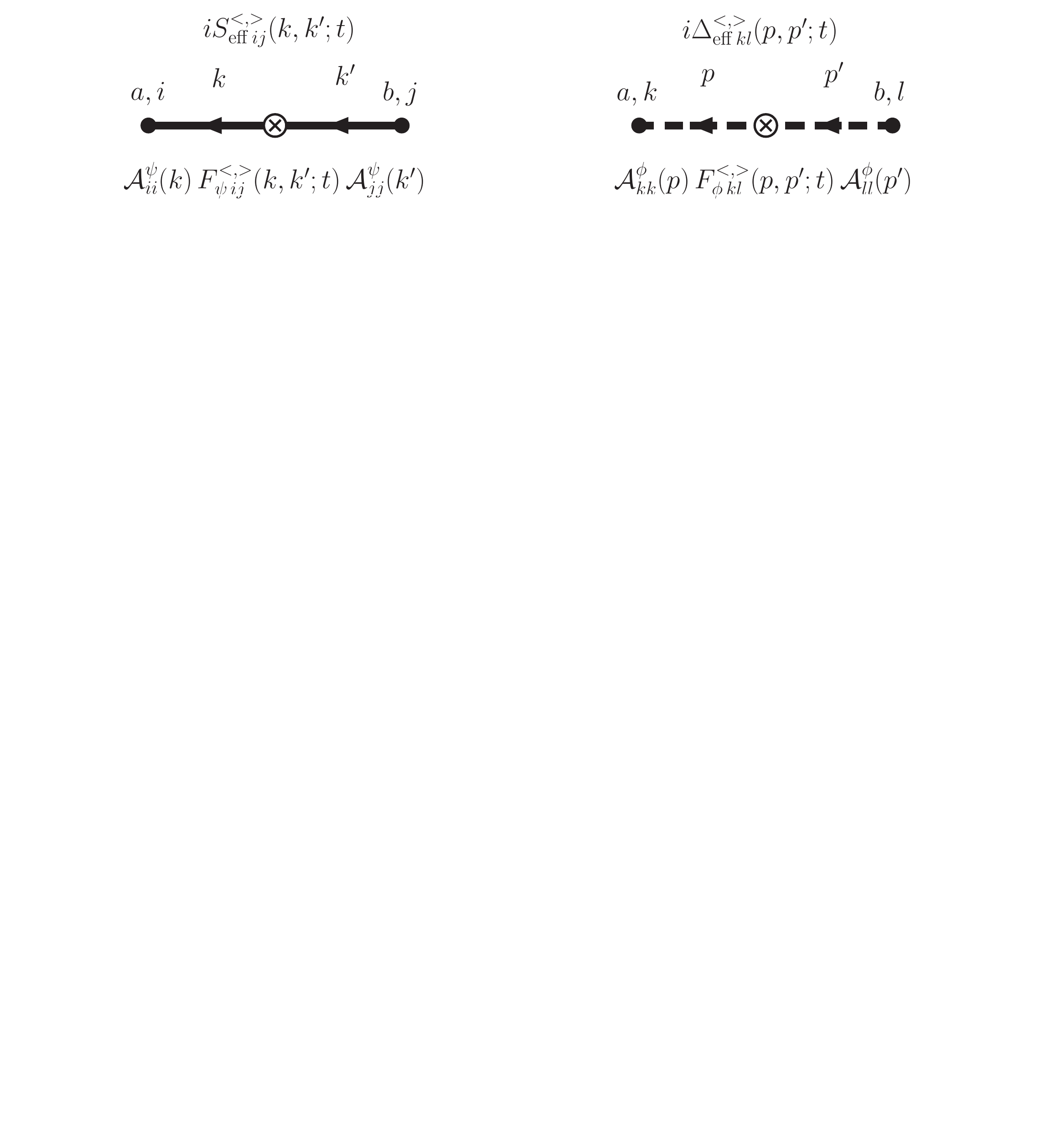}
\vskip -13cm
\caption{The Feynman rules for the composite Wightman propagators $iG^{<,>}$.}
\label{fig:FeynmanRules2}
\end{figure}
These rules are presented graphically in the figures~\ref{fig:FeynmanRules1} and \ref{fig:FeynmanRules2}, where the arrows in the propagator lines indicate the direction of the flow corresponding to the 4-momentum of the positive energy state. Although these rules are specifically designed to compute the self-energy functions $i\Sigma_{\rm eff}(k,t)$ and $i\Pi_{\rm eff}(k,t)$, there is no restriction for computing arbitrary diagrams with arbitrary number of external legs however. In this case, one would need to extend the rules to include in- and out-propagators as well. More details and examples related to cQPA Feynman rules in flavour-covariant formulation are presented in \cite{HKR5}.

%
\section{Application to $CP$-violating flavour mixing in the presence of collisions}
\label{sec:applications}
%

%
\subsection{Setup}
\label{sec:setup}
%

As an application to our formalism, we consider a scenario motivated by electroweak baryogenesis (EWBG): $N \times N$ fermionic mixing through a slowly-varying time-dependent and $CP$-violating mass matrix, in the presence of Yukawa-interactions with thermal bath. To be specific, the interaction Lagrangian is given by
\beq
{\cal L}_{\rm int} = - y_\ell\,\bar\psi_{\ell L} \psi_{qR}\,\phi + h.c.\,,
\eeq
where $\psi_\ell$, $\ell = 1,\ldots,N$, are the mixing fermionic fields in the flavour basis, $\psi_q$ and $\phi$ are a fermionic field and a complex scalar field assumed to be in thermal equilibrium, and $y_\ell$ are flavour-sensitive Yukawa couplings. First, we observe that the leading order (one-loop) self-energies for $\psi_\ell$ do not involve out-of-equilibrium propagators and coherence contributions, and therefore the effective self-energy functions in Eq.~(\ref{eff_sigma}) reduce to standard Wigner representation self-energies $\Sigma^{<,>}(k,t)$ (see discussion in section \ref{sec:fermicollision}).
Apart from the flavour-coupling matrix, these thermal one-loop self-energies are completely identical to the ones for single-flavour case, considered in ref.~\cite{HKR2}:
\beqa 
\Sigma^<_{\ell\ell^\prime}(k,t) &=& i\,y_\ell y_{\ell^\prime}^* \int \frac{{\rm d}^4k'}{(2\pi)^4} P_R S_{q,{\rm eq}}^{<}(k',t) P_L \Delta_{\rm eq}^{<}(k-k',t)
\nonumber\\
&=& y_\ell y_{\ell^\prime}^* \Big(\Sigma^<_0 \,\gamma^0 - \Sigma^<_3 \,\mathbf{\hat{k}}\cdot \vec{\gamma}\Big) P_L \,,  
\label{self1}
\eeqa
where $i\Sigma^<_{0,3}(k_0,\mathbf{k},t)$ are real functions, which eventually need to be evaluated on the mass-shells of the mass-basis quasiparticles (see Eq.~(\ref{Sigma_eff_proj}) and Eq.~(\ref{self3}) below):
\beqa
i\Sigma^<_0(\pm\omega_i,\mathbf{k},t) 
   &=& \frac{T^2}{8\pi |\mathbf{k}|} |I_1(\pm\omega_i,\mathbf{k},t)| \,, 
\\ 
i\Sigma^<_3(\pm\omega_i,\mathbf{k},t) 
   &=& \pm \frac{T^2 \omega_i}{8 \pi \mathbf{k}^2}\left(|I_1(\pm\omega_i,\mathbf{k},t)| 
              - \frac{|\alpha| \,m_i^2}{\omega_i^2} |I_0(\pm\omega_i,\mathbf{k},t)|\right)\,, 
\label{Sigma_0,3}
\eeqa    
with
\beq 
I_n(k_0,\mathbf{k},t) 
  = \theta(\lambda) \int_{\alpha - \delta}^{\alpha + \delta} {\rm d}y \: y^n 
       \frac{1}{(e^y + 1) (e^{k_0/T - y} - 1 )} \,.
\label{I_n}
\eeq  
and
\beqa
\alpha &=& \frac{m_i^2+m_q^2-m_\phi^2}{2 m_i^2} \frac{k_0}{T}
       = \frac{\sqrt{s}E_*k_0}{m_i^2 T}\,,
\nonumber \\
\delta  &=& \frac{\lambda^{1/2}(m_i^2, m_q^2,m_\phi^2)}{2 m_i^2} 
               \frac{|\mathbf{k}|}{T}
       = \frac{\sqrt{s}p_*|\mathbf{k}|}{m_i^2 T} \,,
\eeqa
where $E_*$ and $p_*$ are the energy and the momentum of the decay products in the decay frame, $\sqrt{s}$ is the invariant mass of the decaying (heaviest) particle and $\lambda(a,b,c) = (a+b-c)^2-4bc$ is the usual kinematic phase-space function.\footnote{The time-dependence in Eqs.~(\ref{Sigma_0,3}-\ref{I_n}) comes through time-dependent masses $m_i=m_i(t)$.} Furthermore, the thermal self-energies $\Sigma^{<,>}(k,t)$ are related by KMS relation \cite{KMS}: $\Sigma^>(k,t) = e^{\beta k_0}\Sigma^<(k,t)$, which follows directly from the KMS relations of the thermal correlators $S_{q,{\rm eq}}^{<,>}$ and $\Delta_{\rm eq}^{<,>}$.

After rotation to the mass basis, the self-energy $\Sigma^<$ is given by 
\beq
\Sigma_{\rm d}^<(k,t) = X \Sigma^< Y^\dagger = y_{\rm d}^2 \Big(\Sigma^<_0 \,\gamma^0 - \Sigma^<_3 \,\mathbf{\hat{k}}\cdot \vec{\gamma}\Big) P_L\,,
\label{self2}
\eeq
where the rotated coupling matrix is $y^2_{ij} \equiv (y^2_{\rm d})_{ij} \equiv  U_{i\ell} y_\ell y_{\ell^\prime}^* U^\dagger_{\ell^\prime j}$. Using Eq.~(\ref{self2}) together with the KMS relation for the effective self-energies $i\Sigma^{<,>}_{\rm eff}(k,t)$ it is easy to compute the flavour structure of collision terms from (\ref{Coll_eff_proj_m} - \ref{Coll_eff_proj_c}) to get:
\begin{eqnarray} 
C^{m}_{h\pm}[ f ]_{ij} &=& -\frac{1}{2} \Big( 
   \{ i \Sigma^{>}_{h\pm} , \tfrac{1}{2}(1 \mp h{v}_{{\bf k}}) \, f^{m <}_{h\pm} \}^{m}_{ij} 
 + \{ i \Sigma^{>}_{h\mp} , \tfrac{1}{2}(1 \pm h{v}_{{\bf k}}) \, f^{c <}_{h\mp} \}^{m}_{ij}  
\nonumber\\
&\pm&\,  \frac{h |\mathbf{k}| \Delta\omega_{ij} }{\Omega^{2}_{m ij}} \Big( 
  \big[ i \Sigma^{>}_{h\pm} , \tfrac{1}{2}(1 \mp h{v}_{{\bf k}}) \, f^{m <}_{h\pm} \big]^{m}_{ij} 
+ \big[ i \Sigma^{>}_{h\mp} , \tfrac{1}{2}(1 \pm h{v}_{{\bf k}}) \, f^{c <}_{h\mp} \big]^{m}_{ij} \Big) 
\label{app_Sigma_proj_m}\\
&+&\, \tfrac{1}{2} \big( 1 \mp  \tfrac{h |\mathbf{k}|}{\bar \omega_{ij}} \tfrac{\bar m^2_{ij} }{\Omega^{2}_{m ij} }\big)  \tfrac{\bar \omega_{ij} }{ \bar m_{ij} }
\big( \{ i \Sigma^{>}_{h\pm} , \tfrac{m}{\omega} \, f^{m <}_{h\pm} \}^{m}_{ij} - \{ i \Sigma^{>}_{h\mp} , \tfrac{m}{\omega} \, f^{c <}_{h\mp} \}^{m}_{ij} \big) 
 - \big[ < \leftrightarrow > \big] \Big) \,,
\nonumber
\end{eqnarray}
\begin{eqnarray}
C^{c}_{h\pm}[ f ]_{ij} &=& -\frac{1}{2} \Big( 
   \{ i \Sigma^{>}_{h\pm} , \tfrac{1}{2}(1 \mp h{v}_{{\bf k}}) \, f^{c <}_{h\pm} \}^{c}_{ij}  
 + \{ i \Sigma^{>}_{h\mp} , \tfrac{1}{2}(1 \pm h{v}_{{\bf k}}) \, f^{m <}_{h\mp} \}^{c}_{ij} 
\nonumber\\
&\mp&\,  \frac{h |\mathbf{k}| \bar \omega_{ij} }{\Omega^{2}_{c ij}} \Big( 
    \big[ i \Sigma^{>}_{h\pm} , \tfrac{1}{2}(1 \mp h{v}_{{\bf k}}) \, f^{c <}_{h\pm} \big]^{c}_{ij}
  + \big[ i \Sigma^{>}_{h\mp} , \tfrac{1}{2}(1 \pm h{v}_{{\bf k}}) \, f^{m <}_{h\mp} \big]^{c}_{ij}  \Big) 
\label{app_Sigma_proj_c}\\
&+&\, \tfrac{1}{2} \big( \tfrac{\Delta \omega_{ij}}{\bar \omega_{ij}}  \pm  \tfrac{h |\mathbf{k}|}{\bar \omega_{ij}} \tfrac{\bar m^2_{ij} }{\Omega^{2}_{c ij} }\big) 
\tfrac{\bar \omega_{ij} }{ \bar m_{ij} }
\big( \big[  i \Sigma^{>}_{h\pm} , \tfrac{m}{\omega} \,f^{c <}_{h\pm} \big]^{c}_{ij} - \big[ i \Sigma^{>}_{h\mp} , \tfrac{m}{\omega} \, f^{m <}_{h\mp} \big]^{c}_{ij}  \big) 
 - \big[ < \leftrightarrow > \big] \Big) \,,
\nonumber 
\end{eqnarray}
where the helicity-projected scalar self-energy functions are given by
\begin{equation} 
     i \Sigma^{>}_{ij h\pm}({\bf k},t) =  \frac{y^2_{ij}(t)}{2} \Big( i \Sigma^{>}_{0}(\pm \omega_j,{\bf k},t)  + h \,i \Sigma^{>}_{3}(\pm \omega_j,{\bf k},t) \Big) \,.
\label{self3}
\end{equation}

To proceed further we need to specify the flavour-mixing scenario. We shall restrict ourselves to the case of two 
flavours. The schematic structure of the dispersion relations for two-flavour mixing is presented in Fig~\ref{fig:shells}. We use the following parametrization for the time-dependent 
flavour non-diagonal mass matrix:
\beq
m(t) = \left( \begin{array}{cc}
     m_A & v(t)e^{-i\sigma(t)} \\
     v(t)e^{i\sigma(t)} & m_B 
    \end{array} \right)\,,
\eeq
where the off-diagonal magnitude $v(t)$ and the phase $\sigma(t)$ are assumed to have the following functional forms motivated by EWBG considerations (see ref.~\cite{CLR-MT10}):
\beqa
v(t) &=& \frac{v_0}{2}\Big(1+\tanh\Big(\frac{t}{\tau_w}\Big)\Big) \,,
\nonumber\\
\sigma(t) &=& \frac{\sigma_0}{2}\Big(1+\tanh\Big(\frac{t}{\tau_w}\Big)\Big) \,.
\eeqa
We have chosen here a hermitian mass matrix for simplicity, such that it is diagonalized by a unitary transformation:
\beq
m_{\rm d} = U m U^\dagger = \left( \begin{array}{cc}
     m_1 & 0 \\
     0 & m_2 
    \end{array} \right)\,,
\qquad\quad U(t) = \left( \begin{array}{cc}
     \cos\theta(t) & \sin\theta(t) e^{-i\sigma(t)} \\
     -\sin\theta(t)e^{i\sigma(t)} & \cos\theta(t) 
    \end{array} \right)\,,
\label{2x2_mixing_matrix}
\eeq
with
\beqa
m_{1,2}(t) &=& \frac{1}{2}(m_A + m_B) \pm \frac{1}{2}\sqrt{(m_A - m_B)^2 + 4 v^2(t)}\,,
\nonumber\\
\tan(2\theta(t)) &=& \frac{2 v(t)}{m_A - m_B}\,. 
\eeqa
Using Eq.~(\ref{2x2_mixing_matrix}) with $V=U$ it is now easy to compute the mixing-gradient matrices (\ref{mixing_grad}) appearing 
in the Boltzmann equations (\ref{eom_on-shell1}-\ref{eom_on-shell2}):
\beq
\Xi^{\prime +} = i U \partial_t U^\dagger = i\left( \begin{array}{cc}
     0 & -e^{-i\sigma} \\
     e^{i\sigma} & 0 
    \end{array} \right)
\dot\theta - \left( \begin{array}{cc}
     \sin^2\theta & \frac{1}{2}\sin(2\theta) e^{-i\sigma} \\
     \frac{1}{2}\sin(2\theta) e^{i\sigma} & -\sin^2\theta 
    \end{array} \right)
\dot\sigma\,,
\label{2x2_mixing_grad}
\eeq
while $\Xi^{\prime -} = 0$. We also notice that initially flavour and mass basis coincide. 

Next, we solve the Boltzmann equations (\ref{eom_on-shell1} - \ref{eom_on-shell2}) numerically, using the self-energies (\ref{self3}) in the 
collision integrals (\ref{app_Sigma_proj_m} - \ref{app_Sigma_proj_c}) and Eq.~(\ref{2x2_mixing_grad}) for the mixing-gradient matrix. We assume 
that the distribution functions $f_{\alpha}$ are initially at $t_{\rm in} = -50/T$ in thermal equilibrium with vanishing chemical potential:
\beq
f^m_{ijh\pm}(\mathbf{k},t_{\rm in}) = \pm \frac{\omega_i}{m_i}f_{\rm eq}(\pm \omega_i)\delta_{ij} \,,
\qquad\quad f^c_{ijh\pm}(\mathbf{k},t_{\rm in}) = 0\,.
\eeq
If not stated otherwise, we take $\tau_w = 10/T$ for a representative transition time-scale. For the mixing parameters we use $v_0 = T$, $\sigma_0 = \pi/2$, $m_A = 2.2\,T$, $m_B = 2.0\,T$, and for the interaction parameters we choose massless thermal background fields: $m_q = m_{\phi}= 0 $, and the flavour-basis coupling constants $y_\ell$, $\ell = A,B$, are chosen to be $y_A=1 $ and $y_B=0.5$. Note that the quasiparticle excitations have time-varying interaction strengths with the thermal background: $y^2_{11}(t)$ and $y^2_{22}(t)$, which are in general different.

We solved the system for $60$ momentum $|\mathbf{k}|$ bins between $10^{-2}\,T $ and $30 \, T$, with larger bin density in the IR-region of the phase space. The full fermionic correlator has $8$ real diagonal degrees of freedom and $12$ complex non-diagonal degrees of freedom, which each has an independent phase and amplitude. The overall discretized numerical problem thus consists of $1920$ real coupled ordinary differential equations.

%
\subsection{Numerical results}
\label{sec:results}
%

\begin{figure}[t]
\centering
\includegraphics[width=1.0\textwidth]{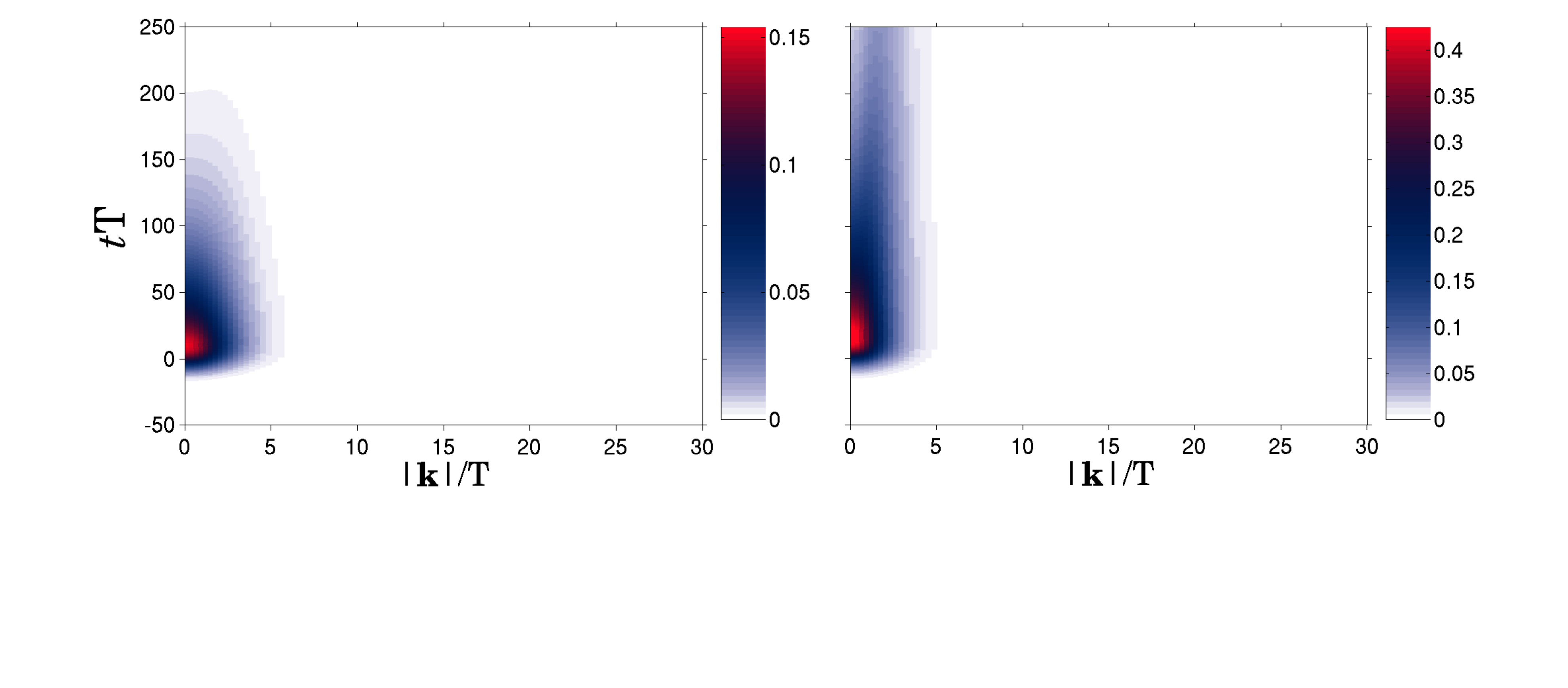}
\vskip-1.5truecm
\caption{Shown is the excess total particle-number density $\delta N_1$ for the heavy field $i=1$ (left) 
and deficiency of the total particle-number density $-\delta N_2$ for the light field $i=2$ (right).} 
\label{fig:results1}
\end{figure}

Let us now study different observables resulting from this sample computation. In figure~\ref{fig:results1} we show the evolution of the total excesses (over thermal equilibrium) of combined particle and antiparticle number densities of mass-states $i=1,2$, given by: 
\beq
\delta N_i(|{\bf k}|,t) \equiv \sum_{h} \Big[
          ( n_{i{\bf k}h} + {\bar n}_{i{\bf k}h} )(t) 
         -( n_{i{\bf k}} + {\bar n}_{i{\bf k}} )_{\rm eq}
         \Big]\,,
\eeq
where the number densities $n_{i{\bf k}h}$ and ${\bar n}_{i{\bf k}h}$ are related to flavour-diagonal distribution functions by Eq.~(\ref{part-number}).
Left panel shows the excess $\delta N_1$ in heavier fermion species 1 and the right panel the corresponding deficiency $-\delta N_2$ of the total particle number for the lighter species 2. The particle number is created at the transition time as a result of coherent mixing and interactions, and it is strongest at small momentum region $|{\bf k}|\lesssim T$. The heavy particle number is depleted by decays when $t T > 200$. The deficiency in the lighter particle sector is depleted by inverse decays about four times slower, because the particle 2 has a weaker Yukawa coupling to the thermal bath.

A similar model for scalar fields has been considered recently in ref.~\cite{CLR-MT10}. However, the formalism used in~\cite{CLR-MT10} only accounts for the direct (particle-particle or antiparticle-antiparticle) flavour-mixing but neglects all particle-antiparticle correlations. The authors of ref.~\cite{CLR-MT10} argued that the latter are irrelevant in slowly-varying backgrounds, as their effects should average out because of the high-frequency ($\sim 2\bar \omega$) oscillations in this sector. Since our formalism includes particle-antiparticle coherence ($f^c$-solutions), we can study this phenomenon in detail. To this end, we performed a series of computations both including and neglecting the $f^c$-solutions and compared the results. Somewhat surprisingly, we find that the $f^c$-correlations cannot in general be neglected even in the case of relatively slowly varying mass profile with $\tau_w \simeq (10-20)/T$, and moreover, the particle production induced by $f^c$-terms can be the dominant contribution to $CP$-violation in spontaneous baryogenesis models.  

The reason behind this observation is that the relevant scale for the $f^c$-induced particle production is the {\em gradient} scale $\sim \tau_w^{-1}$ and not the zitterbewegung scale $\sim 2\bar \omega$ as one might naively expect. Indeed, the structure of the Boltzmann equations (\ref{eom_on-shell1} - \ref{eom_on-shell2}) with several cross-couplings between the on-shell functions $f_\alpha$ implies that the $f^c$-solutions roughly consists of two qualitatively different modes: the {\em homogeneous mode} oscillating at the frequency $\sim 2\bar \omega$, and the non-oscillatory {\em forced mode} due to the ``inhomogeneous'' (gradient) cross-coupling terms to thermal bulk of the particle excitations and to perturbative vacuum.\footnote{Here homogeneous and inhomogeneous terms or solutions refer to classification in differential equation theory, and not to the spacetime symmetry properties of the system.} While it is true that the homogeneous oscillating modes of $f^c$-solutions are barely excited at all, non-negligible forced-mode excitations are typically generated in the transition region with non-vanishing mass-gradients, which give rise to substantial particle-antiparticle coherence.

\begin{figure}[t]
\centering
\includegraphics[width=1.0\textwidth]{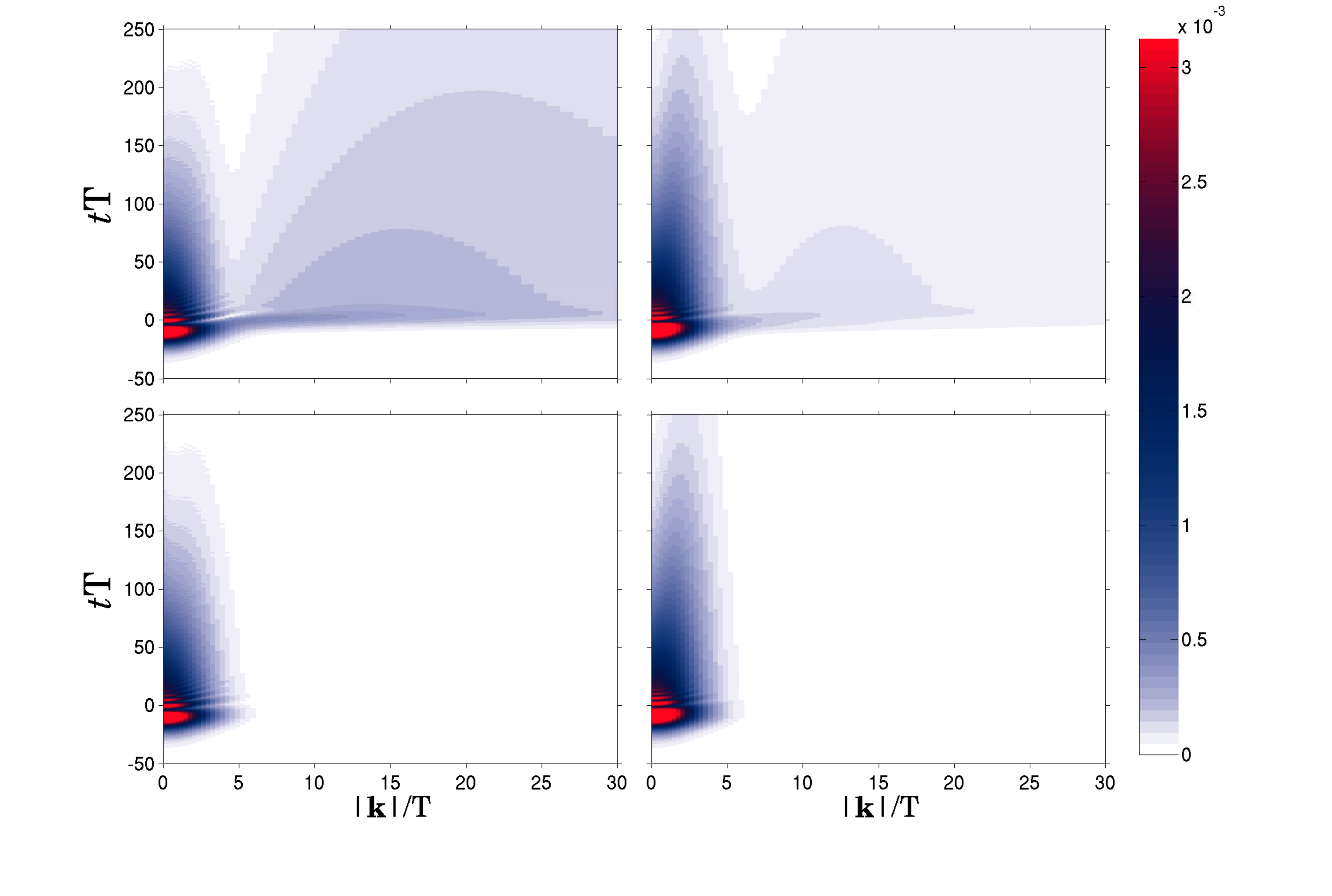}
\caption{Shown are the amplitudes of particle (left panels) and antiparticle (right panels) flavour-coherence correlators $|f^m_{12 h\pm}({\bf k},t)|$ 
for positive helicity $h=1$. The upper panels are computed with the full quantum Boltzmann equations, while the lower panels are computed with the restricted equations where the particle-antiparticle $f^c$-correlators are neglected throughout the calculation.}
\label{fig:results23}
\end{figure}  
\begin{figure}[h]
\centering
\includegraphics[width=1.0\textwidth]{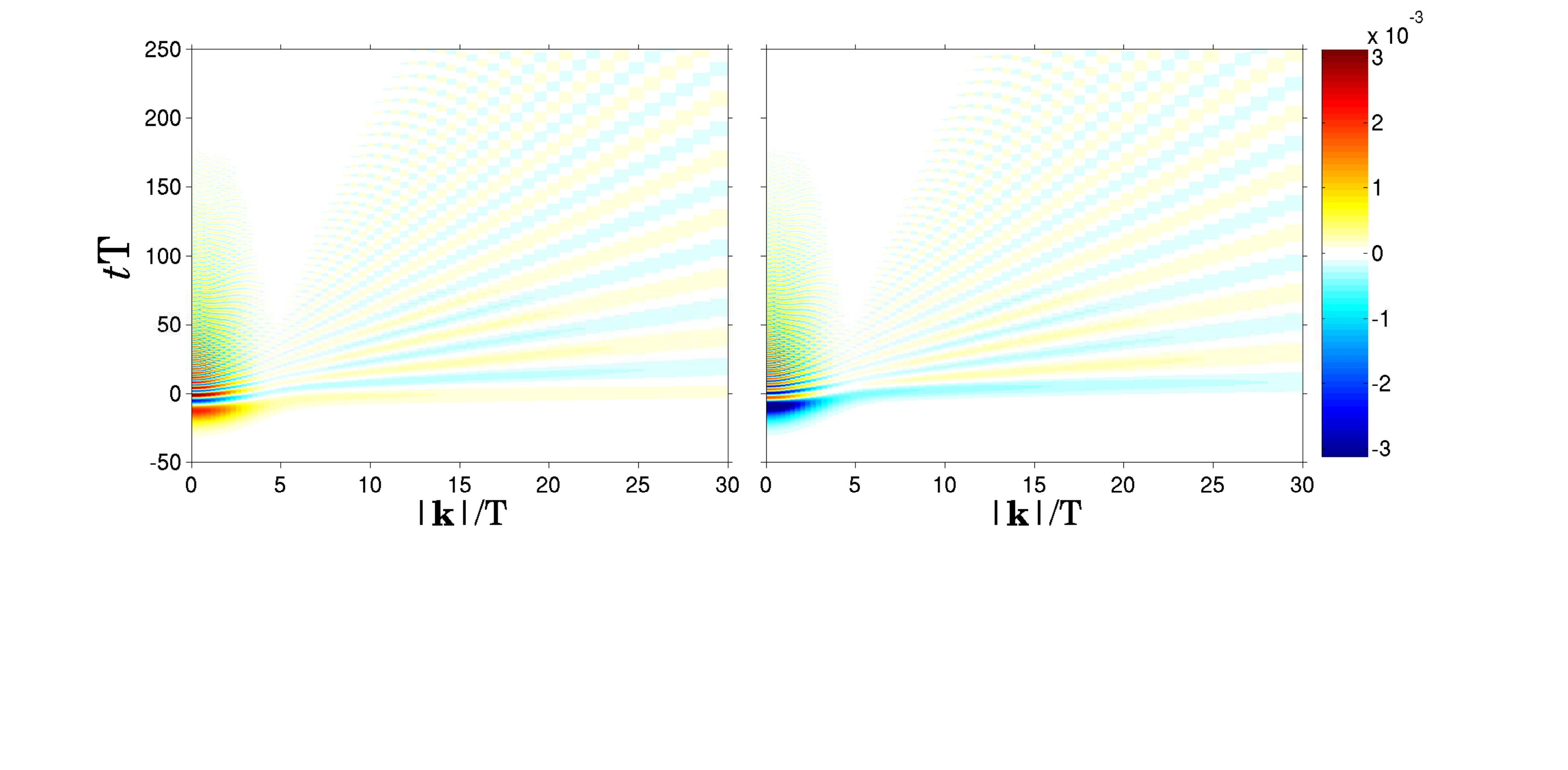}
\vskip-1.5truecm
\caption{Shown are the real (left panel) and imaginary (right panel) parts of the correlator $f^m_{12 h+}({\bf k},t)$ for positive helicity $h=1$, corresponding to the upper left panel in Fig.~\ref{fig:results23}.} 
\label{fig:results23b}
\end{figure} 
\begin{figure}[t]
\centering
\includegraphics[width=1.0\textwidth]{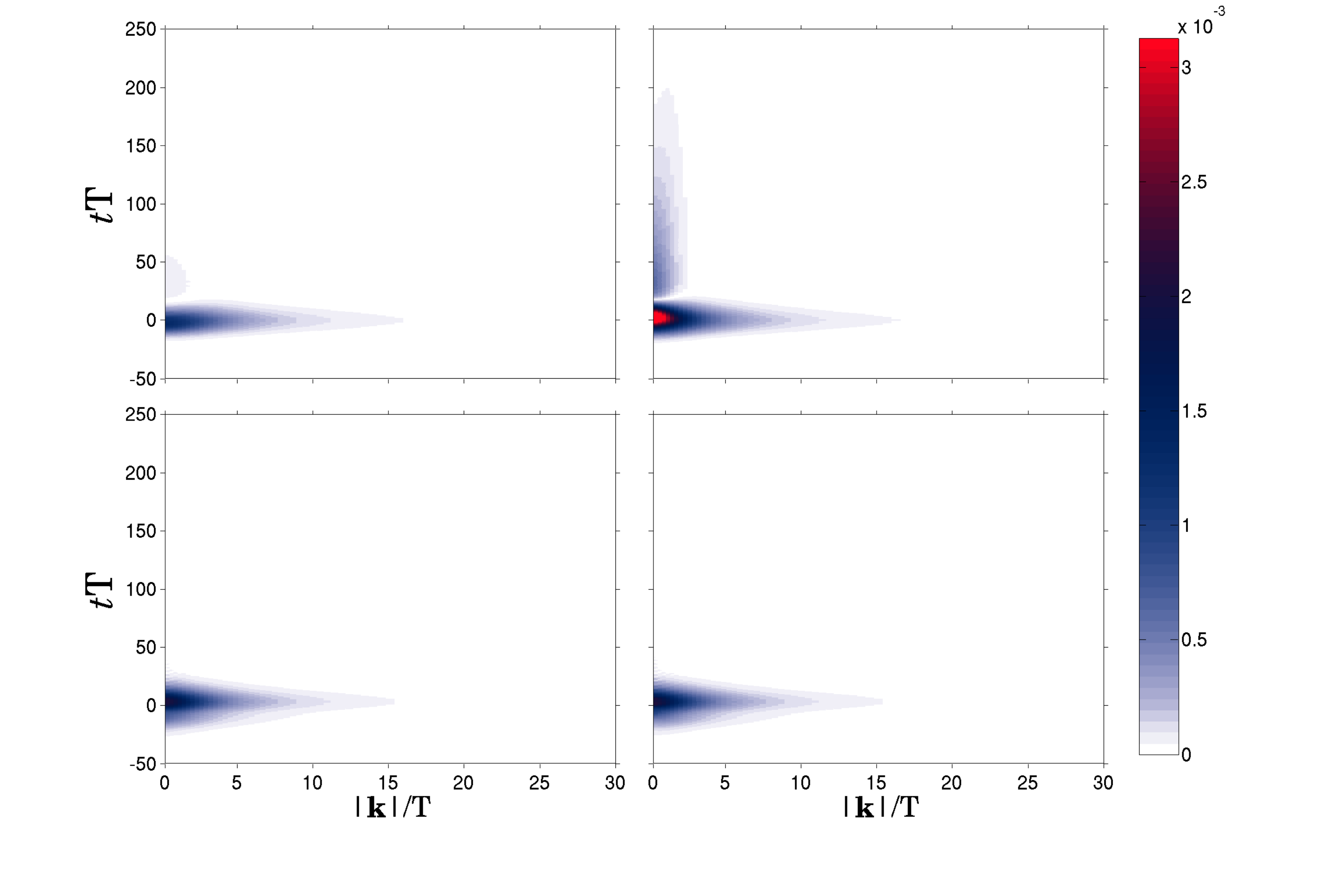}
\caption{Shown are the amplitudes of the particle-antiparticle flavour-coherence correlators for positive helicity $h=1$. The flavour-diagonal components $|f^c_{ii h\pm}({\bf k},t)|$ for the heavy field $i=1$ (upper left) and the light field $i=2$ (upper right), and the flavour off-diagonal components $|f^c_{12 h\pm}({\bf k},t)|$ for state $(+)$ (lower left) and state $(-)$ (lower right). The results are computed with the full quantum Boltzmann equations.}
\label{fig:results56}
\end{figure}
To display the effect of $f^c$-correlators we first consider the solutions for the particle and antiparticle flavour-coherence correlators $|f^m_{12 h\pm}({\bf k},t)|$. These solutions have nontrivial characteristics in the presence of mixing and $CP$-violation, when the time evolution of the system is computed both with and without the particle-antiparticle $f^c$-correlators. In Fig.~\ref{fig:results23} we show the evolution of this correlator for helicity $h=+1$ for the particles (left panels) and for antiparticles (right panels). Upper panels show the results of a full calculation using our complete quantum Boltzmann equations and lower panels the results of a calculation where the $f^c$-correlators have been forced to zero, emulating the calculation performed in 
ref.~\cite{CLR-MT10}. (Note that the amplitude in this figure, despite the same color coding, is two orders of magnitude smaller than in Fig.~\ref{fig:results1}.)
At small momenta $|{\bf k}| \lesssim T$, the results of the full and restricted calculation are very similar; the normal flavour-mixing dynamics dominates the correlator at small momenta. However, in full solution we see also a very wide band of excitations at $|{\bf k}| \gg T$, which are completely missed in pure flavour-mixing dynamics without particle-antiparticle coherence. If one takes the difference of the two results, one further finds that the $f^c$-induced particle-production effect is essentially restricted to large momenta. At small momenta $|{\bf k}| \lesssim T$ the residual effect has about tenth of the amplitude compared to the large momentum region. Indeed, it is clear that the importance of this contribution may only come from its extension over a very large region in phase space.

So far we have shown only amplitudes of the complex off-diagonal correlators. In these plots the oscillatory behaviour of these functions is mostly averaged out. In order to better show the true oscillatory structure,  we display in Fig.~\ref{fig:results23b} the real and imaginary parts of the correlator $f^m_{12 h+}({\bf k},t)$, corresponding to the case shown in the upper left panel in Fig.~\ref{fig:results23}.  The small-scale oscillations, clearly visible both in time and in momentum variable, make this type of calculations very challenging numerically.

\begin{figure}[t]
\centering
\includegraphics[width=1.0\textwidth]{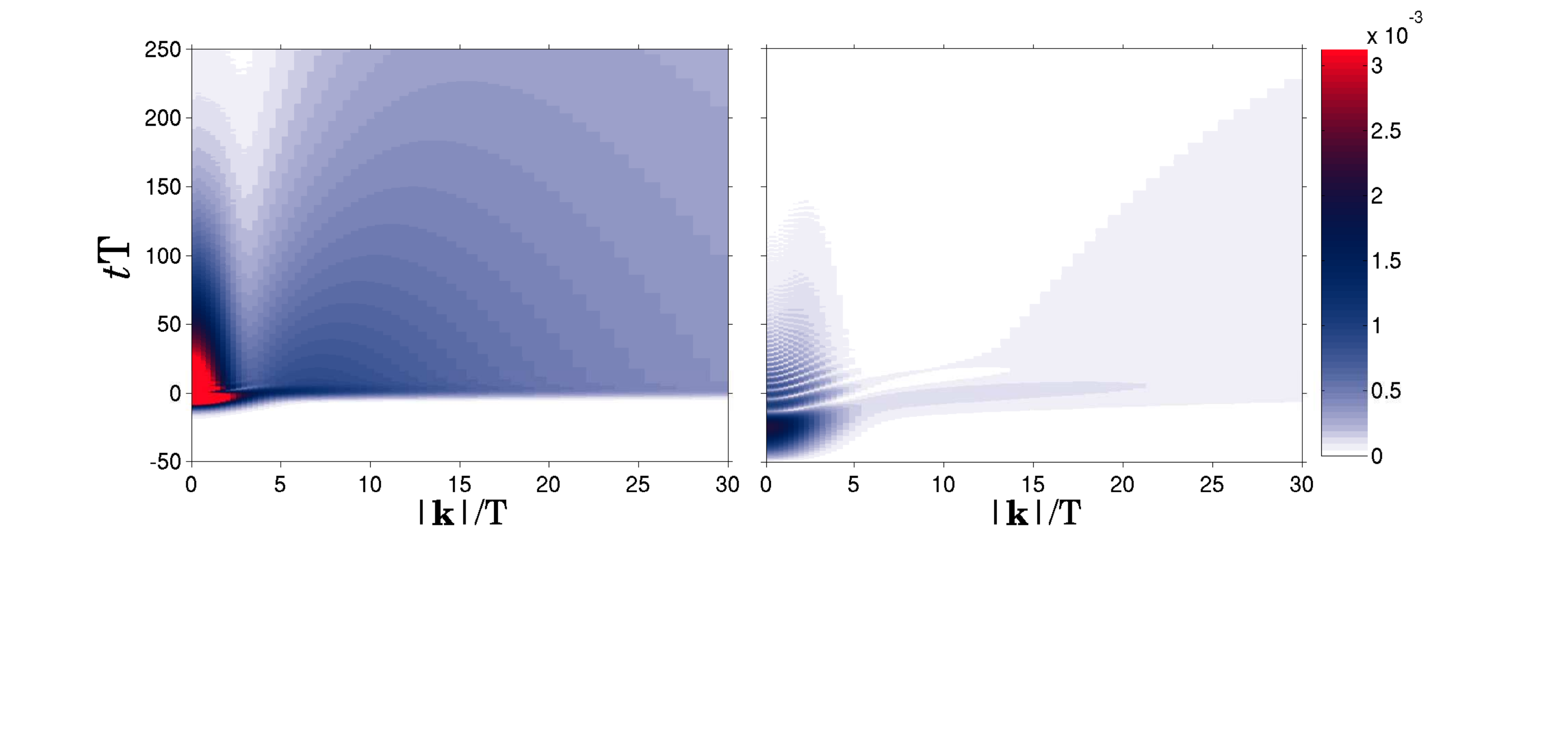}
\vskip-1.5truecm
\caption{Shown is the amplitude of particle flavour-coherence correlator 
$|f^m_{12 h+}({\bf k},t)|$ 
for positive helicity $h=1$ for a faster transition with $\tau_w = 5 /T$ (left) and for a slower transition with $\tau_w = 20 /T$ (right). The results are computed with the full quantum Boltzmann equations.} 
\label{fig:results4}
\end{figure}

In figure \ref{fig:results56} we show the particle-antiparticle $f^c$-correlators for the full calculation. The upper panels show the flavour-diagonal correlators
$|f^c_{ii h\pm}({\bf k},t)|$ with the heavy flavour $i=1$ on the left and the light flavour $i=2$ on the right, and the lower panels show the flavour off-diagonal correlators $|f^c_{12 h\pm}({\bf k},t)|$. Note that the latter represent coherence between particle and antiparticle states of different flavours, \ie the particle-antiparticle flavour coherence. As one would expect these correlators are rather well confined to the transition region. However, in this region they are equally large as the flavour off-diagonal correlators in particle and antiparticle sectors for large momenta, and they extend much farther in momenta than any of the correlators in the restricted calculation neglecting $f^c$-solutions. Clearly, these particle-antiparticle correlators, albeit of their short duration, act as catalysts at the transition time, giving rise to the large-momentum contribution to the flavour off-diagonal $f^m$-correlators.

Taking a closer look at the full solutions in Fig.~\ref{fig:results23}, we observe that the large momentum excitations peak at around $|{\bf k}| \sim 15 T$. The origin of this enhancement is a result of a delicate chain of interactions in the full equation network. However, one can see that this region roughly corresponds to momenta for which the off-diagonal particle or antiparticle flavour-oscillation frequency equals to the inverse transition width\footnote{Remember that while particle and antiparticle flavour-oscillation shells correspond to $k_0 = \pm\bar \omega_{ij}$ the characteristic oscillation frequencies of these correlators are $\pm 2\Delta \omega_{ij}$. The situation is inverted for the particle-antiparticle flavour coherence, living on the shells $\pm \Delta \omega_{ij}$ but oscillating with frequencies $\pm \bar\omega_{ij}$.}: 
$\tau_w^{-1} \approx \Delta \omega_{ij}$. In this case the flavour oscillations are resonant with the rate of change induced by the background variation in the transport equation (\ref{eom_on-shell1}), which can lead to large enhancement of the flavour off-diagonal correlator $f^{m}_{12h\pm}$. Note that this enhancement works typically only for the UV-modes, since the resonance condition yields UV-momenta for relatively large transition widths $\tau_w$. With the parameters used in the current example we have 
\beq
\frac{|{\bf k}|}{T} \Big|_{\rm res} \;\; \approx \;\; \frac{m_1^2-m_2^2}{2T^2}  \; \tau_w T \; \approx \;  4 - 40\,,
\label{eq:resonantcond}
\eeq
where the given range results from the variation of the mass difference $m_1^2(t) - m_2^2(t)$ over the transition region as $v(t)/T$ evolves from zero to one. 

To test the above explanation we have redone our full calculation with two other transition times and show the corresponding flavour off-diagonal correlators in Fig.~\ref{fig:results4}. Left panel corresponds to a relatively rapid transition with $\tau_w = 5/T$ and the right panel to a much broader one with $\tau_w = 20/T$. In the former case the resonant enhancement region is narrower, as expected on the basis of Eq.~(\ref{eq:resonantcond}). In the latter case the overall amplitude is getting very weak. However even in the latter case the result differs significantly from the one where the particle-antiparticle correlators are neglected.

\begin{figure}[t]
\centering
\includegraphics[width=1.0\textwidth]{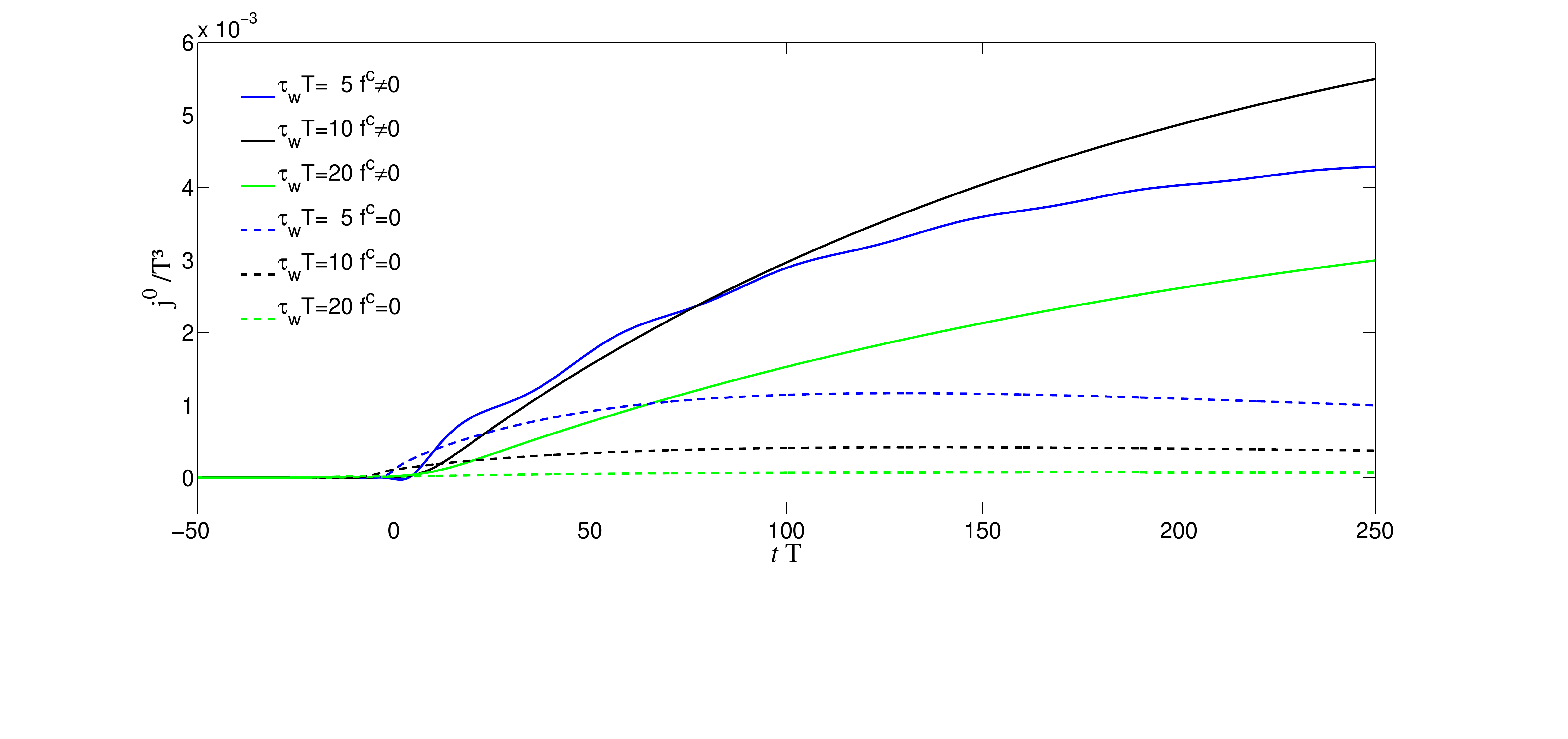}
\vskip-1.5truecm
\caption{Shown is the evolution of the total asymmetry integrated over the phase space $j^0(t)= j^0_1(t) + j^0_2(t)$. The solid lines correspond to the solution of the full equations of motion and the dashed lines to the restricted equations of motion with $f^c = 0$, both respectively with $\tau_W = 5/T, 10/T$ and $20/T$ from top to bottom.} 
\label{fig:results9}
\end{figure}

Finally, let us consider the quantity which is most interesting for many applications, the produced charge asymmetry:
\beq
j^0_i(|{\bf k}|,t) = \sum_{h} (n_{i{\bf k}h} - {\bar n}_{i{\bf k}h})(t) \,.
\eeq
\begin{figure}[t]
\centering
\includegraphics[width=1.0\textwidth]{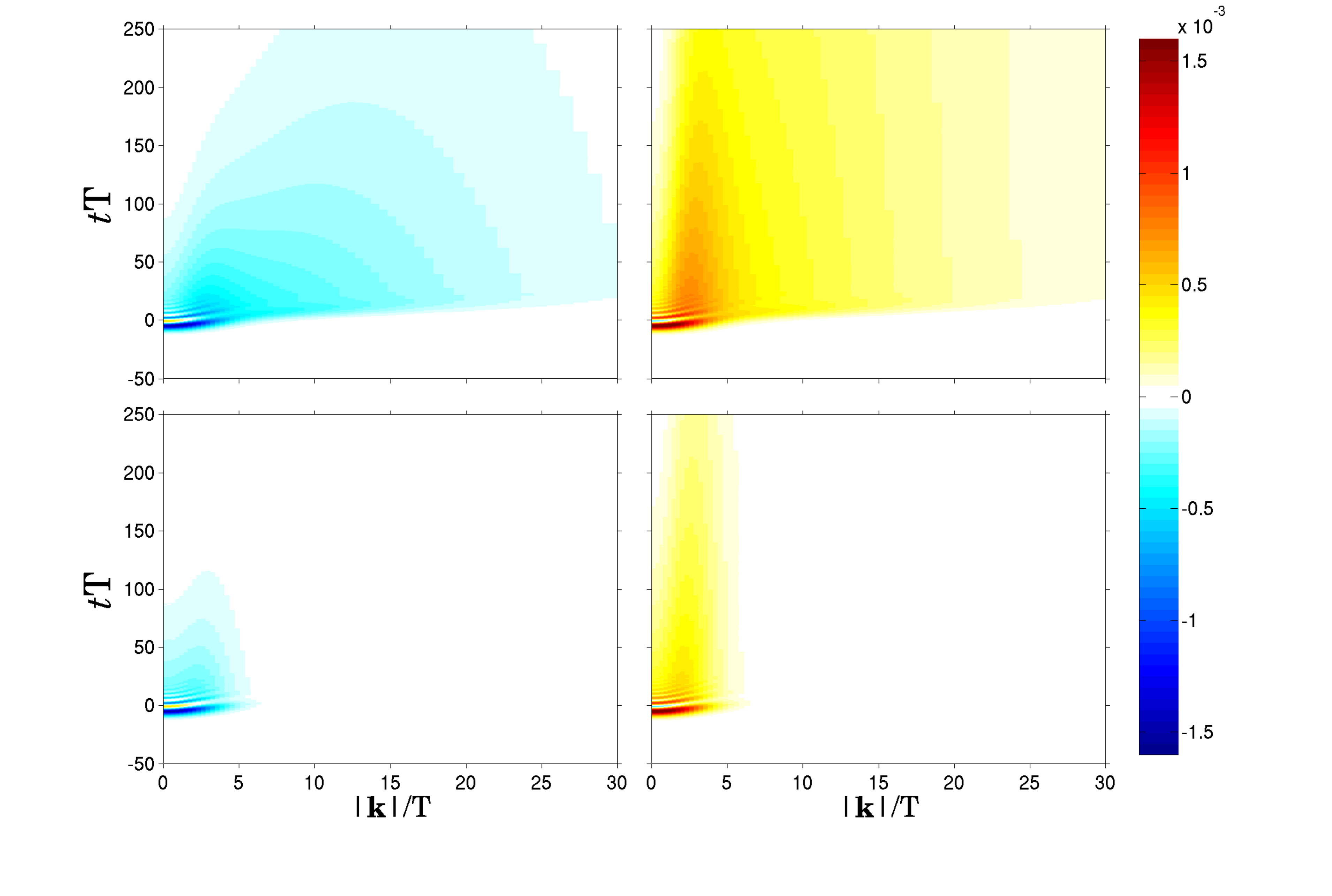}
\caption{Shown are the charge asymmetries $j^0_i(|{\bf k}|,t)$ for heavy field $i=1$ (left panels) and for light field $i=2$ (right panels). The upper panels are computed with the full quantum Boltzmann equations, while the lower panels are computed with the restricted equations where the particle-antiparticle $f^c$-correlators are neglected throughout the calculation.} 
\label{fig:results7}
\end{figure} 
We plot this quantity in Fig.~\ref{fig:results7}, again for the heavy fields (left panels) and the light fields (right panels) and with full (upper panels) and restricted (lower panels) equations of motion. Even for fairly broad transition considered here (we again use $\tau_w = 10/T$), there is a considerable asymmetry for near UV modes above $k \gsim 3 T$ which is completely missed in the restricted calculation where $f^c$-correlators are neglected. Because of the large phase space, this effect dominates the total asymmetry generation. To make this clearly visible, we consider the time evolution of the total asymmetry integrated over the phase space:\footnote{More precisely, this quantity corresponds to the expectation value of the number current density: $\langle j^0(t) \rangle = \sum_i \langle \bar \psi_i \gamma^0\psi_i \rangle = \int \frac{{\rm d}^4 k}{(2\pi)^4}{\rm Tr}[\bar S^<(k,t)]$, where the vacuum part is subtracted and the trace is over both Dirac and flavour indices. Note that this quantity is flavour-basis invariant.}
\beq
j^0(t)= \sum_i j^0_i(t) = \sum_i \int \frac{{\rm d}|{\bf k}| |{\bf k}|^2}{2\pi^2} j^0_i(|{\bf k}|,t) \, .
\eeq
We show this quantity in Fig.~\ref{fig:results9} for three different transition times: $\tau_w = 5/T, 10/T$ and $20/T$. 
The solid curves correspond to the full calculation and the dashed ones to the restricted calculation. In particular for small 
transition times the influence of the particle-antiparticle correlators is crucial and the effect dominates over the pure 
flavour oscillation mechanism. Even for rather smooth transitions $\tau_w = 20/T$ (lowest curves) the effect is quantitatively 
significant. We hence conclude that neglecting particle-antiparticle correlators is in general not warranted. The late time 
$ t \rm T \gsim 100 $ behaviour of the asymmetry is highly dependent on the type of interactions in the model. Using only 
three-body ($ 1 \leftrightarrow 2$) interactions in our toy model leads to a significantly suppressed UV-mode charge transfer 
rate with the external field $\psi_q$, especially for the light field $2$, leading a to maximal asymmetry at very late times 
 $t  \rm T \sim 400-600 $ and to a very slow chemical equilibration rate, for which an exact result can be derived from equations~(\ref{Sigma_0,3} - \ref{self3}):
\beq
\Gamma_{{\rm chem}, i}(|{\bf k}|) =\frac{y_i^2}{32\pi}\frac{m_i^2}{\omega_i}\frac{T}{|{\bf k}|}\ln
\big(\frac{\sinh(\frac{\omega_i + |{\bf k}|}{2T})}{\sinh(\frac{\omega_i - |{\bf k}|}{2T})}\big) \quad \Rightarrow \quad
\Gamma_{{\rm chem}, 2}|_{|{\bf k}| = 20 \rm T}  \approx 4 \times 10^{-4} \rm T \,.  
\eeq
This rate agrees with the ordinary one particle decay rate with the expected
order of magnitude $\Gamma_{{\rm chem}, i} \sim 10^{-2}  \rm\, T $ in the thermal region $|{\bf k}| \sim \rm\, T$. 
The very slow integrated rate seen here reflects the fact that the late time decays are dominated by the weakly interacting 
light state $2$ and that the asymmetry is mostly confined to UV modes at $|{\bf k}| \sim 15-20 \rm\, T$. 
Taking into consideration also the four-body ($ 2 \leftrightarrow 2$) interactions should make the thermalization process much 
faster especially for UV modes. However, 
the considerable difference in asymmetry generation in the transition region $t\rm T  \lsim 50$ 
between the full and the restricted cQPA calculation persists regardless of the model-specific interaction details.

Let us stress that the $f^c$-induced enhancement of the asymmetry generation is an entirely new type of phenomenon, which would not have been discovered without full cQPA transport formalism. We expect this effect to be important for many applications, and in particular to the electroweak baryogenesis type scenarios.  

%
\section{Flavour coherence from the operator formalism}
\label{sec:operator_form}
%

In this section we use standard operator formalism of QFT to study flavour-mixing complex scalar fields in constant background. We show that flavour coherence is emergent provided that the off-diagonal ($i \neq j$) expectation values $\langle a^\dagger_{i\mathbf{k}}\,a_{j\mathbf{k^\prime}}\rangle$, $\langle b^\dagger_{i\mathbf{k}}\,b_{j\mathbf{k^\prime}}\rangle$ or $\langle b_{i\mathbf{k}}\,a_{j\mathbf{k^\prime}}\rangle$ are nonvanishing. In this case, and further satisfying spatial homogeneity and isotropy, we then find that the corresponding 2-point correlator $\Delta^<(k,t)$ has the cQPA-shell structure with $k_0 = \pm\bar\omega$ or $k_0 = \pm\Delta\omega$. We show that the desired nonstandard expectation values can be realized by certain class of exponentiated superposition states, {\em flavoured squeezed states}, obtained from the vacuum state by unitary squeezing operation~\cite{squeezing}. We also show that flavoured squeezing corresponds to flavoured Bogolyubov transformations, providing a generalization of the standard non-flavoured case. We further present a one-to-one identification between the on-shell functions $f^{m,c}_{ij\pm}$ and the independent parameters of the flavoured squeezed state or the corresponding flavoured Bogolyubov transformation.   

We consider mixing complex scalar fields $\phi_i$, $i=1,\ldots,N$, with standard operator expansions
\beq
\phi_i(x) = \int \frac{{\rm d}^3 {\bf k}}{(2\pi)^3 \,2\omega_{i \mathbf{k}}} \left(a_{i\mathbf{k}} e^{-i(\omega_{i \mathbf{k}} x_0 - \mathbf{k} \cdot \mathbf{x})} + b_{i\mathbf{k}}^\dagger e^{i(\omega_{i\mathbf{k}} x_0 - \mathbf{k} \cdot \mathbf{x})}\right)\,,
\label{field_op}
\eeq
where $\omega_{i \mathbf{k}} \equiv \sqrt{{\mathbf{k}}^2 + m_i^2}$ and $a_{i\mathbf{k}}^\dagger$ ($a_{i\mathbf{k}}$) and $b_{i\mathbf{k}}^\dagger$ ($b_{i\mathbf{k}}$) are the creation (annihilation) operators for particles and antiparticles of mass eigenstates $i = 1,\ldots,N$, respectively. These operators satisfy canonical commutation relations
\begin{align}
[a_{i\mathbf{k}}, a_{j\mathbf{k^\prime}}] =& [a_{i\mathbf{k}}^\dagger, a_{j\mathbf{k^\prime}}^\dagger] = [b_{i\mathbf{k}}, b_{j\mathbf{k^\prime}}] = [b_{i\mathbf{k}}^\dagger, b_{j\mathbf{k^\prime}}^\dagger] = 0 \,,
\nonumber\\[2mm]
[a_{i\mathbf{k}}, a_{j\mathbf{k^\prime}}^\dagger] =& [b_{i\mathbf{k}}, b_{j\mathbf{k^\prime}}^\dagger] = 2\omega_{i\mathbf{k}}(2\pi)^3 \delta^3(\mathbf{k}-\mathbf{k^\prime}) \delta_{ij}\,,
\label{can_commutation}
\end{align}
and all commutators between $a_{i\mathbf{k}}$ and $b_{j\mathbf{k^\prime}}$ (and their hermitian conjugates) are vanishing. Let us now assume a state $| \Omega \rangle$ with the expectation values
\beqa
\langle a_{i\mathbf{k}}^\dagger\,a_{j\mathbf{k^\prime}}\rangle &=& \tilde f^{m <}_{ji+}(|\mathbf{k}|)\,2\bar\omega_{ij\mathbf{k}}(2 \pi)^3 \delta^3(\mathbf{k} - \mathbf{k^\prime})\,,
\nonumber\\[1mm]
\langle b_{i\mathbf{k}}^\dagger\,b_{j\mathbf{k^\prime}}\rangle &=& -\big(\delta_{ij} + \tilde f^{m <}_{ij-}(|\mathbf{k}|)\big)\,2\bar\omega_{ij\mathbf{k}} (2 \pi)^3 \delta^3(\mathbf{k} - \mathbf{k^\prime})\,,
\nonumber\\[1mm]
\langle b_{i\mathbf{k}}\,a_{j\mathbf{k^\prime}}\rangle =  \langle a_{j\mathbf{k^\prime}}^\dagger\,b_{i\mathbf{k}}^\dagger \rangle^* &=& \tilde f^{c <}_{ji+}(|\mathbf{k}|))\,2\bar\omega_{ij\mathbf{k}} (2 \pi)^3 \delta^3(\mathbf{k} + \mathbf{k^\prime})\,,
\label{exp_values}
\eeqa
consistent with the requirement of spatial homogeneity and isotropy.
Using the operator expansion (\ref{field_op}) and the relations (\ref{exp_values}) it is straightforward to compute the Wightman function:
\begin{alignat}{3}
i\Delta^<_{ij}(u,v) =& \langle \phi_j^\dagger(v)\phi_i(u) \rangle
\nonumber\\
=& \int \frac{{\rm d}^3 {\bf k}}{(2\pi)^3}\frac{\bar\omega_{ij}}{2 \omega_{i}\omega_{j}} e^{i\mathbf{k} \cdot (\mathbf{u} - \mathbf{v})} \bigg[&&\tilde f^{m <}_{ij+} e^{-i(\omega_{i} u_0 - \omega_{j} v_0)} - \tilde f^{m <}_{ij-} e^{i(\omega_{i} u_0 - \omega_{j} v_0)} 
\nonumber\\
&&+& \tilde f^{c <}_{ij+} e^{-i(\omega_{i} u_0 + \omega_{j} v_0)} + \tilde f_{ji+}^{c < *} e^{i(\omega_{i} u_0 + \omega_{j} v_0)} \bigg]\,,
\end{alignat}
where we have dropped the $\mathbf{k}$-subscripts of $\omega_{i\mathbf{k}}$ and $\bar\omega_{ij\mathbf{k}}$. The Wigner transformation then gives ($t \equiv x_0 = (u_0 + v_0)/2$)
\beqa
i\Delta^<_{ij}(k,x) &=& \int {\rm d}^4(u-v) e^{ik_0(u_0-v_0) - i \mathbf{k} \cdot (\mathbf{u} - \mathbf{v})}i\Delta^<_{ij}(u,v)
\nonumber\\
&=&2\pi \frac{\bar\omega_{ij}}{2 \omega_{i}\omega_{j}}\Big[\tilde f^{m<}_{ij+} e^{-i2\Delta\omega\,t} \delta(k_0 - \bar\omega) - \tilde f^{m <}_{ij-} e^{i2\Delta\omega\,t} \delta(k_0 + \bar\omega)
\nonumber\\
&& \qquad\qquad\;\,+ \tilde f^{c <}_{ij+} e^{-i2\bar\omega\,t} \delta(k_0 - \Delta\omega) + \tilde f^{c < *}_{ji+} e^{i2\bar\omega\,t} \delta(k_0 + \Delta\omega)\Big]\,,  
\eeqa
which is readily identified as the singular cQPA Wightman function of Eq.~(\ref{Dist}) by 
\begin{equation}
f^{m <}_{ij\pm}(\mathbf{k},t) = \tilde f^{m <}_{ij\pm}(\mathbf{k})\,e^{\mp i2\Delta\omega_{ij}\,t}\,,\qquad\quad f^{c <}_{ij\pm}(\mathbf{k},t) = \tilde f^{c <}_{ij\pm}(\mathbf{k})\,e^{\mp i2\bar\omega_{ij}\,t}\,,
\end{equation}
using the relation $\tilde f^{c <}_{ij-} =  \tilde f^{c < *}_{ji+}$. Indeed, we have found that the correlator $\Delta^<(k,t)$ involves flavour coherence and all four singular shells $k_0 = \pm\bar\omega$ and $k_0 = \pm\Delta\omega$ are excited, provided that the expectation values of Eq.~(\ref{exp_values}) are nonvanishing (and consistent with the spatial homogeneity and isotropy).

Next, we would like to find a state $| \Omega \rangle$ with the nonstandard expectation values (\ref{exp_values}). It is clear that any number state: $(\prod\limits_m a_m^\dagger)(\prod\limits_n b_n^\dagger)| 0 \rangle$, with a finite number of quanta cannot possibly satisfy the relations (\ref{exp_values}), because the flavour off-diagonal expectation values would necessarily vanish. It turns out, however, that certain superposition states with unspecified (infinite) number of quanta indeed have the desired properties (\ref{exp_values}). To see this, we use the methods of squeezed state formalism~\cite{squeezing}.

We begin by defining a unitary {\em two-mode squeeze operator} for multiple mixing fields:
\beq
S_2(r_{ij})  \equiv e^A  \equiv \exp\bigg[\sum_{ij} \int \frac{{\rm d}^3 {\bf k}}{(2\pi)^3} \left(r_{ij} \hat b_{i\mathbf{k}} \hat a_{j-\mathbf{k}} - r_{ij}^* \hat a_{j\mathbf{k}}^\dagger \hat b_{i-\mathbf{k}}^\dagger\right)\bigg] \,,  
\eeq
where $r_{ij}(|\mathbf{k}|)$ is a complex matrix in flavour indices, and we denote $\hat a_{i\mathbf{k}} \equiv a_{i\mathbf{k}}/\sqrt{2 \omega_{i\mathbf{k}}}$ and $\hat b_{i\mathbf{k}} \equiv b_{i\mathbf{k}}/\sqrt{2 \omega_{i\mathbf{k}}}$. Next, we show how the creation and annihilation operators transform upon acting on $S_2$, \ie we want to compute $S_2 a_{i\mathbf{k}} S_2^{\dagger}$ and its hermitian conjugate. First, we find that
\beq
\big[\hat a_{i\mathbf{k}}, A \big] = - \sum_{j} r_{ij}^\dagger\,\hat b_{j-\mathbf{k}}^\dagger\,,\qquad\qquad \big[\hat b_{i\mathbf{k}}^\dagger, A \big] = - \sum_{j} r_{ij}\,\hat a_{j-\mathbf{k}}\,.
\eeq
Furthermore, $r_{ij}$ can be diagonalized by a biunitary transformation: $r = U^\dagger r_{\rm d} V$ with $(r_{\rm d})_{ij} = r_i \delta_{ij}$, where $r_i(|\mathbf{k}|) \geq 0$ and $U(|\mathbf{k}|)$ and $V(|\mathbf{k}|)$ are unitary matrices. This suggests to define the following linear combinations:
\beq
\hat c_{i\mathbf{k}\pm} = \sum_j \big( V_{ij} \hat a_{j\mathbf{k}} \pm U_{ij} \hat b_{j-\mathbf{k}}^\dagger \big)\,,
\label{aux_op_def}
\eeq
which by construction satisfy simple diagonal commutation relations
\beq
\big[\hat c_{i\mathbf{k}\pm}, A \big] = \mp r_i \hat c_{i\mathbf{k}\pm}\,.  
\eeq
By recursive use of these relations we find that $\hat c_{i\mathbf{k}\pm} A^n = (A \mp r_i)^n \hat c_{i\mathbf{k}\pm}$, and furthermore
\begin{equation}
S_2 \,\hat c_{i\mathbf{k}\pm}S_2^\dagger = e^{\pm r_i} \hat c_{i\mathbf{k}\pm}\,. 
\end{equation}
From this equation and the inverse of Eq.~(\ref{aux_op_def}) we directly obtain the transformation law for $\hat a_{i\mathbf{k}}$:
\begin{align}
\tilde{\hat a}_{i\mathbf{k}} \equiv S_2\,\hat a_{i\mathbf{k}}S_2^\dagger =& \sum_j \frac{1}{2} V_{ij}^{\dagger}\big( S_2 \,\hat c_{j\mathbf{k}+} S_2^\dagger + S_2 \,\hat c_{j\mathbf{k}-} S_2^\dagger \big)
\nonumber\\
=& \sum_j \big( \alpha_{ij} \hat a_{j\mathbf{k}} + \beta_{ij}^\dagger \hat b_{j-\mathbf{k}}^\dagger \big)\,,
\nonumber\\[3mm]
\tilde{\hat b}_{i-\mathbf{k}}^\dagger \equiv S_2\,\hat b_{i-\mathbf{k}}^\dagger S_2^\dagger =& \sum_j \frac{1}{2} U_{ij}^{\dagger}\big( S_2 \,\hat c_{j\mathbf{k}+} S_2^\dagger - S_2 \,\hat c_{j\mathbf{k}-} S_2^\dagger \big)
\nonumber\\
=& \sum_j \big( \beta_{ij} \hat a_{j\mathbf{k}} + \gamma_{ij} \hat b_{j-\mathbf{k}}^\dagger \big)\,,
\label{fla_bogol}
\end{align}
with
\begin{align}
\alpha_{ij}(|\mathbf{k}|) \equiv& \big[ V^{\dagger} \cosh(r_{\rm d}) V \big]_{ij}\,,\qquad\qquad \beta_{ij}(|\mathbf{k}|) \equiv \big[ U^\dagger \sinh(r_{\rm d}) V \big]_{ij}\,,
\nonumber\\
\gamma_{ij}(|\mathbf{k}|) \equiv& \big[ U^{\dagger} \cosh(r_{\rm d}) U \big]_{ij}\,.
\end{align}
We see that the transformation law (\ref{fla_bogol}) has the form of a {\em flavoured Bogolyubov transformation}. For the inverse transformation we get 
\begin{align}
\hat a_{i\mathbf{k}} \equiv& S_2^\dagger \,\tilde{\hat a}_{i\mathbf{k}} S_2 = \sum_j \big( \alpha_{ij}^\dagger \tilde{\hat a}_{j\mathbf{k}} - \beta_{ij}^\dagger \tilde{\hat b}_{j-\mathbf{k}}^\dagger \big)\,,
\nonumber\\
\hat b_{i-\mathbf{k}}^\dagger \equiv& S_2^\dagger \,\tilde{\hat b}_{i-\mathbf{k}}^\dagger S_2 = \sum_j \big( -\beta_{ij} \tilde{\hat a}_{j\mathbf{k}} + \gamma_{ij}^\dagger \tilde{\hat b}_{j-\mathbf{k}}^\dagger \big)\,.
\label{fla_bogol_inv}
\end{align}
Because of the unitarity of the transformation (\ref{fla_bogol}), it is clear that the transformed operators $\tilde{\hat a}_{i\mathbf{k}}$, $\tilde{\hat b}_{i\mathbf{k}}$ satisfy the same commutation algebra as $\hat a_{i\mathbf{k}}$, $\hat b_{i\mathbf{k}}$. Also, we instantly note that the state $| \tilde 0 \rangle = S_2 | 0 \rangle$ is a vacuum state for the transformed operators, \ie $\tilde{\hat a}_{i\mathbf{k}}|\tilde 0 \rangle = 0$.  

It is now easy to construct a state $| \Omega \rangle$ with the desired properties (\ref{exp_values}). It turns out that a first guess: 
\beq
| \Omega \rangle = | \tilde 0 \rangle = S_2 | 0 \rangle =  \exp\bigg[\sum_{ij} \int \frac{{\rm d}^3 {\bf k}}{(2\pi)^3} \left(r_{ij} \hat b_{i\mathbf{k}} \hat a_{j-\mathbf{k}} - r_{ij}^* \hat a_{j\mathbf{k}}^\dagger \hat b_{i-\mathbf{k}}^\dagger\right)\bigg] | 0 \rangle
\eeq
is indeed correct, since a direct computation gives
\beqa
\langle a_{i\mathbf{k}}^\dagger \,a_{j\mathbf{k^\prime}} \rangle &=& 2\sqrt{\omega_{i\mathbf{k}} \omega_{j\mathbf{k^\prime}}} \sum_{i^\prime j^\prime} \langle \tilde 0 | \big( \alpha_{i^\prime i} \tilde{\hat a}_{i^\prime\mathbf{k}}^\dagger - \beta_{i^\prime i} \tilde{\hat b}_{i^\prime-\mathbf{k}} \big) \big( \alpha_{jj^\prime}^\dagger \tilde{\hat a}_{j^\prime\mathbf{k^\prime}} - \beta_{jj^\prime}^\dagger \tilde{\hat b}_{j^\prime-\mathbf{k^\prime}}^\dagger \big) | \tilde 0 \rangle 
\nonumber\\
&=& (\beta^\dagger \beta)_{ji} \,2\sqrt{\omega_{i} \omega_{j}}\,(2\pi)^3 \delta^3(\mathbf{k}-\mathbf{k^\prime}) 
\eeqa
and similarly
\beqa
\langle b_{i\mathbf{k}}^\dagger \,b_{j\mathbf{k^\prime}} \rangle &=& \quad (\beta \beta^\dagger)_{ij} \,2\sqrt{\omega_{i} \omega_{j}}\,(2\pi)^3 \delta^3(\mathbf{k}-\mathbf{k^\prime})\,,
\nonumber\\
\langle b_{i\mathbf{k}} \,a_{j\mathbf{k^\prime}} \rangle &=& -(\beta^\dagger \gamma)_{ji}\,2\sqrt{\omega_{i} \omega_{j}}\, (2\pi)^3 \delta^3(\mathbf{k}+\mathbf{k^\prime})\,.
\eeqa
These relations readily provide a one-to-one identification between the on-shell functions and the parameters of the flavoured Bogolyubov transformations (or equivalently the parameters $r_{ij}$ of the corresponding squeezed state): 
\begin{align}
\tilde f^{m <}_{ij+} =& \frac{\sqrt{\omega_{i} \omega_{j}}}{\bar\omega_{ij}}(\beta^\dagger \beta)_{ij}\,,\quad\quad 
&&\tilde f^{c <}_{ij+} = -\frac{\sqrt{\omega_{i} \omega_{j}}}{\bar\omega_{ij}}(\beta^\dagger \gamma)_{ij}\,,
\nonumber\\
\tilde f^{m <}_{ij-} =& -\delta_{ij} - \frac{\sqrt{\omega_{i} \omega_{j}}}{\bar\omega_{ij}}(\beta \beta^\dagger)_{ij}\,,\quad\quad 
&& \tilde f^{c <}_{ij-} = -\frac{\sqrt{\omega_{i} \omega_{j}}}{\bar\omega_{ij}}(\gamma^\dagger \beta)_{ij}\,.
\end{align}
%

%
\section{Conclusions and outlook}
\label{sec:discussion}
%

In this paper we have derived quantum transport equations and perturbative Feynman rules for flavour-mixing quantum fields in spatially homogeneous systems, including both flavour- and particle-antiparticle coherence. We have considered both fermionic and bosonic fields using flavour-mixing Yukawa-couplings as a model for interactions. Our formalism is based on the coherent quasiparticle approximation (cQPA)~\cite{HKR1,HKR2,HKR3,Thesis_Matti,Glasgow,HKR4,HKR5}, which is a distributive expansion of the 2-point correlation functions in the limit of small background gradients and weak interactions. In cQPA the coherence information, both of the flavour- and particle-antiparticle mixing, resides on distinct coherence shells in the phase space, located at time-like average energies $\bar \omega_{ij} = \sfrac{1}{2}(\omega_i+\omega_j)$ for direct flavour-mixing and at space-like average energies (between positive and negative energy states) $\Delta \omega_{ij} = \sfrac{1}{2}(\omega_i-\omega_j)$ for particle-antiparticle flavour-mixing.

We have derived explicit, generic forms of flavour-coherent transport equations for fermion and scalar fields, which are valid for any type of interactions. These transport equations reduce to the usual quantum Boltzmann equations when the coherence correlators are neglected. We have also derived a simple set of Feynman rules for perturbative calculations of the self-energy functions appearing in the collision integrals of the transport equations including non-local coherence effects.
For the single-flavour case the Feynman rules derived in this paper are equivalent but simpler than the ones introduced in ref.~\cite{HKR4}. 
In a companion paper~\cite{HKR5} we present an alternative, flavour-covariant formulation of the cQPA transport equations and Feynman rules.  
This formulation also provides a straightforward generalization of the formalism to the case of nonzero dispersive self-energy $\Sigma_h$ with nontrivial quasiparticle dispersion relations.

As a numerical example, we have studied a two-flavour mixing scenario, where the mixing fermions interact with a thermal bath. A similar model for mixing scalar fields was recently considered in ref.~\cite{CLR-MT10}, where the particle-antiparticle coherence solutions were neglected in the dynamics based on physical arguments. However, in our calculations we have found that neglecting particle-antiparticle coherence is not in general justified even when the background field is slowly varying in comparison with the zitterbewegung frequency $\sim 2\omega$. Indeed, particle-antiparticle coherence correlators were explicitly shown to be excited in such cases. Moreover, these correlators can trigger a resonant growth of flavour coherence at large momenta over a large part of the phase space. Comparing the full calculations with the restricted model calculations, where the particle-antiparticle coherence correlators were neglected, we have found that these new modes can dominate over the pure flavour dynamics. These results may suggest a new, more efficient way of generating an asymmetry in the EWBG-type models, even for relatively slowly-varying mass profiles such as in the case of electroweak phase transition in MSSM. 

We have also shown that cQPA coherence correlators are related to {\em squeezed states} in the operator formalism language. Indeed, we have found that the on-shell coherence functions cannot be related to any number states, but a one-to-one connection exists between the coherence functions and the parameters of a squeezed state. We also showed that flavoured squeezing corresponds to flavoured Bogolyubov transformations, generalizing the usual squeezing formalism to the flavour-mixing case.

Our formalism has several interesting applications. First, a study of thermal leptogenesis in the resonant regime using cQPA formalism is in progress. Second, we are working on to derive transport equations for the usual neutrino-mixing scenarios from first principles using our formalism.
Third, it will be interesting to generalize the simple toy-model for the quenching EWBG-transition introduced in~\cite{HKR4} to the more realistic flavour-mixing case.  Fourth, we are working on to extend our results to stationary, planar symmetric problems, which are more directly relevant for traditional EWBG models.

%
\section*{Acknowledgments}
%

This work is supported by the Alexander von Humboldt Foundation and by the Gottfried Wilhelm 
Leibniz programme of the Deutsche Forschungsgemeinschaft.

%
\begin{appendix}
%

%
\section{Gradient corrections to the phase space structure of $\Delta^{<,>}(k,t)$}
\label{sec:appendix}
%

In this section we examine the first order gradient corrections $\sim \partial_t m$ to the phase space structure of the scalar Wightman functions $\Delta^{<,>}(k,t)$. We also find generalizations of the relations (\ref{RelationEq}) between the moments $\rho_n$ and the on-shell functions $f_\alpha$, which can be used to obtain order ${\cal O}(\Gamma \partial_t m)$ corrections to the collision integrals (\ref{Scalar_real_CT1}-\ref{Scalar_real_CT3}). 

Using both KB-equations (\ref{ConstraintEq}-\ref{EvoEq}) and keeping the first order mass gradients we obtain a generalization of the (algebraic) zeroth-order constraint equation (\ref{shell_constraint}):
\begin{align} \nonumber
0 =& k_0 \left((k_0^2-\mathbf{k}^2-M^2)k_0^2 + (\Delta m \bar m )^2 \right)i\Delta^{<,>}_{\rm d} \\ \nonumber
&-\frac{1}{2}\left(k_0^2(\Delta\omega' \bar\omega + \bar\omega' \Delta\omega) + \Delta\omega \bar\omega (\Delta\omega'\Delta\omega +\bar\omega' \bar\omega)\right)\Delta^{<,>}_{\rm d} \\
&- \left(k_0^3(\Delta\omega' \bar\omega + \bar\omega' \Delta\omega) - k_0\Delta\omega \bar\omega (\Delta\omega'\Delta\omega +\bar\omega' \bar\omega)\right)\partial_{k_0}\Delta^{<,>}_{\rm d}\,.
\label{constraint_corr}
\end{align}
Because of the $k_0$-derivatives $\partial_{k_0}\Delta^{<,>}_{\rm d}(k,t)$, this equation cannot be solved by a simple singular phase-space structure  like Eq.~(\ref{Dist}). To proceed, we integrate over $k_0$ to obtain the following relations for the moments $\rho_n$:
\begin{eqnarray}\label{momentshell} \nonumber
\rho_n &=& \rho_{n-2}(\Delta\omega^2 + \bar\omega^2) - \rho_{n-4}\Delta\omega^2\bar\omega^2 \\
&& +\frac{i (2n-5)}{2}\rho_{n-3}(\Delta\omega'\bar\omega + \bar\omega'\Delta\omega) -\frac{i(2n-7)}{2}\rho_{n-5}\Delta\omega\bar\omega(\Delta\omega'\Delta\omega + \bar\omega'\bar\omega)\,.
\end{eqnarray}
Clearly, the moments of the singular (zeroth order) correlator (\ref{Dist}) satisfy the zeroth-order terms of this equation. To solve Eq.~(\ref{momentshell}) up to first order, we make an ansatz for the Wightman functions $\Delta^{<,>}_{\rm d}$, where a linear correction is parametrized as an expansion in derivatives of (zeroth order) phase-space delta functions:
\begin{equation}\label{gradientcorr}
\Delta^{<,>}_{\rm d}(k,t) = \Delta^{<,>}_{d0}(k,t) + \sum \limits_{n\pm} \partial_{k_0}^n (\pm a_{n\pm} 
\delta(k_0 \mp \bar{\omega}) + b_{n\pm}\delta(k_0 \mp \Delta\omega))\,,
\end{equation}
where $\Delta^{<,>}_{d0}(k,t)$ is the zeroth-order solution (\ref{Dist}), and $a_{n\pm}$ and $b_{n\pm}$ are of order $\partial_t m$. Parameters $a_{n\pm}$ and $b_{n\pm}$ can be determined by computing the moments $\rho_n$ from the ansatz (\ref{gradientcorr}) and using them in Eq.~(\ref{momentshell}). We find that only first- and second-order $k_0$-derivatives of the delta functions are required to parametrize the first-order gradient corrections:
\begin{align}\nonumber \label{linearshell}
 i\Delta^{<,>}_{ij}(k,t) = \frac{\pi \bar\omega_{ij}}{\omega_i \omega_j} \sum_\pm \bigg[&\pm f^{m <,>}_{\pm}
\left( 1 + \frac{i}{2}\frac{\Delta\omega'\bar\omega-\bar\omega'\Delta\omega}{\bar\omega^2-\Delta\omega^2}\partial_{k_0} \pm \frac{i}{4}\Delta\omega'\partial^2_{k_0}\right) \delta(k_0 \mp \bar\omega) 
\\
+& f^{c <,>}_{\pm}\left( 1 + \frac{i}{2}\frac{\Delta\omega'\bar\omega-\bar\omega'\Delta\omega}{\bar\omega^2-\Delta\omega^2}\partial_{k_0} \pm \frac{i}{4}\bar\omega'\partial^2_{k_0}\right) \delta(k_0 \mp \Delta\omega)\bigg]_{ij} \,.
\end{align}
As a distribution, \ie if Eq.~(\ref{linearshell}) is multiplied by a smooth test function, it is possible to replace the derivatives of the delta-functions by a Gaussian distribution function with a complex width and a shift in the expectation value. This can be shown straightforwardly by comparing the resulting moment-integrals up to the first-order terms. Therefore, we can parametrize the first-order Wightman function equivalently as   
\begin{align}\nonumber\label{linearshell_gaussian}
 i\Delta^{<,>}_{ij}(k,t) = \frac{\pi \bar\omega_{ij}}{\omega_i \omega_j} \sum_\pm \bigg[\pm& f^{m <,>}_{\pm}\,\mathcal{N}\left(\pm \bar\omega - \frac{i}{2}\frac{\Delta\omega'\bar\omega-\bar\omega'\Delta\omega}{\bar\omega^2-\Delta\omega^2}, \pm \frac{i}{2}\Delta\omega'\right) \\
+& f^{c <,>}_{\pm}\,\mathcal{N}\left(\pm \Delta\omega - \frac{i}{2}\frac{\Delta\omega'\bar\omega-\bar\omega'\Delta\omega}{\bar\omega^2-\Delta\omega^2}, \pm \frac{i}{2}\bar\omega'\right) \bigg]_{ij}\,,
\end{align}
where
\begin{equation}
 \mathcal{N}(\mu,\sigma^2) \equiv \frac{1}{\sqrt{2\pi}\sigma}\rm{exp}\left(-\frac{(k_0-\mu)^2}{2\sigma^2}\right)\,.
\end{equation}
is the (Gaussian) normal distribution function. According to the expectations, we have found that the first-order gradient terms give rise to a {\em finite width} for the Wightman functions $\Delta^{<,>}_{\rm d}(k,t)$, which is directly proportional to $\partial_t m$. Also, a shift to the positions of the ``shells'' (also proportional $\partial_t m$) is induced.  

The extended phase-space structure (\ref{linearshell}) (or (\ref{linearshell_gaussian})) for $\Delta^{<,>}(k,t)$ allows us to compute gradient corrections to the relations (\ref{RelationEq}) and consequently to the collision integrals (\ref{Scalar_real_CT1}-\ref{Scalar_real_CT3}).
First, by using Eq.~(\ref{linearshell}) in the kinetic equation (\ref{EvoEq}) and taking any four independent moments (or equivalently using the moment equations (\ref{EvoEqMoments})) we obtain the equations
\begin{eqnarray}
\partial_t f^{m <,>}_{\pm} \pm 2i\Delta\omega f^{m <,>}_{\pm} + \frac{\Delta\omega ( \Delta\omega' \bar\omega-\bar\omega'\Delta\omega )}{\bar\omega(\bar\omega^2-\Delta\omega^2)}f^{m <,>}_{\pm} &=& 0,
\nonumber\\
\partial_t f^{c <,>}_{\pm} \pm 2i\bar\omega f^{c <,>}_{\pm} + \frac{\Delta\omega ( \Delta\omega' \bar\omega-\bar\omega'\Delta\omega )}{\bar\omega(\bar\omega^2-\Delta\omega^2)}f^{c <,>}_{\pm} &=& 0\,,
 \label{firstorder}
\end{eqnarray}
which are valid up to order ${\cal O}(\partial_t m)$ (here we neglect the gradients of the mixing matrix $U$ for simplicity).
By taking the moments (\ref{linearshell}) and using Eqs.~(\ref{firstorder}) to relate $\partial_t \rho_0$ to $f_\alpha$ we then obtain the following generalization of the invertible relations (\ref{RelationEq}):
\begin{equation} 
 \left(\begin{array}{c}
	\rho_0 \\ \partial_0\rho_0 \\ \rho_1 \\ \rho_2
 \end{array}\right)_{ij}
 = \frac{\bar\omega_{ij}}{2 \omega_i \omega_j} \left[
 \left(\begin{array}{cccc}
	1 & -1 & 1 & 1 \\
	-2i\Delta\omega & -2i\Delta\omega & -2i\bar\omega & 2i\bar\omega \\
	\bar\omega & \bar\omega & \Delta\omega & -\Delta\omega \\
	\bar\omega^2 & -\bar\omega^2 & \Delta\omega^2 & \Delta\omega^2 
 \end{array}\right) + \chi^{\prime} \,
\right]_{ij}
\left(\begin{array}{c}
	f^m_{+} \\ f^m_{-} \\ f^c_{+} \\ f^c_{-}
\end{array}\right)_{ij}
\end{equation}
with
\begin{equation}
\chi^{\prime}_{ij} \equiv  
 \left(\begin{array}{cccc}
	0&0&0&0 \\
	D_1&-D_1&D_1&D_1\\
	D_2&-D_2&D_2&D_2\\
	\,\,2D_2\bar\omega+\frac{i}{2}\Delta\omega'\,\,&\,\,2D_2\bar\omega+\frac{i}{2}\Delta\omega'\,\,&\,\,2D_2\Delta\omega+\frac{i}{2}\bar\omega'\,\,&\,	\,-2D_2\Delta\omega-\frac{i}{2}\bar\omega'\,\,
 \end{array}\right)_{ij}\,, 
\end{equation}
where 
\begin{equation}
D_1 \equiv -\frac{\bar\omega'\bar\omega-\Delta\omega'\Delta\omega}{\bar\omega^2-\Delta\omega^2}\qquad {\rm and}\qquad D_2 \equiv -i\frac{\Delta\omega'\bar\omega - \bar\omega'\Delta\omega}{2(\bar\omega^2-\Delta\omega^2)}\,.
\end{equation}
These relations can be used to obtain order ${\cal O}(\Gamma \partial_t m)$ corrections to the collision integrals (\ref{Scalar_real_CT1}-\ref{Scalar_real_CT3}). However, the corrections of this order involve also direct derivative terms $\sim \partial_t \Gamma \sim \Gamma \partial_t m$ and the mixing gradient terms $\sim \Theta^\prime \equiv i U\partial_t U^\dagger$, as well as corrections to the resummation of the oscillatory coherence terms in the collision integrals, which have not been considered here. Note that the flow terms of Eqs.~(\ref{EvoEqMoments}) obtain no gradient corrections as they are written in terms of the moments.

\end{appendix}
%

%
%

\end{document}